\documentclass[journal,draftclsnofoot,onecolumn,12pt]{IEEEtran}
\ifCLASSINFOpdf
  \usepackage[pdftex]{graphicx}
\else
  \usepackage[dvips]{graphicx}
\fi
\usepackage{float}
\usepackage{amsmath,amsfonts}
\usepackage{algorithm}
\usepackage{algpseudocode}
\usepackage{lipsum}

\makeatletter
\newenvironment{breakablealgorithm}
  {
   \begin{center}
     \refstepcounter{algorithm}
     \hrule height.8pt depth0pt \kern2pt
     \renewcommand{\caption}[2][\relax]{
       {\raggedright\textbf{\ALG@name~\thealgorithm} ##2\par}%
       \ifx\relax##1\relax 
         \addcontentsline{loa}{algorithm}{\protect\numberline{\thealgorithm}##2}%
       \else 
         \addcontentsline{loa}{algorithm}{\protect\numberline{\thealgorithm}##1}%
       \fi
       \kern2pt\hrule\kern2pt
     }
  }{
     \kern2pt\hrule\relax
   \end{center}
  }
\makeatother

\usepackage{array}
\usepackage[caption=false]{subfig}
\usepackage{textcomp}
\usepackage{stfloats}
\usepackage{amsthm}
\usepackage{amssymb}
\newtheorem{lemma}{Lemma}
\newtheorem{theorem}{Theorem}
\newtheorem{definition}{Definition} 
\newtheorem{proposition}{Proposition}
\theoremstyle{definition}
\newtheorem{example}{Example}
\usepackage{url}
\usepackage{verbatim}
\usepackage{mathrsfs}
\usepackage{cite}

\usepackage{bm}
\usepackage{tikz}
\usepackage{pgfplots}
\pgfplotsset{compat=newest}
\usepackage{ifthen}
\usetikzlibrary{arrows.meta, shapes.geometric, decorations.pathmorphing, backgrounds, positioning, fit, petri, calc}
\usepackage{pgffor}
\usetikzlibrary{plotmarks}
\usepgfplotslibrary{patchplots}
\usepackage{grffile}
\usepackage{tablefootnote}
\usepackage{nicematrix}
\usepackage[multiple]{footmisc}

\begin{document}

\title{A Universal List Decoding Algorithm with Application to Decoding of Polar Codes}

\author{Xiangping Zheng and Xiao Ma,~\IEEEmembership{Member,~IEEE}
\thanks{Corresponding author: Xiao Ma.}
\thanks{The authors are with the School of Computer Science and Engineering, and also with
the Guangdong Key Laboratory of Information Security Technology, Sun Yat-sen University,  Guangzhou 510006, China (e-mail: zhengxp23@mail2.sysu.edu.cn; maxiao@mail.sysu.edu.cn).}}


\markboth{Journal of \LaTeX\ Class Files,~Vol.~1, No.~2, December~2023}%
{Shell \MakeLowercase{\rmit{et al.}}: A Sample Article Using IEEEtran.cls for IEEE Journals}


\maketitle

\begin{abstract}
This paper is concerned with a guessing codeword decoding~(GCD) of linear block codes, which is optimal and typically requires a fewer number of searches than the naive exhaustive search decoding~(ESD). Compared with the guessing noise decoding~(GND), which is only efficient for high-rate codes, the GCD is efficient for not only high-rate codes but also low-rate codes. We prove that the GCD typically requires a fewer number of queries than the GND. Compared with the conventional ordered statistics decoding~(OSD), the GCD does not require the online Gaussian elimination~(GE). In addition to limiting the maximum number of searches, we suggest limiting the radius of searches in terms of soft weights or tolerated performance loss to further reduce the decoding complexity, resulting in the so-called truncated GCD. The performance gap between the truncated GCD and the optimal decoding can be upper bounded approximately by the saddlepoint approach or other numerical approaches. The derived upper bound captures the relationship between the performance and the decoding parameters, enabling us to balance the performance and the complexity by optimizing the decoding parameters of the truncated GCD. We also introduce a parallel implementation of the (truncated) GCD algorithm to reduce decoding latency without compromising performance. Another contribution of this paper is the application of the GCD to the polar codes.
We propose a multiple-bit-wise decoding algorithm over a pruned tree for the polar codes, referred to as the successive-cancellation list (SCL) decoding algorithm by GCD. First, we present a strategy for pruning the conventional polar decoding tree based on the complexity analysis rather than the specific bit patterns. Then we apply the GCD algorithm in parallel aided by the early stopping criteria to the leaves of the pruned tree. Simulation results show that, without any performance loss as justified by analysis, the proposed decoding algorithm can significantly reduce the decoding latency of the polar codes. 
\end{abstract}


\section{Introduction}




\IEEEPARstart{T}he ultra-reliable low-latency communication~(URLLC), enabling applications that require high reliability and very low latency, is an important use case of beyond 5G and 6G communication networks~\cite{ref2ebmm,ref2URLLC,ITU_R},
where URLLC in 6G requires a significantly lower end-to-end latency~(25$\mu s \sim$ 1$ms$) and a high level of transmission reliability~(block error rates of $10^{-5}$ to $ 10^{-7}$), compared with the 5G new radio~(NR)~\cite{20216Greview}. To meet the stringent requirements of URLLC, the utilization of short block codes with efficient decoding algorithms has rekindled a great deal of interest~\cite{ShortImpact2019}. 

As a near maximum-likelihood~(ML) decoding algorithm for short linear block codes, ordered statistics decoding~(OSD)~\cite{DorshOSD, Fossorier1995} produces a list of codewords by querying and re-encoding the patterns in the most reliable basis~(MRB), where the querying process is implemented in an order of non-decreasing soft weight~\cite{DorshOSD} or Hamming weight~\cite{Fossorier1995}. The OSD is universal and applicable to any short linear block codes~(from low rates to high rates). To reduce the complexity in terms of the number of queries, several improved OSD algorithms have been proposed, such as  segmentation-discarding OSD~(SD-OSD)~\cite{yue2019segmentation}, probability-based OSD~(PB-OSD)~\cite{yue2021probability} and linear-equation OSD~(LE-OSD)~\cite{yue2022linear}. Recently, a variant of OSD called OSD with local constraints~(LC-OSD) has been proposed in~\cite{LC_OSD2022,LC_OSDljf2023} to further reduce the number of queries. The basic idea of LC-OSD is to select $K+\delta$ bits as the extended MRB. However, Gaussian elimination~(GE) is required by OSD for each reception of block. Even worse, the GE is a serial algorithm and no efficient parallel implementation is available for a general matrix. The inevitable decoding latency and complexity caused by the GE hinder the application of OSD in URLLC. One way to solve this issue is to skip the GE process by precalculating and storing multiple systematic generator matrices~\cite{choi2021fast} or invoking specific decoding conditions~\cite{yue2022ordered}.  Actually, by pre-storing a systematic generator matrix, low-complexity reduced GE is sufficient for performing modified OSD~\cite{fossorier2024modified}. Another way is to relax the requirement of the MRB, leading to the representative OSD~(ROSD) of the staircase matrix codes~\cite{ROSD}~\cite{wang2024random}, where the GE can be implemented in parallel, and the quasi-OSD~(QOSD) of Reed-Solomon~(RS) codes~\cite{QuasiOSD}, where the GE can be replaced by the parallel Lagrange interpolation\footnote{
The Lagrange interpolation was also used to parallelize the GE in the OSD of Bose-Chaudhuri-Hocquenghem~(BCH) codes~\cite{ChenBCH2022}, where extended systematic generator matrix are formed not for BCH codes themselves but for the corresponding RS codes.}.





In contrast to the OSD, the guessing random additive noise decoding~(GRAND) algorithm~\cite{ref7Gran,ref8gran} guesses the error patterns from the most likely to the least likely to find the correct codeword, which was also mentioned in the introductory paragraph of~\cite{Fossorier2001}. Several variants were proposed for improving the GRAND algorithm, such as the soft-GRAND~(SGRAND)~\cite{SGRAND}, GRAND with symbol reliability information~(SRGRAND)~\cite{ref9SRgrand}, ordered reliability bits GRAND~(ORBGRAND)~\cite{orbgrand}, and partially constrained-GRAND~(PC-GRAND)~\cite{wang2023PCGRAND}. However, the GRAND algorithms are effective~(in terms of complexity) only for the short-length and high-rate linear block codes~\cite{orbgrand}. 

As a class of capacity-achieving codes for binary-input output-symmetric discrete-memoryless channels~(BIOS-DMCs), polar codes~\cite{ref1Arikan} have been employed for 5G systems for use in control channels~\cite{ref2ebmm,ref2URLLC}.
Under the successive cancellation (SC) decoding algorithm,  polar codes can approach the maximum likelihood~(ML) performance when the code length tends to infinity~\cite{ref1Arikan}. To enhance the decoding performance in the finite code length regime, the SC list~(SCL) decoding algorithm~\cite{ref3SCL} and the cyclic redundancy check~(CRC) aided SCL~(CA-SCL) decoding algorithm~\cite{ref3SCLIT, ref4CASCL} were proposed, which can achieve satisfactory performance. The main issue of the SCL decoding algorithm is the decoding latency since it is a sequential bit-by-bit decoding algorithm. This becomes clear if one is aware that a polar code can be represented by a rooted binary tree, over which the SCL decoding can be implemented by a pre-order traversal and a post-order traversal~\cite{AlamdarSSC}. One main approach to reducing the decoding latency is pruning
the tree by merging multiple leaf nodes with a predefined bit pattern into a single node~(a subcode) for decoding, such as rate-0/rate-1 nodes~\cite{AlamdarSSC, HashemiSSCL}, and single-parity-check~(SPC)/repetition~(REP) nodes~\cite{SarkisFSSC}. Later, Sarkis \emph{et al.}~\cite{SarkisFASTsclL} proposed a fast list decoder for the rate-1 and SPC nodes, which is an empirical approach and does not guarantee the same performance as the conventional SCL decoding. To achieve the same performance as the conventional SCL decoder, Hashemi \emph{et al.}~\cite{HashemiFSSCL} proposed a fast list decoder for the rate-1 and SPC nodes, where the required number of path splitting~(related to the list size) is derived. To further reduce the decoding latency, five new types of nodes, namely, Type-I, Type-II, Type-III, Type-IV, and Type-V nodes in the decoding tree were introduced in~\cite{HanifFSSC} and the corresponding fast list decoders were designed in~\cite{2019ArdakaniFastSCL} based on the works in~\cite{HashemiFSSCL}. 
Most of the aforementioned nodes~\cite{AlamdarSSC,HashemiSSCL,SarkisFSSC,SarkisFASTsclL,HashemiFSSCL,HanifFSSC,2019ArdakaniFastSCL} can be treated as special cases of the generalized REP~(G-REP) nodes and generalized parity-check~(G-PC) nodes~\cite{CondoFSSC}, which can be further generalized to the sequence repetition (SR) nodes~\cite{Zheng2020,ZhengSRnodes2021}. Recently, Ren \emph{et al.}~\cite{RenFastSCL2022} proposed an SR-list decoding algorithm that can be divided into two phases: enumerating all information bit possibilities for the low-rate part and sequential node-based list decoding for the high-rate part.


In this paper, the OSD-like algorithms are referred to as the guessing codeword decoding~(GCD) algorithms, while the GRAND-like algorithms are referred to as the guessing noise decoding~(GND) algorithm since they are also applicable to other noisy channels after transformation. Recently, it has been proved in~\cite{ma2024guessing} that the GCD algorithm is an ML decoding algorithm and can be more efficient than the GND algorithm, where the GCD produces a list of codewords by re-encoding patterns in increasing soft weight order over an information set~(selected offline) instead of the MRB. For this reason, we focus on the GCD algorithm without online GE and develop an optimal list decoding~\cite{elias1957list} that lists the $L$ most likely codewords as output. We present three conditions for truncation, which can be optimized to reduce the complexity in terms of the number of queries with negligible performance loss, referred to as truncated GCD. Moreover, we propose a parallel (truncated) GCD to reduce the decoding latency without sacrificing performance. By the use of the GCD, we propose a multiple-bit-wise SCL decoding algorithm for polar codes by embedding GCD into a pruned tree to reduce the decoding latency. The main contributions of this paper are summarized as follows.
\begin{enumerate}
    \item We prove by analysis that the GCD typically requires a fewer number of queries than the GND, indicating that the GCD is more efficient than the GND since the complexity per query is comparable.
    \item Theoretically, we estimate the upper bound on the performance gap between the truncated GCD and the GCD, which does not rely on any specific code. We also propose to utilize the saddlepoint technique~\cite{saddle2011,saddle2018} to speed up the calculation of the performance gap.   
    \item  The estimated performance gap relates the performance
of the truncated GCD to the decoding parameters, such as the maximum query number, providing guidelines  on the choices of the proposed three conditions. To further reduce the decoding latency, we consider performing the (truncated) GCD in parallel.
    \item For practical application to polar decoding, we first present a strategy for pruning the polar decoding tree based on the complexity analysis. Distinguished from existing strategies, which depend on certain specific bit patterns, our strategy does not depend on any specific bit patterns. Then we present the multiple-bit-wise successive cancellation list~(SCL) decoding algorithm by GCD, where we apply the GCD algorithm in parallel aided by the early stopping criteria to the low/high-rate sub-codes. The proposed decoding algorithm is analyzed in comparison with the SCL decoding algorithm, suggesting no performance loss. 
\end{enumerate}

The rest of this paper is organized as follows. In Section~II, we illustrate the system model and analyze the efficiency of the GCD. In Section~III, we provide the complexity analysis and present four variants of GCD to reduce the complexity and latency. In Section~IV, we apply the GCD in polar decoding and present a multiple-bit-wise SCL decoding algorithm for polar codes over a pruned tree, namely the SCL decoding algorithm by GCD. Section~V concludes this paper.















\section{Guessing Codeword Decoding}\label{OESD}
\subsection{Problem Statement}
Let $\mathbb{F}_2=\{0,1\}$ be the binary field and $\mathscr{C}[N,K]$ be a binary linear block code of dimension $K$ and length $N$. The binary linear block code $\mathscr{C}[N,K]$ can be specified either by a generator matrix $\mathbf{G}$ of size $K \times N$ or a parity-check matrix $\mathbf{H}$ of size $(N-K) \times N$. Associated with an information vector $\boldsymbol{u} \in \mathbb{F}_2^{K}$ is a codeword $\boldsymbol{c} = \boldsymbol{u} \mathbf{G}$, which must satisfy $\boldsymbol{c}\mathbf{H}^{T}=\mathbf{0}$, where $\mathbf{H}^{T}$ is the transpose of $\mathbf{H}$. Now suppose that $\boldsymbol{c} \in \mathbb{F}_2^{N}$ is transmitted over a binary input discrete-time memoryless channel~(B-DMCs), resulting in  $\boldsymbol{y} \in \mathcal{Y}^N$, where $\mathcal{Y}$ is the alphabet of the channel outputs.  In this paper, the components of a vector, say $\boldsymbol{y}$, are indexed from 1 to its length. We interchangeably use $y_i$ and $\boldsymbol{y}[i]$ to represent the $i$-th component of $\boldsymbol{y}$.

Upon receiving  $\boldsymbol{y}$, the log-likelihood ratio (LLR) vector $\boldsymbol{r}$ is calculated as
\begin{equation}\label{eq1}
    r_i = \ln{\frac{P_{Y|C}(y_i|c_i=0)}{P_{Y|C}(y_i|c_i = 1)}},\ 1 \leq i \leq N,
\end{equation}
where $P_{Y|C}(\cdot|\cdot)$ is the conditional probability mass~(or density) function specifying the channel. Given the LLR vector $\boldsymbol{r}$, the hard-decision vector $\boldsymbol{z} \in \mathbb{F}_2^{N}$ is calculated as
\begin{equation}\label{harddecison}
	z_i=
	\begin{cases} 
		0, & \mbox{if }r_i\geq0\\
		1, & \mbox{if }r_i<0\\
	\end{cases}
	,~1\leq i \leq N.
\end{equation}

The ML decoding is to find a codeword $\boldsymbol{v}^*$ such that
\begin{equation}
 \boldsymbol{v}^*= \mathop{\rm{argmax}}\limits_{\boldsymbol{v}\in \mathscr{C}} P_{Y|C}(\boldsymbol{y}|\boldsymbol{v}),
\end{equation}
which is equivalent to
\begin{equation}
 \boldsymbol{v}^*= \mathop{\rm{argmin}}\limits_{\boldsymbol{v}\in \mathscr{C}} \log \frac{P_{Y|C}(\boldsymbol{y}|\boldsymbol{z})}{P_{Y|C}(\boldsymbol{y}|\boldsymbol{v})}.
\end{equation}
For a test vector $\boldsymbol{v} \in \mathbb{F}_2^{N}$, we can define its corresponding test error pattern~(TEP) $\boldsymbol{e}\in \mathbb{F}_2^N$ as
\begin{equation}
    \boldsymbol{e} \triangleq \boldsymbol{z}-\boldsymbol{v}.
\end{equation}
This can be written as $\boldsymbol{z}= \boldsymbol{v}+\boldsymbol{e}$ and hence the channel is transformed into an additive noise channel, which accepts the codeword as input and delivers the hard-decision vector as output. Defining the soft weight\footnote{This can be viewed as a kind of discrepancy~\cite{GallagerLDPC1963}, which was referred to as confidence level~\cite{DorshOSD} and ellipsoidal weight~\cite{valembois2002comparison}.} of a TEP $\boldsymbol{e}$, denoted by $\gamma(\boldsymbol{e})$, as
\begin{equation}\label{softweight}
\gamma(\boldsymbol{e}) \triangleq \log  \frac{P_{Y|C}(\boldsymbol{y}|\boldsymbol{z})}{P_{Y|C}(\boldsymbol{y}|\boldsymbol{z}-\boldsymbol{e})} = \sum\limits_{i=1}^{N}e_i |r_i|,
\end{equation}
we can see that $\gamma(\boldsymbol{e}) \geq 0$ for any vector $\boldsymbol{e} \in \mathbb{F}_2^{N}$ and $\gamma(\boldsymbol{e}) = 0$ for $\boldsymbol{e} = \boldsymbol{0}$. In contrast to the Hamming weight, the soft weight of $\boldsymbol{e}$, as a weighted sum, is determined not only by its non-zero components but also by the corresponding reliabilities $|r_i|$ with $e_i = 1$.  Then the ML decoding is equivalent to the lightest soft weight decoding. That is,  the ML decoding is equivalent to
\begin{equation}
\begin{aligned}
    \min\limits_{\boldsymbol{e}\in \mathbb{F}_2^N} \quad &\gamma(\boldsymbol{e})\\
    \mbox{s.t.}\quad & \boldsymbol{e}\mathbf{H}^T=\boldsymbol{s},
\end{aligned}
\end{equation}
where $\boldsymbol{s}=\boldsymbol{z}\mathbf{H}^T$ is the computable syndrome.

Without loss of generality, we assume that the first $N-K$ columns of $\mathbf{H}$ are linearly independent. That is, $\mathbf{H}$ can be transformed by elementary row operations into a systematic form,
\begin{equation}
    \mathbf{H} \rightarrow [\mathbf{I}\quad \mathbf{P}],
\end{equation}
where $\mathbf{I}$ is the identity matrix of order $N-K$ and $\mathbf{P}$ is a matrix of size $(N-K) \times K$. Then a TEP $\boldsymbol{e}$ can be written as $\boldsymbol{e}= (\boldsymbol{e}_I,\boldsymbol{e}_P)$, where $\boldsymbol{e}_I \in \mathbb{F}_2^{N-K}$ and $\boldsymbol{e}_P \in \mathbb{F}_2^K$. Similarly, $\boldsymbol{r}= (\boldsymbol{r}_I,\boldsymbol{r}_P)$ and $\boldsymbol{z}= (\boldsymbol{z}_I,\boldsymbol{z}_P)$. We see that, for any valid TEP $\boldsymbol{e}$, $\boldsymbol{e}_I$ is uniquely determined by $\boldsymbol{e}_P$ since $\boldsymbol{e}_I + \boldsymbol{e}_P\mathbf{P}^T = \boldsymbol{s}$. Hereafter, by a valid TEP $\boldsymbol{e}$, we mean a vector $\boldsymbol{e}$ such that $\boldsymbol{z}-\boldsymbol{e}$ is a codeword. A valid TEP $\boldsymbol{e}$ is referred to as the true TEP~(TrTEP) if $\boldsymbol{z}-\boldsymbol{e}$ is the transmitted codeword. 

Now, for any $1 \leq L \leq 2^K$, we consider the optimal list decoding that finds the $L$ lightest valid TEPs, collectively denoted as $\mathcal{L}$. Not surprisingly, this can be achieved by exhaustive search decoding~(ESD), which sorts all $2^K$ TEPs and has a complexity of order $\mathcal{O}(2^K)$. 
One approach to reducing the complexity is limiting the search space. If so,  we need to answer the following three questions:
\begin{enumerate}
    \item Search space: for what TEPs to search?
    \item Search order: in which order to search?
    \item Termination: under which condition to terminate the search?
\end{enumerate}
Intuitively, the search space should be as small as possible but contain the TrTEP as often as possible. Given the search space, the search order~(known as strategy~\cite{valembois2002comparison}) should query the TrTEP as early as possible. Given a search order, the search should be terminated when further searches are unnecessary. 

We aim to design a list decoding with low complexity such that the error probability of the list decoding is as small as possible. Here, following~\cite{elias1957list}~\cite[Exercise 5.20]{Gallager1968}, we say that  a list-decoding error has occurred if the TrTEP is not in the list of the decoding output. For a list decoding with list size $L$, we denote by $\epsilon(L)$ the probability of a list-decoding error. Specifically, $\epsilon(1)$ is the frame error rate~(FER) of the ordinary decoding algorithm.

In this paper,  we present the GCD algorithm and then introduce three conditions for truncation to further reduce the complexity. For the time being, we assume that an ordered~(partial) TEP generator is available for all partial TEPs $\boldsymbol{e}_P \in \mathbb{F}_2^K$ that delivers $\boldsymbol{e}_P^{(i)}$ before $\boldsymbol{e}_P^{(j)}$ if $\gamma\left(\boldsymbol{e}_P^{(i)}\right) < \gamma\left(\boldsymbol{e}_P^{(j)}\right)$ or $\gamma\left(\boldsymbol{e}_P^{(i)}\right) = \gamma\left(\boldsymbol{e}_P^{(j)}\right)$ but $\boldsymbol{e}_P^{(i)}$ is prior to $\boldsymbol{e}_P^{(j)}$ in the lexicographic order.  Then a sequence of  $\boldsymbol{e}_P \in \mathbb{F}_2^K$ can be produced~(on demand) such that 
\begin{equation}\label{GCDSort}
 \gamma\left(\boldsymbol{e}_P^{(1)}\right) \leq \gamma\left(\boldsymbol{e}_P^{(2)}\right) \leq \cdots \leq \gamma\left(\boldsymbol{e}_P^{(\ell)}\right) \leq \cdots \leq \gamma\left(\boldsymbol{e}_P^{(2^K)}\right).   
\end{equation}

In principle, the ordered TEP generator can be implemented, say, with the aid of the flipping pattern tree~(FPT)~\cite{FPTtang}\cite{SGRAND}~\cite{yue2021probability}~\cite{LC_OSDljf2023}, which will be described for integrity in Section III-A. Given the sorted partial TEPs~\eqref{GCDSort}, the GCD as described in Algorithm 1 finds the $L \geq 1$ lightest valid TEPs~(corresponding to the $L$ best codewords). We also illustrate in Fig.~1 the update process of the linked list $\mathcal{L}$ in Algorithm~1. 

\begin{algorithm}[!t]
\renewcommand{\algorithmicrequire}{\textbf{Input:}}
    \renewcommand{\algorithmicensure}{\textbf{Output:}}
\caption{GCD}\label{alg:alg1}
\begin{algorithmic}[1]
\Require The parity-check matrix $[\mathbf{I}\quad \mathbf{P}]$, the LLR vector $\boldsymbol{r}$, the list size $L$, and the maximum query number $\ell_{\rm{max}}=2^K$.
\State Initialization: $\mathcal{L}=\{\boldsymbol{f}^{(i)},1\leq i \leq L\}$ is a linked list of size $L$ which is maintained in order~(non-decreasing soft weight) during the query process and the initial $L$ elements in $\mathcal{L}$ are set to NULL TEPs with soft weight $\infty$.
\For{$\ell = 1,\dots,\ell_{\rm{max}}$}
\State Generate the $\ell$-th lightest partial TEP $\boldsymbol{e}_P^{(\ell)}$.
\If{$\gamma\left(\boldsymbol{e}_P^{(\ell)}\right) \geq \gamma\left(\boldsymbol{f}^{(L)}\right)$}
\State \textbf{break}.
\Else
\State $\boldsymbol{e}_I^{(\ell)} = \boldsymbol{s}- \boldsymbol{e}_P^{(\ell)}\mathbf{P}^T$.
\State $\boldsymbol{e}^{(\ell)} = \left(\boldsymbol{e}_I^{(\ell)} , \boldsymbol{e}_P^{(\ell)}\right)$.
\If{$\gamma\left(\boldsymbol{e}^{(\ell)}\right)<\gamma\left(\boldsymbol{f}^{(L)}\right)$}
\State Update the list $\mathcal{L}$ by removing $\boldsymbol{f}^{(L)}$ and inserting $\boldsymbol{e}^{(\ell)}$.
\EndIf
\EndIf
\EndFor
\Ensure The list $\mathcal{L}$ and the corresponding codewords $\{\boldsymbol{c}|\boldsymbol{c}=\boldsymbol{z}-\boldsymbol{e}, \boldsymbol{e}\in \mathcal{L} \}$.
 \end{algorithmic}
\label{alg1}
\end{algorithm}

\begin{figure}[!t]
\centering
\includegraphics[width=5.0in]{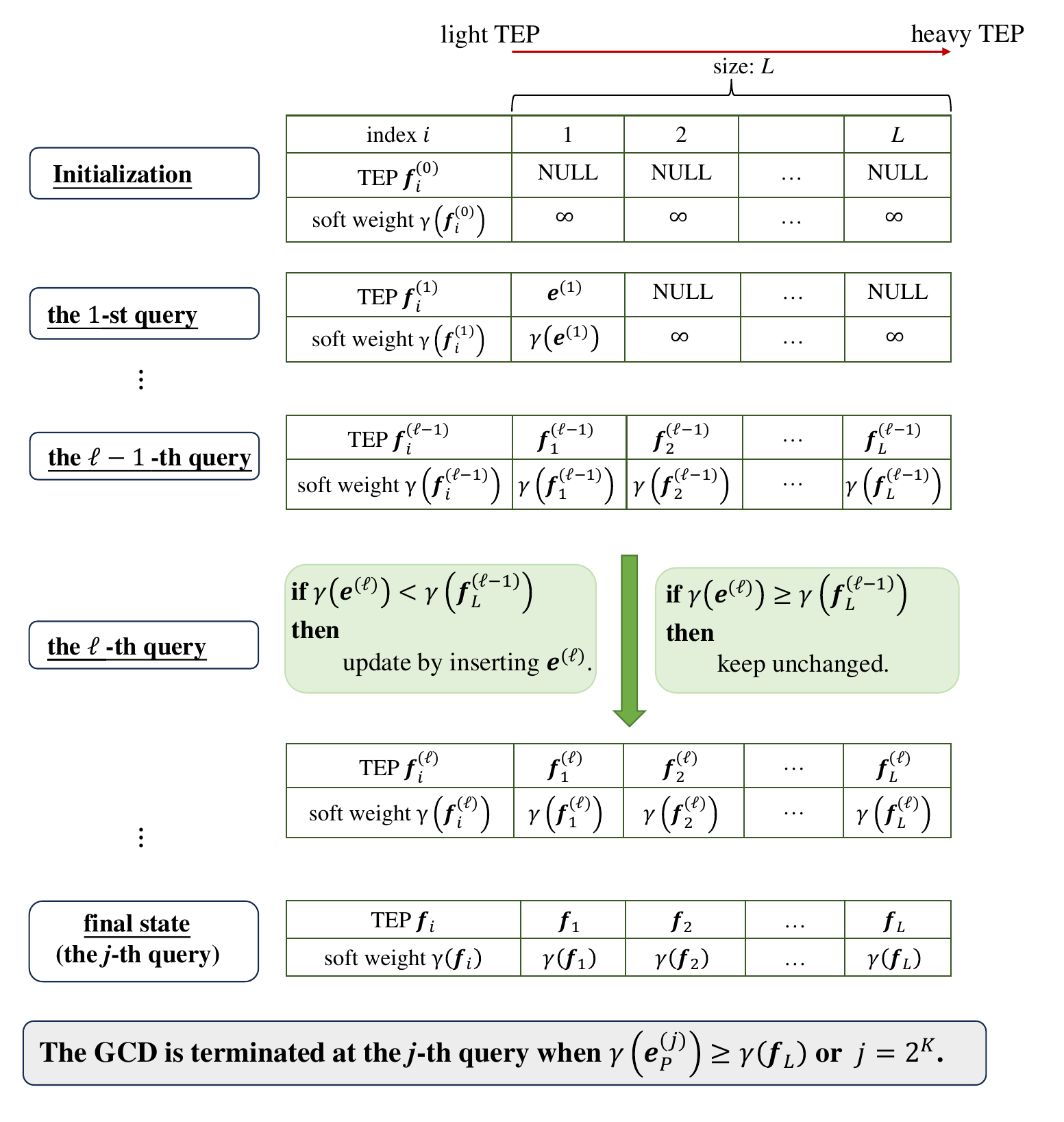}
\caption{The update process of the linked list $\mathcal{L}$.}
\label{Fig__SCLproof_example}
\end{figure}

\begin{theorem}
   The TEPs in $\mathcal{L}$ are the $L$ lightest ones and hence the output codewords in Algorithm~1 are the $L$ most likely ones. 
\end{theorem}
\begin{IEEEproof}
 Let $\boldsymbol{e}^{(\ell)} = \left(\boldsymbol{e}_I^{(\ell)}, \boldsymbol{e}_P^{(\ell)}\right)$ be the $\ell$-th queried TEP in the GCD and denote by $\mathcal{L}^{(\ell)}=\{\boldsymbol{f}^{(\ell,i)},1\leq i\leq L\}$ the linked list consisting of the $L$ lightest TEPs among the queried $\ell$ TEPs. Let $\gamma_L^{(\ell)}$ be the soft weight of the heaviest TEP in $\mathcal{L}^{(\ell)}$. That is,
\begin{equation}
\gamma_L^{(\ell)} =\max\limits_{\boldsymbol{f}\in \mathcal{L}^{(\ell)}}\gamma(\boldsymbol{f})=\gamma\left(\boldsymbol{f}^{(\ell,L)}\right).
\end{equation}

The GCD terminates at the $j$-th query in either of the following two cases, $\gamma\left(\boldsymbol{e}_P^{(j)}\right) \geq \gamma_L^{(j-1)}$ or $j=2^K$. The latter case occurs only when $\gamma\left(\boldsymbol{e}_P^{(\ell)}\right) \geq \gamma_L^{(\ell-1)}$ is not activated for all $\ell \leq 2^K$, which is actually an ESD algorithm. In this case, the delivered list $\mathcal{L}= \mathcal{L}^{\left(2^K\right)}$, which naturally consists of the $L$ lightest valid TEPs. The former case occurs when $\gamma\left(\boldsymbol{e}_P^{(j)}\right) \geq \gamma_L^{(j-1)}$ for some $j < 2^K$, which is the case of interest in terms of the complexity. In this case, the delivered $\mathcal{L}= \mathcal{L}^{(j-1)}$, which consists of the $L$ lightest valid TEPs as proved below. We have, for all unqueried TEP $\boldsymbol{e}^{(\ell)}$~($\ell \geq j$),
    \begin{equation}
    \begin{aligned}
      \gamma_L^{(j-1)} &\leq \gamma\left(\widetilde{\boldsymbol{e}}_P^{(j)}\right) && (\text{by assumption})\\
        &\leq \gamma\left(\widetilde{\boldsymbol{e}}_P^{(\ell)}\right) && (\widetilde{\boldsymbol{e}}_P^{(\ell)}\text{is non-decreasing})\\
        & \leq \gamma\left(\widetilde{\boldsymbol{e}}^{(\ell)}\right). && (\widetilde{\boldsymbol{e}}^{(\ell)} = (\widetilde{\boldsymbol{e}}_I^{(\ell)}, \widetilde{\boldsymbol{e}}_P^{(\ell)}))  
    \end{aligned}
    \end{equation}

On the other hand, the definition of $\mathcal{L}^{(j-1)}$ ensures that $\gamma_L^{(j-1)} \leq \gamma\left(\boldsymbol{e}^{(\ell)}\right)$ holds for all queried TEPs $\boldsymbol{e}^{(\ell)}$~($\ell <j$) satisfying $\boldsymbol{e}^{(\ell)} \notin \mathcal{L}^{(j-1)}$.

\end{IEEEproof}

\subsection{Guessing Codeword Versus Guessing Noise}
Assuming that an ordered TEP generator is available, the GCD algorithm generates and re-encodes partial TEPs, while the GND~(SGRAND) algorithm~\cite{SGRAND} produces the whole TEPs from most likely to least likely until a valid TEP has been found. Intuitively, the GND algorithm can be extended to an optimal list decoding algorithm, delivering the $L$ most likely codewords as output. The differences between the GND and the GCD along with their complexity per query are analyzed below.
\begin{itemize}
    \item At the $\ell$-th query, the GND generates the $\ell$-th lightest TEP $\boldsymbol{e}^{(\ell)} \in \mathbb{F}_2^N$, while the GCD generates the $\ell$-th lightest partial TEP $\boldsymbol{e}_{P}^{(\ell)} \in \mathbb{F}_2^K$. The complexity is comparable for $K \approx N$.
    \item For the $\ell$-th TEP $\boldsymbol{e}^{(\ell)}$, the GND calculates $\mathbf{H}\left(\boldsymbol{e}^{(\ell)}\right)^T $ for checking with a complexity of order $\mathcal{O}((N-K)N)$.  In contrast, the GCD calculates  $\boldsymbol{e}_{I}^{(\ell)} = \boldsymbol{s}-\boldsymbol{e}_{P}^{(\ell)}\mathbf{P}^T$ with a complexity of order $\mathcal{O}((N-K)K)$, delivering a valid TEP  $\boldsymbol{e}^{(\ell)} = (\boldsymbol{e}_{I}^{(\ell)},\boldsymbol{e}_{P}^{(\ell)})$. Since the size of the matrix $\mathbf{P}$ is smaller than that of the matrix $\mathbf{H}$, the complexity of the re-encoding in the GCD is usually lower than the complexity of the checking in the GND unless $\mathbf{H}$ is a very sparse matrix but $\mathbf{P}$ is a dense matrix.
    \item  The checking in the GND compares $\mathbf{H}\boldsymbol{e}^T $ and $\boldsymbol{s}^T$, while the checking in the GCD compares  $\gamma(\boldsymbol{e}_{P})$ and $\gamma\left(\boldsymbol{f}^{(L)}\right)$. The complexity is comparable.
    \item The GND checks TEPs with non-decreasing soft weights but generates only $L$ valid TEPs just before its termination. In contrast, the GCD re-encodes partial TEPs with non-decreasing soft weights but generates at least $L$ valid TEPs with $\gamma\left(\boldsymbol{f}^{(L)}\right)$ non-increasing.
    
\end{itemize}

The total complexity can be roughly measured by the operations per query multiplied by the number of queries. We have seen that the complexity per query for the GCD is not higher than that of the GND. Then an immediate question arises: Can a GCD be more efficient than a GND? The answer is positive, and the key is the early stopping criterion $\gamma(\boldsymbol{e}_{P})\geq \gamma(\boldsymbol{e}^*)$.

\begin{theorem}
 The number of queries for the GCD is less than or equal to the number of queries for the GND.   
\end{theorem}
\begin{IEEEproof}
Assume that $\boldsymbol{e}^* = (\boldsymbol{e}_{I}^*,\boldsymbol{e}_{P}^*)$ is the $L$-th lightest valid TEP, which is not known in advance but exists. The GND terminates eventually, and checks a list $\mathcal{L}_{\text{GND}} = \mathcal{P}_{\text{GND}} \cup \mathcal{Q}_{\text{GND}}$, where $\mathcal{P}_{\text{GND}} = \{\boldsymbol{e} \in \mathbb{F}_2^N \ |\  \gamma(\boldsymbol{e}) < \gamma(\boldsymbol{e}^*) \}$ and $\mathcal{Q}_{\text{GND}}$ is a subset of $\{\boldsymbol{e}\in \mathbb{F}_2^N \ |\  \gamma(\boldsymbol{e})=\gamma(\boldsymbol{e}^*)\}$.  In contrast, the GCD terminates with $\boldsymbol{f}^{(L)}=\boldsymbol{e}^*$ and re-encodes a list 
$\mathcal{L}_{\text{GCD}} = \mathcal{P}_{\text{GCD}} \cup \mathcal{Q}_{\text{GCD}}$, where $\mathcal{P}_{\text{GCD}}=\{\boldsymbol{e}_{P}\in \mathbb{F}_2^K \ |\ \gamma(\boldsymbol{e}_{P}) < \gamma(\boldsymbol{e}^*) \}$ and $\mathcal{Q}_{\text{GCD}}=\{\boldsymbol{e}_{P} \in \mathbb{F}_2^K\ |\  \gamma(\boldsymbol{e}_{P})  = \gamma(\boldsymbol{e}^*) \}$. The set $\mathcal{Q}_{\text{GND}}$~(if non-empty) consists of those (invalid) TEPs  $\boldsymbol{e}\in \mathbb{F}_2^N$  that  satisfy  $\gamma(\boldsymbol{e})=\gamma(\boldsymbol{e}^*)$ but are prior to $\boldsymbol{e}^*$ in the lexicographic order. In contrast, the set $\mathcal{Q}_\text{GCD}$~(if non-empty) consists of those partial TEPs $\boldsymbol{e}_{P}\in \mathbb{F}_2^K$ that satisfy $\gamma(\boldsymbol{e}_{P})=\gamma(\boldsymbol{e}^*)$ but are prior to $\boldsymbol{e}_{P}^*$ in the lexicographic order. This occurs only when $\gamma(\boldsymbol{e}_{I}^*) = 0$ and hence $\gamma(\boldsymbol{e}_{P}) = \gamma(\boldsymbol{e}_{P}^*) = \gamma(\boldsymbol{e}^*)$ since, otherwise, $\gamma(\boldsymbol{e}_{P}) \leq \gamma(\boldsymbol{e}_{P}^*) < \gamma(\boldsymbol{e}^*)$.

For any $\boldsymbol{e}_{P}\in \mathcal{L}_{\text{GCD}}$, we construct a TEP $\boldsymbol{e}=(\boldsymbol{0},\boldsymbol{e}_{P})$ with $\boldsymbol{0}\in \mathbb{F}_2^{N-K}$. We have $\boldsymbol{e} \in \mathcal{L}_{\text{GND}}$ since either $\gamma(\boldsymbol{e}) < \gamma(\boldsymbol{e}^*)$ or $\gamma(\boldsymbol{e}) = \gamma(\boldsymbol{e}^*)$ but is prior to $\boldsymbol{e}^*$ in the lexicographic order. The latter case is true since $\boldsymbol{e}_{P}$ is prior to $\boldsymbol{e}_{P}^*$ and hence $\boldsymbol{e}$ is prior to $\boldsymbol{e}^*$ in the lexicographic order. Thus we have constructed an injective mapping $\boldsymbol{e}_{P} \rightarrow \boldsymbol{e}=(\boldsymbol{0},\boldsymbol{e}_{P})$ from $\mathcal{L}_{\text{GCD}}$ into $\mathcal{L}_{\text{GND}}$. This completes the proof that $|\mathcal{L}_{\text{GCD}}| \leq |\mathcal{L}_{\text{GND}}|$.
\end{IEEEproof}

\begin{example}{(A toy example)}
 Consider the Hamming code $\mathscr{C}_{\text{Hamming}}[7,4]$ over a binary symmetric channel~(BSC) with cross error probability $p < 1/2$. In this case, the soft weight is equivalent to Hamming weight. No matter what codeword is transmitted and what vector is received, the GND with $L=1$ will find the lightest TEP $\boldsymbol{e}^*$ with at most 8 queries, one for the all zero TEP and 7 for the TEPs with Hamming weight one. The first query is successful if and only if the true error pattern is a codeword, which occurs with a probability $p_0 = (1-p)^7 + 7p^3(1-p)^3 + p^7$. Hence, the average number of queries for the GND is given by 
 \begin{equation}
     p_0+2p_1+3p_1+4p_1+5p_1+6p_1+7p_1+8p_1 = p_0 + 35p_1,
 \end{equation}
where $p_1 = (1-p_0)/7$. In contrast, the maximum number of queries for the GCD with $L=1$ is 5, one for the all-zero partial TEP and 4 for the partial TEPs with Hamming weight one. The first query is successful if and only the TEP~(obtained from $\boldsymbol{e}_{P}=\boldsymbol{0}$ by re-encoding) has a Hamming weight zero or one. In either case, further queries are not necessary because all the remaining queries are for $\boldsymbol{e}_{P}$ with $W_H(\boldsymbol{e}_{P}) \geq 1$ and must deliver $\boldsymbol{e}$ with $W_H(\boldsymbol{e}) \geq W_H(\boldsymbol{e}_{P}) \geq 1$. The probability that the first query is successful is given by $p_0 + 3p_1$. The average number of queries for the GCD is given by 
\begin{equation}
   (p_0+3p_1)+2p_1+3p_1+4p_1+5p_1=p_0 + 17p_1,
\end{equation}
 which is strictly less than the average number of queries for the GND.    
\end{example}

\begin{example}
    Consider a binary linear block code $\mathscr{C}[N,K]$ over a BSC. In this case, the soft weight is equivalent to Hamming weight. Suppose that $\boldsymbol{e}^*=(\boldsymbol{e}_{I}^*,\boldsymbol{e}_{P}^*)$ is the $L$-th lightest valid TEP, which is not known in advance but exists. There are two cases. One is $W_H(\boldsymbol{e}_{I}^*) >0 $ and the other is $W_H(\boldsymbol{e}_{I}^*) =0 $.
 \begin{itemize}
        \item  For $W_H(\boldsymbol{e}_{I}^*) >0 $, the GCD will definitely find $\boldsymbol{f}^{(L)}=\boldsymbol{e}^*$ within $\sum_{i=0}^{W_H(\boldsymbol{e}_{P}^*)}\binom{K}{i} $ queries. The GCD continues the query process since it cannot check whether $\boldsymbol{f}^{(L)}$ is the $L$-th lightest one or not. As the query process proceeds, $W_H(\boldsymbol{e}_{P})$ increases but $\boldsymbol{f}^{(L)}$~$(=\boldsymbol{e}^*)$ keeps unchanged. Once all $\boldsymbol{e}_{P} \in \mathbb{F}_2^K$ with $W_H(\boldsymbol{e}_{P}) < W_H(\boldsymbol{e}^*)$ have been re-encoded, the GCD can safely confirm that $\boldsymbol{f}^{(i)}$~($1\leq i\leq L$) are the $L$ lightest valid TEPs. Therefore, the total number of queries for the GCD is $\min\left\{2^K, \sum_{i=0}^{W_H(\boldsymbol{e}^*)-1}\binom{K}{i}\right\}$, which is strictly less than $\sum_{i=0}^{W_H(\boldsymbol{e}^*)-1}\binom{N}{i}$, a lower bound on the number of queries for the GND.
        \item  For $W_H(\boldsymbol{e}_{I}^*)=0$, the number of queries for the GCD is $\sum_{i=0}^{W_H(\boldsymbol{e}^*)-1}\binom{K}{i} + T$, where $T$ is the rank of $\boldsymbol{e}_{P}^*$~(according to the lexicographic order)    
        in the set $\{\boldsymbol{e}_{P}\in \mathbb{F}_2^K| W_H(\boldsymbol{e}_{P}) = W_H(\boldsymbol{e}_{P}^*)\}$. Again, this number is strictly less than $\sum_{i=0}^{W_H(\boldsymbol{e}^*)-1}\binom{N}{i} + T'$, the number of queries for the GND, where $T'$ is the rank of $\boldsymbol{e}^*$ in the set $\{\boldsymbol{e}\in \mathbb{F}_2^N| W_H(\boldsymbol{e}) = W_H(\boldsymbol{e}^*)\}$.
    \end{itemize}

\end{example}

\begin{figure}[!t]
\centering
\includegraphics[width=3.0in]{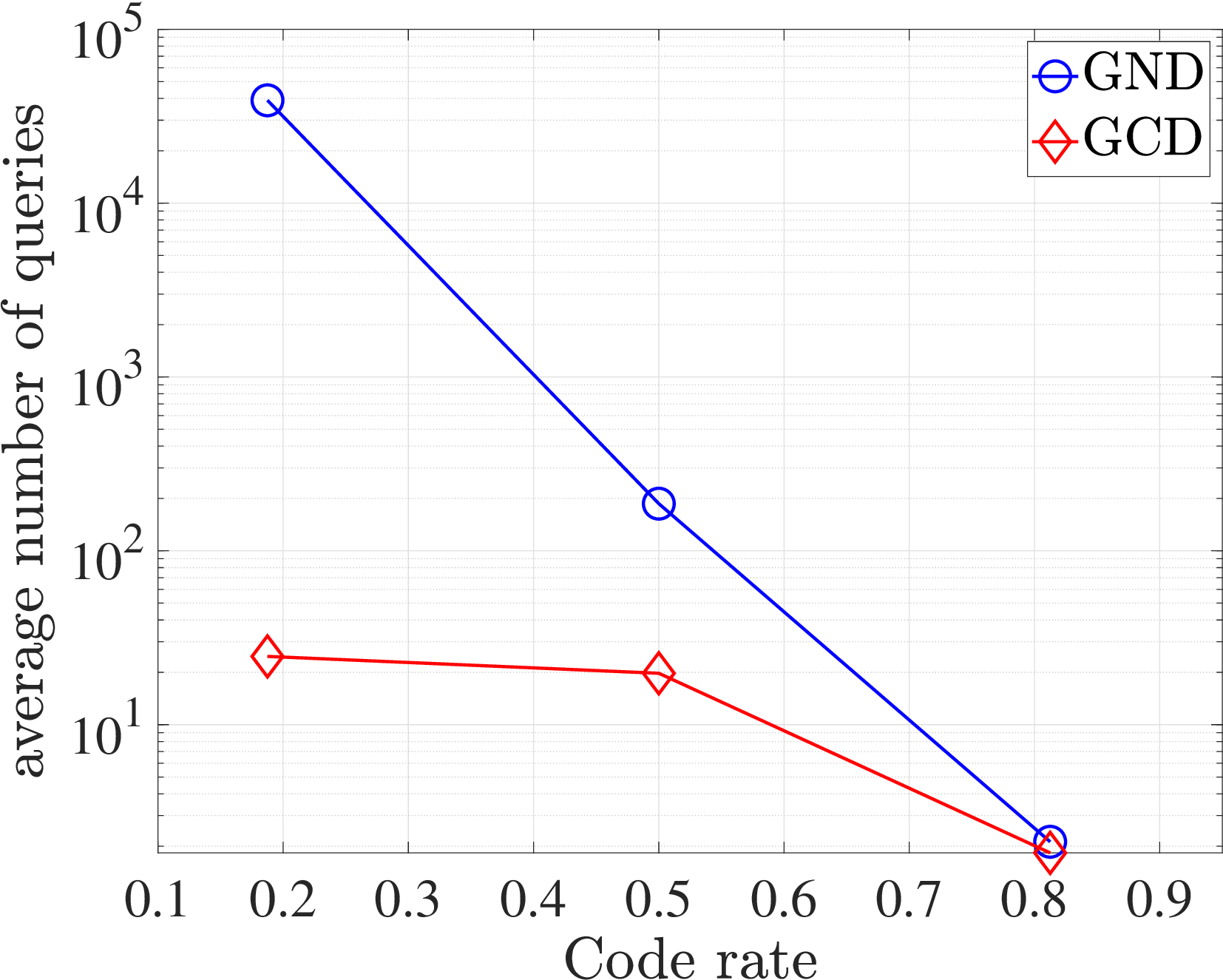}
\caption{Average number of queries for the GND~(SGRAND) and the GCD of the RM codes of length $N = 32$ and dimension $K \in \{6, 16, 26\}$. Here, the channel is BPSK-AWGN, the target $\text{FER}=10^{-3}$ and $L =1$.}
\label{Query}
\end{figure}

\begin{example}
 Consider three Reed-Muller~(RM) codes, $\mathscr{C}_{\text{RM}}[32,6]$, $\mathscr{C}_{\text{RM}}[32,16]$ and $\mathscr{C}_{\text{RM}}[32,26]$, over an additive white Gaussian noise channel~(AWGN) with binary phase shift keying~(BPSK) modulation. Shown in Fig.~\ref{Query} are the average numbers of queries per reception of noisy codeword for the GCD and the GND at target FER $10^{-3}$~(corresponding to different signal-to-noise ratios~(SNRs) for different code rates). We see that the GCD requires a fewer number of queries than the GND, validating our analysis. We also see that the  gap between the number of queries is narrowed as the code rate increases. This suggests that, compared with the GND, the GCD is more universal and applicable to codes with a wide range of code rates.     
\end{example}

\textbf{Remarks}. 
We have shown that the GCD typically requires a fewer number of queries than the GND and that the two algorithms have comparable complexity per query, indicating that the GCD is more efficient than the GND. For practical applications, we need to find efficient ways to generate the TEPs, possibly with tolerated performance loss. We will present in detail an ordered TEP generator aided by the FTP, while any TEP generator developed for the GND can be transferred as a partial TEP generator.  For example, the hardware-efficient generator implemented in the ORBGRAND~\cite{orbgrand,ORBGRANDhard,orbgrabdhardCondo} has been employed for generating partial TEPs in the partial OSD~(POSD) algorithm~\cite{POSD}, 
achieving lower complexity and better performance than the GND~(with a limited maximum list size) algorithm.


\section{An Ordered TEP Generator and Complexity Analysis}
\subsection{Flipping Pattern Tree}\label{FPT}

For any partial TEP $\boldsymbol{e}_P \in \mathbb{F}_2^K$, we define its support set as $\mathcal{S}(\boldsymbol{e}_P)=\{j: \boldsymbol{e}_P[j] \neq 0\}$, whose cardinality $|\mathcal{S}(\boldsymbol{e}_P)|$ is the Hamming weight $W_H(\boldsymbol{e}_P)$ of $\boldsymbol{e}_P$. Notice that a vector $\boldsymbol{f} \in \mathbb{F}_2^K$ can be uniquely specified by its support set $\mathcal{S}(\boldsymbol{f})$. Without loss of generality, we assume that the partial reliability vector is non-decreasing. That is, $$|\boldsymbol{r}_P[1]|\leq |\boldsymbol{r}_P[2]|\leq \cdots \leq |\boldsymbol{r}_P[K]|.$$
All possible partial TEPs $\boldsymbol{e}_P \in \mathbb{F}_2^K$ are arranged as vertexes into an ordered rooted tree, denoted by $\mathcal{T}$, as described below.
\begin{itemize}
    \item The root of the tree is the all-zero vector $\boldsymbol{0}$, located at the 0-th level.
    \item For $i \geq 1,$ the $i$-th level of the tree consists of all partial TEPs with Hamming weight $i$. 
    \item For a vertex $\boldsymbol{f}$ with $ \min\mathcal{S}\left(\boldsymbol{f}\right)>1$ at the $i$-th level, we 
    define its left-most child, denoted as $\boldsymbol{f}_{\downarrow}$,  by $\mathcal{S}\left(\boldsymbol{f}_{\downarrow}\right)=\mathcal{S}\left(\boldsymbol{f}\right)\cup \{1\}$. By convention, we assume for the empty set $\Phi$ that $\min \Phi =\infty.$
    \item For a vertex $\boldsymbol{f}$ at the $i$-th level~($i\geq 1$) with $\min\mathcal{S}(\boldsymbol{f})<K$ and $\min\mathcal{S}(\boldsymbol{f})+1\notin \mathcal{S}(\boldsymbol{f})$, we define its adjacent right-sibling, denoted as $\boldsymbol{f}_{\rightarrow}$, by $\mathcal{S}\left(\boldsymbol{f}_{\rightarrow}\right)=\mathcal{S}(\boldsymbol{f}) \setminus \{\min\mathcal{S}(\boldsymbol{f})\} \cup \{\min\mathcal{S}(\boldsymbol{f})+1\} $.
    \item The root has $K$ children. For $i\geq 1$, a vertex $\boldsymbol{f}$ has $\min\mathcal{S}(\boldsymbol{f})-1$ children.
\end{itemize}

\begin{figure}[!t]
\centering
\includegraphics[width=4.0in]{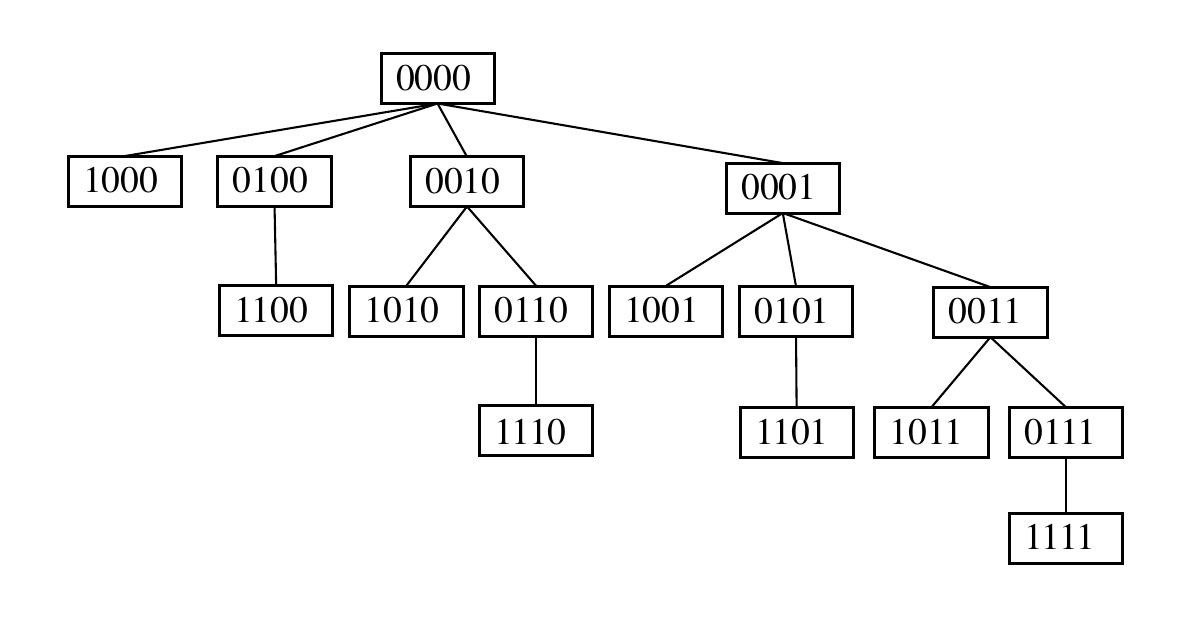}
\caption{A rooted tree with $K=4$.}
\label{roottree}
\end{figure}

\begin{example}
Consider $K=4$. A rooted tree $\mathcal{T}$ which takes $(0000)$ as root is presented in Fig.~\ref{roottree}.    
\end{example}

To define the total order of all vertexes, we write $\boldsymbol{f} \prec \boldsymbol{g}$ 
if one of the following conditions holds:
\begin{enumerate}
    \item $\gamma(\boldsymbol{f})<\gamma(\boldsymbol{g})$,
    \item $\gamma(\boldsymbol{f})=\gamma(\boldsymbol{g})$ and $W_H(\boldsymbol{f})<W_H(\boldsymbol{g})$,
    \item $\gamma(\boldsymbol{f})=\gamma(\boldsymbol{g})$ and $W_H(\boldsymbol{f})=W_H(\boldsymbol{g})$ but $\boldsymbol{f}$ is prior to $\boldsymbol{g}$ in the lexicographic order.
\end{enumerate}
We can see that the root of $\mathcal{T}$ has soft weight $\gamma(\boldsymbol{0})=0$. Given a vertex $\boldsymbol{f}$ with soft weight $\gamma(\boldsymbol{f})$, its leftmost child $\boldsymbol{f}_{\downarrow}$~(if it exists) has soft weight $\gamma(\boldsymbol{f}_{\downarrow})=\gamma(\boldsymbol{f}) +  |\boldsymbol{r}_P[1]|$, while its adjacent right-siblings $\boldsymbol{f}_{\rightarrow}$~(if it exists) has soft weight $\gamma(\boldsymbol{f}_{\rightarrow})= \gamma(\boldsymbol{f}) -  |\boldsymbol{r}_P[i]| +  |\boldsymbol{r}_P[i+1]|$, where $i = \min\mathcal{S}(\boldsymbol{f})$. Obviously, we have the following lemma.
\begin{lemma}
For each vertex $\boldsymbol{f}$ of $\mathcal{T}$, we have $\boldsymbol{f} \prec \boldsymbol{f}_{\downarrow}$~(if $\boldsymbol{f}_{\downarrow}$ exists) and $\boldsymbol{f} \prec \boldsymbol{f}_{\rightarrow}$~(if $\boldsymbol{f}_{\rightarrow}$ exists). 
\end{lemma}
Therefore, it is not necessary to consider  $\boldsymbol{f}_{\downarrow}$ and $\boldsymbol{f}_{\rightarrow}$ before $\boldsymbol{f}$ is queried. Such a structure of the tree $\mathcal{T}$ is critical to design the FPT algorithm, which generates one-by-one upon request all partial TEPs in the following order
\begin{equation}
\begin{aligned}
\centering
\mathcal{M} \triangleq \boldsymbol{e}_P^{(1)} \prec \boldsymbol{e}_P^{(2)} \prec \cdots \prec \boldsymbol{e}_P^{(i)}\prec \cdots ,
\end{aligned}
\end{equation}
where $\mathcal{M}$ is the linked list maintained in order. The FPT algorithm is described in Algorithm~2, from which we observe that once a partial TEP $\boldsymbol{e}_P$ is generated, $\boldsymbol{e}_P$ is removed from $\mathcal{M}$ and (at most) two new partial TEPs are inserted into $\mathcal{M}$. One is its left-most child $\boldsymbol{e}_{P\downarrow}$~(if it exists) by flipping the first bit from 0 to 1, and
the other is its adjacent right-sibling $\boldsymbol{e}_{P\rightarrow}$~(if it exists) by flipping two bits indexed by $\min\mathcal{S}\left(\boldsymbol{e}_P\right)$~(from 1 to 0) and $\min\mathcal{S}\left(\boldsymbol{e}_P\right)+1$~(from 0 to 1).


\begin{algorithm}[!t]
\renewcommand{\algorithmicrequire}{\textbf{Input:}}
    \renewcommand{\algorithmicensure}{\textbf{Output:}}
\caption{FPT algorithm}\label{alg:alg1}
\begin{algorithmic}[1]
\Require The ascending sorted reliability vector $|\boldsymbol{r}_P|\in \mathbb{R}^K$, and the search order $\ell \leq \ell_{\rm{max}}\leq 2^K$.
\Ensure The $\ell$-th lightest partial TEP $\boldsymbol{e}_P^{(\ell)}$.
\State $\mathcal{M} \gets \boldsymbol{0}$. \Comment{A linked list which is maintained in order during the query process.}
 \Function{FPT}{$|\boldsymbol{r}_P|,\ell$}
 \Comment{This should be called in order of $\ell = 1, 2, \ldots, \ell_{\rm{max}}$.}
\State $\boldsymbol{e}_P^{(\ell)} \gets$ the head partial TEP in $\mathcal{M}$.
\If{$\min\mathcal{S}\left(\boldsymbol{e}_P^{(\ell)}\right)>1$}
\Comment{Left-most child.}
\State $\boldsymbol{e}_{P\downarrow} = \boldsymbol{e}_P^{(\ell)}$.
\State $\boldsymbol{e}_{P\downarrow}[1] = 1$. 
\State  Update the linked list $\mathcal{M}$ by inserting $\boldsymbol{e}_{P\downarrow}$.
\EndIf
\If{$\min\mathcal{S}\left(\boldsymbol{e}_P^{(\ell)}\right) < K$ and $\min\mathcal{S}\left(\boldsymbol{e}_P^{(\ell)}\right)+1\notin \mathcal{S}\left(\boldsymbol{e}_P^{(\ell)}\right)$}
\Comment{Adjacent right sibling.}
\State $\boldsymbol{e}_{P\rightarrow} = \boldsymbol{e}_P^{(\ell)}$.
\State $\boldsymbol{e}_{P\rightarrow}\left[\min\mathcal{S}\left(\boldsymbol{e}_P^{(\ell)}\right)\right] = 0$.
\State $\boldsymbol{e}_{P\rightarrow}\left[\min\mathcal{S}\left(\boldsymbol{e}_P^{(\ell)}\right)+1\right] = 1$.
\State Update the linked list $\mathcal{M}$ by inserting $\boldsymbol{e}_{P\rightarrow}$.
\EndIf
\State Update the linked list $\mathcal{M}$ by removing $\boldsymbol{e}_P^{(\ell)}$.
\EndFunction
\end{algorithmic}
\label{alg1}
\end{algorithm}

\begin{figure}[!t]
\centering
\includegraphics[width=5.8in]{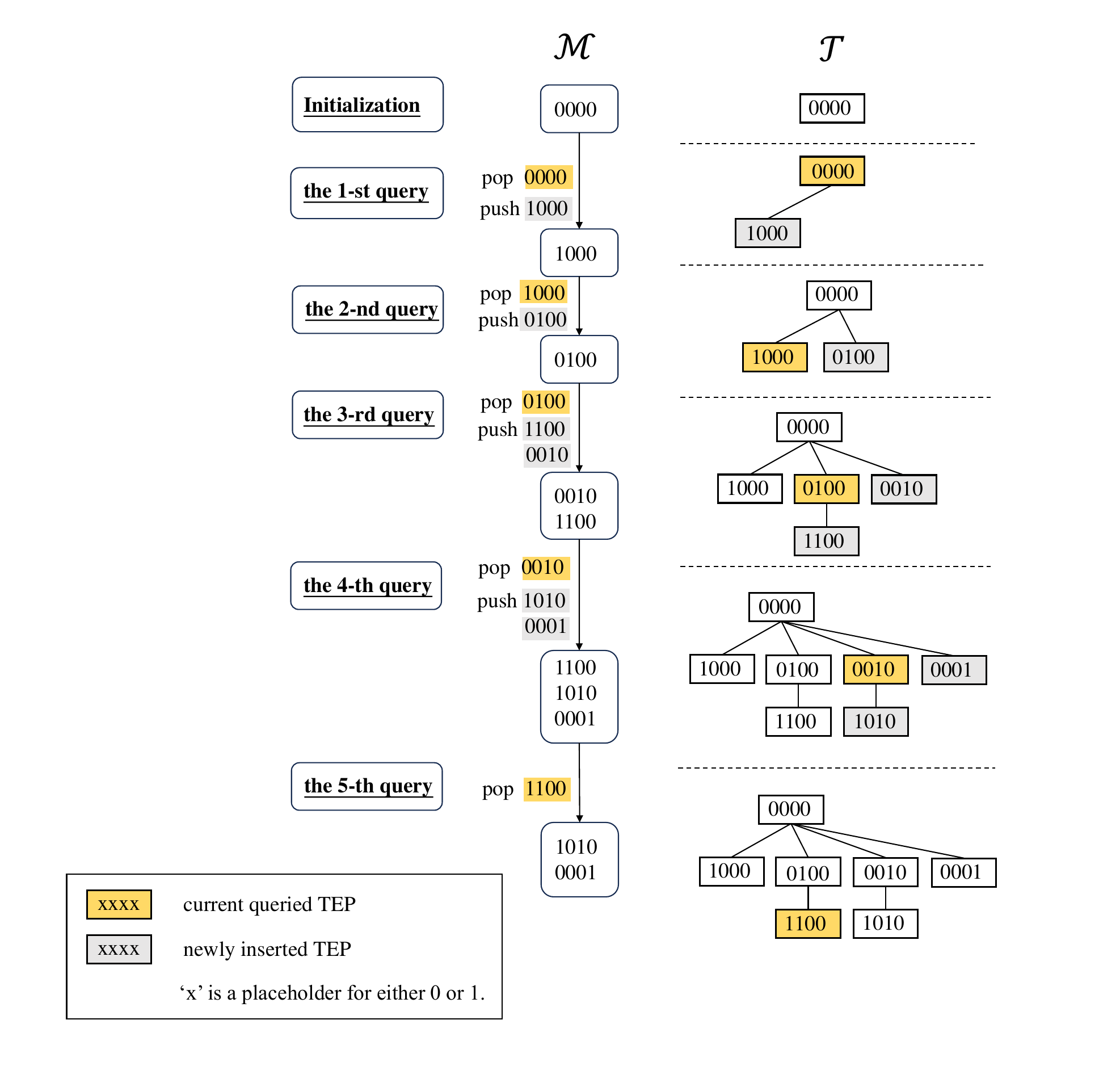}
\caption{An example of the FPT algorithm to generate an ordered list of the five best TEPs $\{0000,1000,0100,0010,1100\}$. Here, $\mathcal{M}$ is a linked list and $\mathcal{T}$ is an ordered rooted tree.}
\label{FPT4}
\end{figure}

\begin{example}
Consider a linear block code $\mathscr{C}[4, 4]$. 
Let $\boldsymbol{y}$ be the received vector from a memoryless channel that specifies the LLR vector $\boldsymbol{r}=\{0.5, 1.0, -1.2, 1.9\}$. The process of the FPT algorithm with a   linked list $\mathcal{M}$ to generate an ordered list of the five best TEPs is shown in Fig.~\ref{FPT4}, along with the progressive construction of the ordered rooted tree $\mathcal{T}$, from which we can see that the ordered list of the five best TEPs is $\{0000,1000,0100,0010,1100\}$.
\end{example}

\subsection{Complexity Analysis}
The computational complexity can be evaluated by the number of operations (OPs). As shown in Algorithm~2, the complexity of insertion and deletion of new partial TEPs in a linked list is proportional to $\log(|\mathcal{M}|)$, while generating new partial TEPs only requires $\mathcal{O}(K)$ OPs by some naive sequence manipulations. Denote by $\ell_{\rm{avg}}$ the average number of queries per received vector. Thus the computational complexity of the FPT algorithm can be evaluated as
\begin{equation}\label{FPT}
\begin{aligned}
\centering
 C_{\text{avg}}^{\text{FPT}}= & \underbrace{\mathcal{O}\left(\ell_{\text{avg}}\log \ell_{\text{avg}}\right)}_{\text{linked list $\mathcal{M}$}} +\underbrace{\mathcal{O}\left({\ell}_\text{avg}K\right)}_{\text{sequence manipulation}}.   
\end{aligned}
\end{equation}

The computational complexity of the GCD involves four dominant parts:
\begin{enumerate}
    \item Sorting. Sorting the partial reliability vector $|\boldsymbol{r}_P|\in \mathbb{R}^K$ as the input to the FPT algorithhm requires $\mathcal{O}(K\log K)$ OPs.
    \item FPT. The complexity of FPT is measured by $ C_{\rm{avg}}^{\rm{FPT}}$.
    \item Re-encoding. The GCD calculates $\boldsymbol{s}=\boldsymbol{z}\mathbf{H}^T$ with a complexity of order $\mathcal{O}((N-K)N)$, delivering the first valid TEP $\boldsymbol{e}^{(1)} = (\boldsymbol{s},\boldsymbol{0})$.  For $\ell>1$, if $\boldsymbol{e}_P^{(\ell)}$ is generated by its parent, denoted as $\boldsymbol{e}_P^{(j)}$~($j<\ell$) by flipping one bit, the corresponding $\boldsymbol{e}_I^{(\ell)}$ can be simply calculated by $\boldsymbol{e}_I^{(\ell)}=\boldsymbol{e}_I^{(j)}-\left(\mathbf{P}^T\right)_1$, where $\left(\mathbf{P}^T\right)_1$ stands for the first row of the matrix $\mathbf{P}^T$, delivering a new TEP $\boldsymbol{e}^{(\ell)}=\left(\boldsymbol{e}_I^{(\ell)}, \boldsymbol{e}_P^{(\ell)}\right)$, which avoids many unnecessary computations and only requires $(N-K)$ OPs. Similarly, if $\boldsymbol{e}_P^{(\ell)}$ is generated by its adjacent left-sibling, denoted as $\boldsymbol{e}_P^{(k)}$~($k<\ell$) by flipping two bits, $\boldsymbol{e}_I^{(\ell)}=\boldsymbol{e}_I^{(k)}+\left(\mathbf{P}^T\right)_i-\left(\mathbf{P}^T\right)_{i+1}$ with $2(N-K)$ OPs, where $i=\min\mathcal{S}\left(\boldsymbol{e}_P^{(k)}\right)$ and $\left(\mathbf{P}^T\right)_i$ stands for the $i$-th row of the matrix $\mathbf{P}^T$.   
    \item The complexity of insertion and deletion in the linked list $\mathcal{L}$ with size $L$ is at most $\log L$ for each query.
    
\end{enumerate}
Thus the overall computational complexity of the GCD can be evaluated as
 \begin{equation}\label{LIST-GRAND}
\begin{aligned}
\centering
 C_{\text{avg}}^{\text{GCD}}= & \underbrace{\mathcal{O}\left(K\log K\right)}_{\text{sorting}}
+\underbrace{\mathcal{O}\left({\ell}_\text{avg}(N-K)\right)}_\text{re-encoding} \\
&+  \underbrace{\mathcal{O}\left(\ell_{\text{avg}}\log L\right)}_{\text{linked list $\mathcal{L}$}}+ C_{\text{avg}}^{\text{FPT}}.   
\end{aligned}
\end{equation}

\textbf{Remarks}. It is worth pointing out that the GCD performs the GE offline, which is distinguished from the OSD. Consequently, the complexity of transforming $\mathbf{H}$ into $[\mathbf{I}\quad \mathbf{P}]$ is not taken into account in the above analysis.

Now, we compare the GCD with the existing optimal decoding algorithms, including ESD and OSD~\cite{Fossorier1995,DorshOSD}, in terms of the decoding complexity.
\begin{enumerate}
    \item The differences in the number of queries.
    \begin{itemize}
        \item Typically, due to the use of a sufficient condition for termination, the GCD requires a fewer number of queries than the ESD since the maximum query number of the GCD is set to be $2^K$.
        \item The number of queries for the OSD is usually  less than the number of queries for the GCD since the OSD queries the TEPs in the MRB and the TrTEP is expected to be  queried earlier.
    \end{itemize}
    \item The differences in the complexity.
    \begin{itemize}
        \item  Compared with the ESD, the GCD requires a fewer number of queries but sorts the partial TEPs. Therefore, the decoding complexity of the GCD is in general lower than that of the ESD unless the code dimension $K$ is very small, in which case the ESD is preferable.
        \item Compared with the GCD, the OSD has a fewer number of queries but requires online GE operation. The complexity comparison between the GCD and the OSD depends on the code rates and the channels. For low/high-rate codes, the GCD is more efficient than the OSD. For the channels with bad quality, the number of queries dominates the complexity and the OSD is a preferable choice, while for the channels with good quality, the GE dominates the complexity. In the latter case, the GCD is more competitive. 
    \end{itemize}
\end{enumerate}

\begin{figure}[!t] 
  \centering
  \subfloat[Performance. Here, from left to right, the curves are marked by $(K)$ with the dimension $K$ inside.\label{fig:rmfer}]{\includegraphics[width=3.1in]{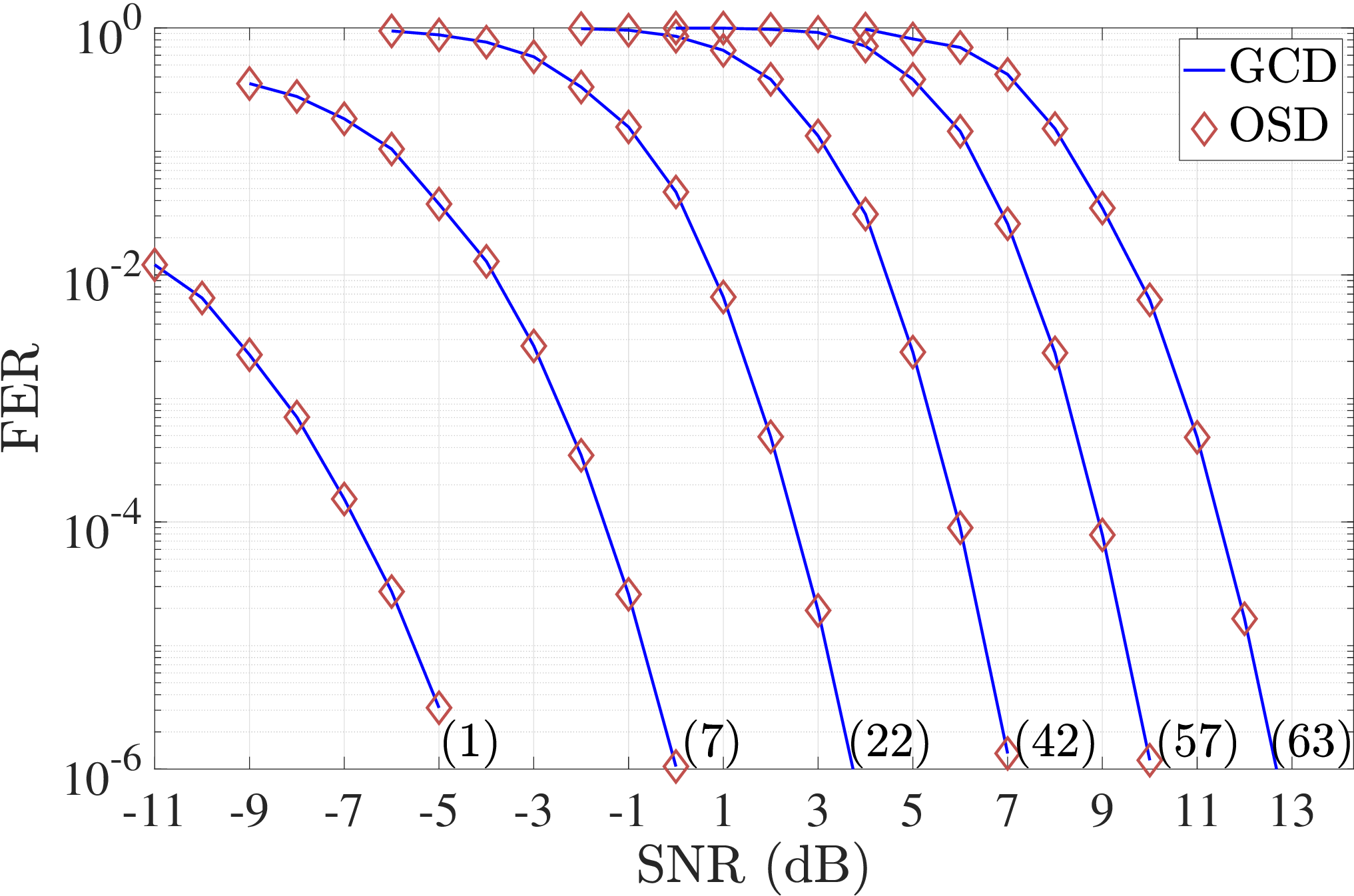}}
   \hfill
  \subfloat[Average number of queries at the target FER~$10^{-5}$.\label{fig:rmQuery}]{\includegraphics[width=3.1in]{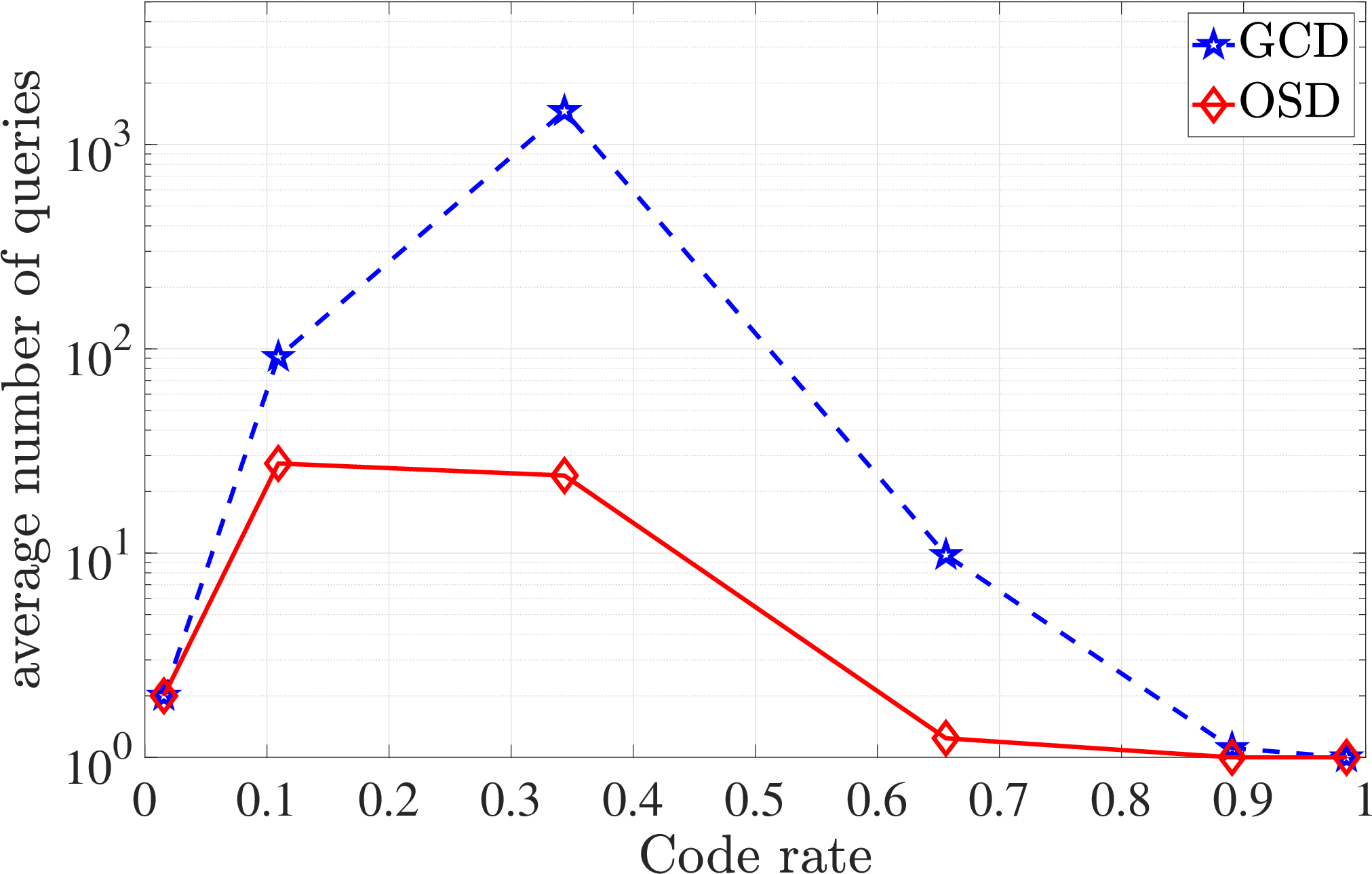}}
  \caption{
    ML decoding of the RM codes $\mathscr{C}_{\rm{RM}}[64,K]$ with OSD and GCD~($L=1$).
  }
  \label{Fig_OESD}
\end{figure}

\begin{example}
 We have simulated the RM codes with different code rates over BPSK-AWGN channels. The simulation results are shown in Fig.~\ref{Fig_OESD}. From Fig.~\ref{fig:rmfer}, we see that both the GCD and the OSD exhibit the ML decoding performance. From Fig.~\ref{fig:rmQuery}, we observe that for moderate code rates, the GCD requires more queries than the OSD. For low code rates and high code rates, we see that the average number of queries can be reduced to ten or even a few both for the GCD and the OSD. Taking into account the complexity of the online GE, we can safely conclude that, compared with OSD, GCD is preferable for the low/high-rate codes. For the moderate/high-rate codes, we see that the number of queries is much less than that of the naive ESD.
\end{example}

\subsection{Truncated GCD}\label{TOESD}


As an optimal list decoding, the GCD algorithm queries the partial TEPs in descending order of their probability until the $L$ most lightest valid TEPs are identified. Recalling that a list decoding error has occurred if the TrTEP~(corresponding to the sent codeword) is not in the list $\mathcal{L}$, we have the error probability $\epsilon(L,\rm{GCD})$ of the GCD given by 
\begin{equation}
 \epsilon(L,\rm{GCD}) = \mathbb{P}\{\text{the TrTEP is not in the list $\mathcal{L}$}\}.   
\end{equation}

\begin{figure}[!t]
\centering
\includegraphics[width=3.4in]{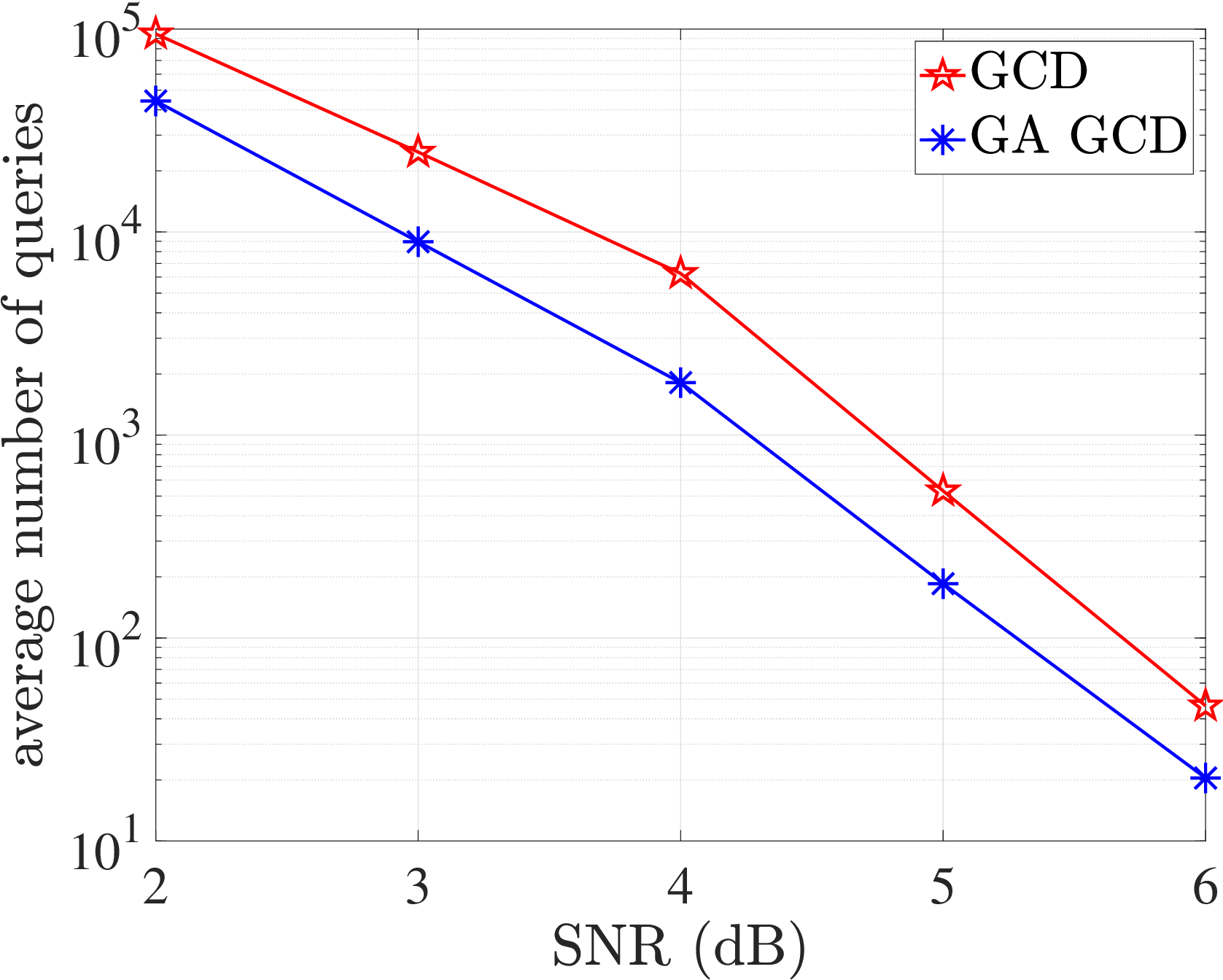}
\caption{Average number of queries of the GCD and the GA GCD for the RM code $\mathscr{C}_{\rm{RM}}[64,42]$. Here, $L=1$.}
\label{fig_GAquery} 
\end{figure}

\begin{example}
Consider an RM code $\mathscr{C}_{\rm{RM}}[64,42]$. We compare the genie-aided~(GA) GCD\footnote{The GA GCD is similar to the GCD with the only difference that the GA GCD terminates immediately once when the TrTEP is queried.} and the GCD in terms of the average number of queries over BPSK-AWGN channels, as shown in Fig.~\ref{fig_GAquery}, from which we see that the GA GCD achieves a fewer average number of queries compared with the GCD. This suggests that the average number of queries can be reduced without too much performance degradation. 
\end{example}

To avoid unnecessary queries, we propose a sub-optimal list decoding algorithm, which truncates the ordered search whenever certain conditions are fulfilled even if the $L$ lightest TEPs have not been identified. Such a sub-optimal list decoding is referred to as a truncated GCD and the 
resulting $L$ lightest valid TEPs are collectively
denoted as $\mathcal{L}_{\rm{truncated}}$.
The error probability of the truncated GCD, denoted as $\epsilon(L,\rm{truncated})$, is then defined as the probability that the TrTEP is not in $\mathcal{L}_{\rm{truncated}}$. That is,
\begin{equation}
\begin{aligned}
  \epsilon(L,\rm{truncated})
  &= \mathbb{P}\{\text{the TrTEP is not in the list } \mathcal{L}_{\rm{truncated}}\}\\
  &=  \mathbb{P}\{\boldsymbol{e}_P \text{ is not queried}\}+\mathbb{P}\{\boldsymbol{e}_P \text{ is queried but the TrTEP is not in the list $\mathcal{L}_{\rm{truncated}}$}\}\\   
 &\leq \mathbb{P}\{\boldsymbol{e}_P \text{ is not queried}\} + \mathbb{P}\{ \text{the TrTEP is not in the list $\mathcal{L}$}\}\\
&=\mathbb{P}\{\boldsymbol{e}_P \text{ is not queried}\} + \epsilon(L,\rm{GCD}),   
\end{aligned}
\end{equation}
Hence, the performance gap between the truncated GCD and the GCD can be bounded by
\begin{equation} \label{TOESD-FER}
 0\leq  \epsilon(L,\text{truncated})-\epsilon(L,\text{GCD}) \leq \mathbb{P}\{\boldsymbol{e}_{P} \text{ is not queried}\}.
\end{equation}
In the case when the probability $\mathbb{P}\{\boldsymbol{e}_P \text{ is not queried}\}$ is negligible, the truncated GCD performs
almost the same as the GCD. We propose three conditions for truncation, which attempt to reduce the average number of queries but keep $\mathbb{P}\{\boldsymbol{e}_P \text{ is not queried}\}$ at a tolerable level. 



\begin{definition}[Truncated GCD with maximum query number $\ell_{\rm{max}}$] If the query process reaches the $\ell_{\rm{max}}$-th query, terminate the GCD.
This truncated GCD is denoted as $\ell_{\rm{max}}$-GCD, where $\ell_{\rm{max}}\geq 1$ is preset to be around the average rank of the partial TrTEP~(typically much less than $2^K$).
\end{definition}

\begin{definition}[Truncated GCD with soft weight threshold $\tau_s$]
If for some $j$,
\begin{equation}
    \gamma\left(\boldsymbol{e}_P^{(j)}\right) \geq \tau_s,
\end{equation}
terminate the GCD. This truncated GCD is denoted as $\tau_s$-GCD, where $\tau_s > 0$ is preset to be around the average soft weight of the partial TrTEP~(typically far less than the maximum soft weight).

\end{definition}

\begin{definition}[Truncated GCD with tolerated error probability loss $\tau_p$]
If for some $j$, 
\begin{equation}
\begin{aligned}
 \sum\limits_{i=1}^{j}P\left(\boldsymbol{e}_P^{(i)}|\boldsymbol{r}_P\right) \geq 1-\tau_p,
\end{aligned}
\end{equation}
terminate the GCD. This truncated GCD is denoted as $\tau_p$-GCD, where $\tau_p$ is  typically preset
to be lower than the target error probability by an order of magnitude.

\end{definition}


\subsubsection{$\ell_{\rm{max}}$-GCD} We have the following proposition.
\begin{proposition}
 The performance gap between the $\ell_{\rm{max}}\text{-GCD}$ and the GCD is upper bounded by 
 $\mathbb{P}\{D>\ell_{\rm{max}}\}$. That is,
 \begin{equation}
\epsilon(L,\ell_{\mathrm{max}}) \leq    \epsilon(L,\mathrm{GCD}) +  \mathbb{P}\{D>\ell_{\mathrm{max}}\} ,
\end{equation}
where $D$ represents the number of the partial TEPs lighter than the partial TrTEP $\boldsymbol{e}_P$.
\end{proposition}
\begin{IEEEproof}
Define
\begin{equation}
    \mathcal{D}\triangleq \{\boldsymbol{f}\in \mathbb{F}_2^K:\gamma(\boldsymbol{f})\leq \gamma(\boldsymbol{e}_P)\}. 
\end{equation}    
We have $D = |\mathcal{D}|$. From \eqref{TOESD-FER}, the error probability of the $\ell_{\rm{max}}$-GCD can be bounded  by
\begin{equation}
\begin{aligned}
\epsilon(L,\ell_{\rm{max}}) &\leq \epsilon(L,\mathrm{GCD}) + \mathbb{P}\{\boldsymbol{e}_P \text{ is not queried}\} \\
&= \epsilon(L,\mathrm{GCD}) +  \mathbb{P}\{D>\ell_{\mathrm{max}}\}.
\end{aligned}
\end{equation}
\end{IEEEproof}

To provide the guideline on the choice of the parameter $\ell_{\rm{max}}$, we turn to $\mathbb{P}\{D>\ell_{\rm{max}}\}$, the complementary cumulative distribution function~(CCDF) of the random variable $D$. Without loss of generality, we assume that the all-zero codeword $\boldsymbol{0}\in \mathbb{F}_2^N$ is sent. Since the randomness of $D$ comes from the LLR vector $\boldsymbol{r}_P$, we write $D(\boldsymbol{r}_P)$ and have 
\begin{equation}
   \mathbb{P}\{D>\ell_{\rm{max}}\} = \mathbb{E}_{\boldsymbol{r}_P}[\mathbb{I}\{D(\boldsymbol{r}_P)>\ell_{\rm{max}}\}], 
\end{equation}
where $\mathbb{I}$ is the indicator function. One direct way to calculate $D(\boldsymbol{r}_P)$ is to count the exact number by listing all sequences in the set $\mathcal{D}$. This can be achieved by the FPT algorithm which is efficient for the codes with $K \leq 64$. More generally, we may employ the saddlepoint approximation~\cite{saddle2011}~\cite{saddle2018} to calculate $D(\boldsymbol{r}_P)$.

Suppose that $\boldsymbol{f}$ is a random vector distributed uniformly over $\mathbb{F}_2^K$. Then $D(\boldsymbol{r}_P)$ can be written as
\begin{equation}\label{Dr}
\begin{aligned}
D(\boldsymbol{r}_P) &= 2^K \mathbb{P}\{\gamma(\boldsymbol{f})\leq \gamma(\boldsymbol{e}_P)|\boldsymbol{r}_P\} \\
&=  2^K\mathbb{P}\left\{\sum\limits_{i=1}^{K}f_i|\boldsymbol{r}_P[i]|-\sum\limits_{i=1}^{K}\boldsymbol{e}_P[i]|\boldsymbol{r}_P[i]| < 0\bigg|\boldsymbol{r}_P\right\}\\
&= 2^K\mathbb{P}\left\{\sum\limits_{i=1}^{K}(f_i-\boldsymbol{e}_P[i])|\boldsymbol{r}_P[i]|<0\bigg|\boldsymbol{r}_P\right\}\\
&= 2^K\mathbb{P}\left\{\sum\limits_{i=1}^{K}W_i|\boldsymbol{r}_P[i]|<0\bigg|\boldsymbol{r}_P\right\}\\
&= 2^K\mathbb{P}\left\{W<0|\boldsymbol{r}_P\right\},
\end{aligned}
\end{equation}
in which $\boldsymbol{e}_P[i] = \mathbb{I}[\boldsymbol{r}_P[i]<0]$ with the indicator function $\mathbb{I}[\cdot]$, $W_i = (f_i-\boldsymbol{e}_P[i])|\boldsymbol{r}_P[i]|$, and $W = \sum\limits_{i=1}^{K}W_i$. $W_1, \dots, W_K$ are independent random variables with distribution
\begin{equation}
    \mathbb{P}\{W_i=0|\boldsymbol{r}_P\}=\mathbb{P}\{f_i=\boldsymbol{e}_P[i]|\boldsymbol{r}_P\} = \frac{1}{2},
\end{equation}
\begin{equation}
    \mathbb{P}\{W_i=r_i|\boldsymbol{r}_P\}=\mathbb{P}\{f_i\neq \boldsymbol{e}_P[i]|\boldsymbol{r}_P\} = \frac{1}{2}.
\end{equation}
From~\eqref{Dr}, we turn to estimate the tail probability of the random variable $W$ by employing saddlepoint approximation~\cite{saddle2011,saddle2018}. Define the cumulant generating function of $W$ as
\begin{equation}
\begin{aligned}
\kappa(s)&=\log\mathbb{E}\left[e^{sW}\right]\\
&= \sum\limits_{i=1}^{K}\log\mathbb{E}\left[e^{sW_i}\right]\\
&= \sum\limits_{i=1}^{K}\log\mathbb{E}\left[\frac{1}{2}+\frac{1}{2}e^{s \boldsymbol{r}_P[i]}\right].
\end{aligned}
\end{equation}
Then according to~\cite{saddle2018}, the probability density function $P(w)$ is given by
\begin{equation}
\begin{aligned}
P(w) \approx e^{\kappa(\hat{s})-\hat{s}w}\cdot \frac{1}{\sqrt{2\pi \kappa''(\hat{s})}} e^{-\frac{(w-\kappa'(\hat{s}))^2}{2\kappa''(\hat{s})}},  
\end{aligned}
\end{equation}
where $\kappa'$ and $\kappa''$ denote the first and the second order derivative of $\kappa(s)$, respectively, and $\hat{s}$ is the solution of $\kappa'(s)=0$.
Thus given $\boldsymbol{r}_P$, the $D(\boldsymbol{r}_P)$ can be estimated by
\begin{equation}\label{saddleDr}
\begin{aligned}
D(\boldsymbol{r}_P) &= 2^K\cdot \mathbb{P}\left\{W<0|\boldsymbol{r}_P\right\}\\
&=2^K \cdot \int_{-\infty}^{0}P(w)\rm{d}w\\
&\approx 2^K\cdot \frac{1}{2}e^{\kappa'(\hat{s})-\hat{s}\kappa'(\hat{s})+\frac{1}{2}\hat{s}^2\kappa''(\hat{s})}\cdot \rm{erfc}\left(\frac{\kappa'(\hat{s})-\hat{s}^2\kappa''(\hat{s})}{\sqrt{2\kappa''(\hat{s})}}\right),
\end{aligned}
\end{equation}
in which ${\rm{erfc}}(x)$ is the complementary error function. Averaging $\mathbb{I}\{D(\boldsymbol{r}_P)>\ell_{\rm{max}}\}$ with respect to $\boldsymbol{r}_P$ from \eqref{saddleDr}, we obtain the approximation of the $\mathbb{P}\{D>\ell_{\rm{max}}\}$.

\begin{figure}[!t]
\centering
\includegraphics[width=3.3in]{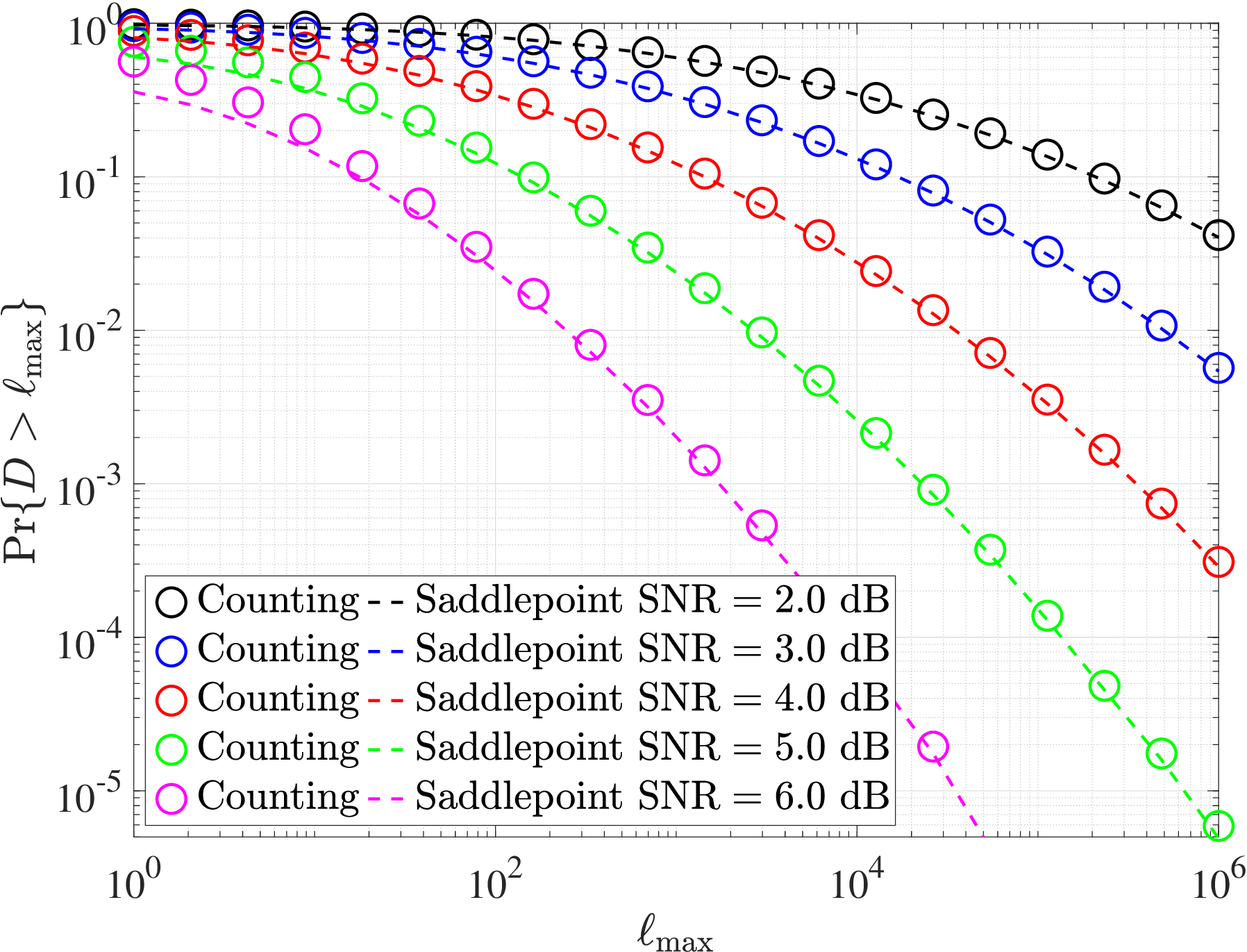}
\caption{The CCDF of $D$. Here, $K=42$.}
\label{Fig_Dr} 
\end{figure}

\begin{example}
In this example, we employ the saddlepoint approximation~\eqref{saddleDr} to estimate $\mathbb{P}\{D>\ell_{\rm{max}}\}$. We also utilize the FPT algorithm to count the exact number $D(\boldsymbol{r}_P)$. The simulation results are shown in Fig.~\ref{Fig_Dr}, from which we see that the results estimated by the saddlepoint method match well with those calculated by counting. For any given $\ell_{\rm{max}}$, the $\mathbb{P}\{D>\ell_{\rm{max}}\}$ decreases as the SNR increases. Specifically, the rank of $\boldsymbol{e}_P$ is less than $10^3$ for about $90\%$ realization of $\boldsymbol{r}$ at $\rm{SNR}=4.0~\rm{dB}$ while about $98\%$ realization at $\rm{SNR}=5.0~\rm{dB}$.  We also observe that for a large $\ell_{\rm{max}}$, the $\mathbb{P}\{D>\ell_{\rm{max}}\}$ can be sufficiently small. As $\ell_{\rm{max}}$ increases, the computational complexity increases but the performance gain is marginal. This motivates us to adaptively select maximum query numbers at different SNRs to trade off performance and complexity.
\end{example}


\begin{figure}[!t]
  \centering
  \subfloat[Performance.\label{fer_saddle_42}]{\includegraphics[width=3.1in]{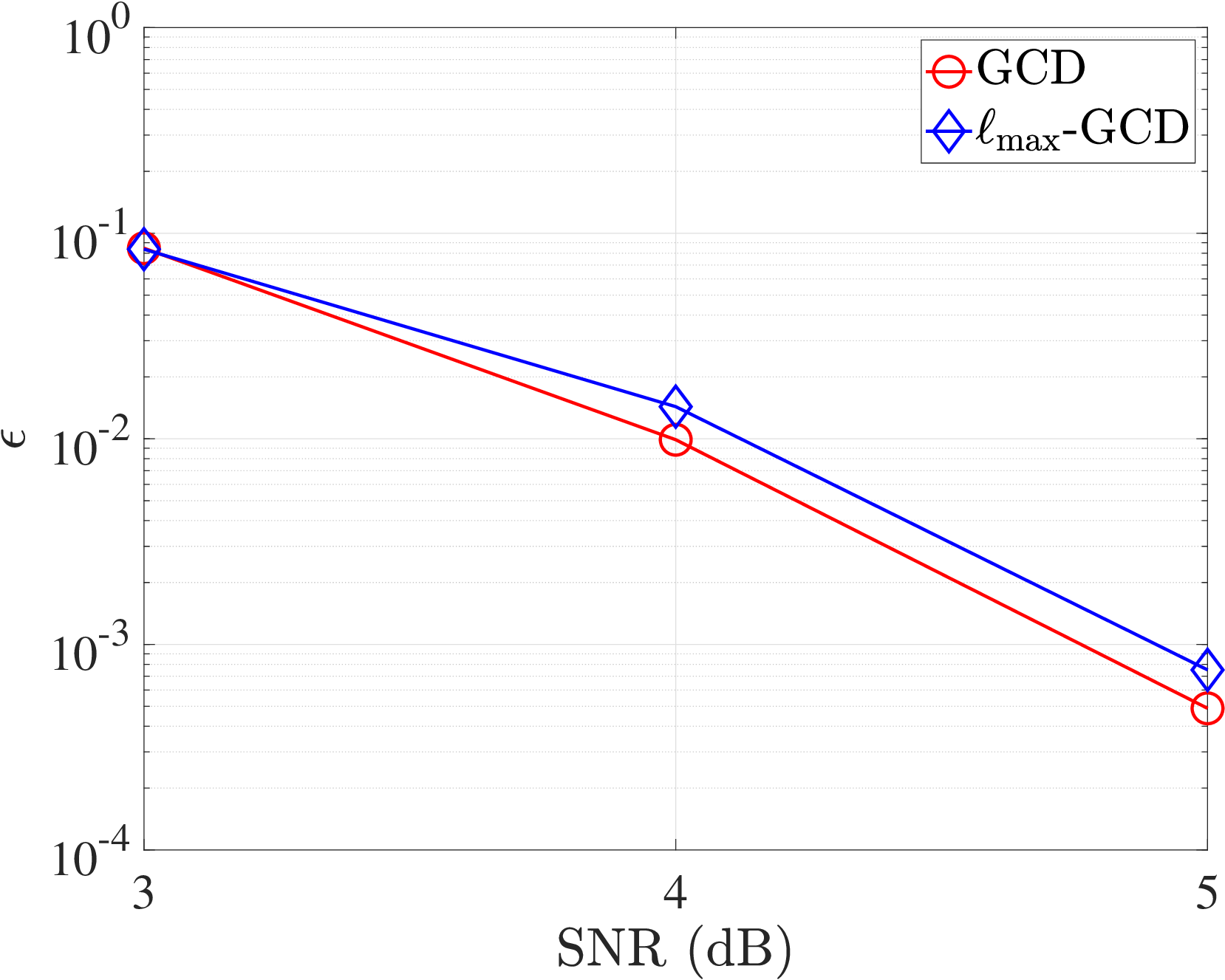}}
  \hfill
  \subfloat[Average number of queries.\label{query_saddle_42}]{\includegraphics[width=3.1in]{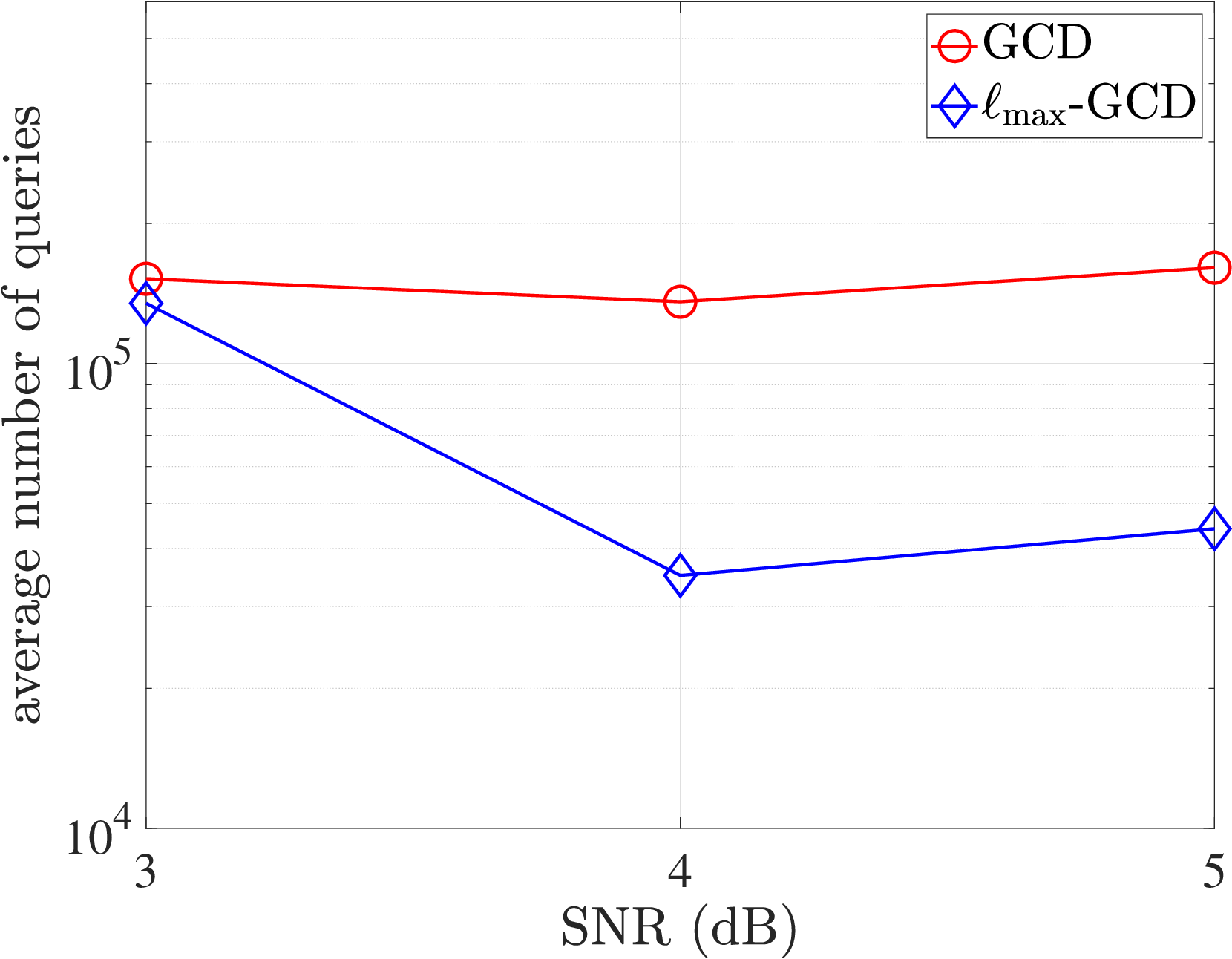}}
  \caption{
   Simulation results of the RM code $\mathscr{C}_{\rm{RM}}[64,42]$ for GCD and $\ell_{\rm{max}}$-GCD. Here, $L=2$.
  }
   \label{Fig_OESD_DifferentL}  
\end{figure}



\begin{example} Consider an RM code $\mathscr{C}_{\rm{RM}}[64,42]$.  We present the performance and average number of queries of the $\ell_{\rm{max}}$-GCD and the GCD  over BPSK-AWGN channels, as shown in Fig.~\ref{Fig_OESD_DifferentL}, from which we see that the $\ell_{\rm{max}}$-GCD exhibits comparable performance to the GCD but requires a fewer average number of queries.
\end{example}



\subsubsection{$\tau_s$-GCD}
We have the following proposition.
\begin{proposition}
 The performance gap between the $\tau_s$-GCD and the GCD is upper bounded by 
 $\mathbb{P}\{\Gamma>\tau_s\}$. That is,
 \begin{equation}
\epsilon(L,\tau_s) \leq \epsilon(L,\rm{GCD}) +  \mathbb{P}\{\Gamma>\tau_s\} , 
\end{equation}
where  $\Gamma$ represents the soft weight of the partial TrTEP $\boldsymbol{e}_P$.
\end{proposition}

\begin{IEEEproof}
From \eqref{TOESD-FER}, the error probability of the $\tau_s$-GCD is bounded  by
\begin{equation}
\begin{aligned}
\epsilon(L,\tau_s) &\leq \epsilon(L,{\rm{GCD}}) +  \mathbb{P}\{\boldsymbol{e}_P \text{ is not queried}\} .
\end{aligned}
\end{equation}
Since $\mathbb{P}\{\boldsymbol{e}_P \text{ is not queried}\} =\mathbb{P}\{\Gamma>\tau_s\}$, we have
\begin{equation}
\begin{aligned}
\epsilon(L,\tau_s) &\leq \epsilon(L,\rm{GCD}) +  \mathbb{P}\{\Gamma>\tau_s\}.
\end{aligned}
\end{equation}
\end{IEEEproof}

To provide the guideline on the choice of the parameter $\tau_s$, we turn to $\mathbb{P}\{\Gamma>\tau_s\}$, the CCDF of the random variable $\Gamma$. Without loss of generality, we assume that the all-zero codeword $\boldsymbol{0}\in \mathbb{F}_2^N$ is sent.
Since the randomness of $\Gamma$  comes from the LLR vector $\boldsymbol{r}_P$, we write $\Gamma(\boldsymbol{r}_P)$ and have 
\begin{equation}
    \Gamma(\boldsymbol{r}_P) = \sum\limits_{i=1}^{K}\boldsymbol{z}_P[i]|\boldsymbol{r}_P[i]|.
\end{equation} 
Therefore, we can estimate $\mathbb{P}\{\Gamma>\tau_s\}$ by the Monte Carlo simulation. We can also approximate $\mathbb{P}\{\Gamma>\tau_s\}$ with the Chernoff bound since $\Gamma$ is a sum of $K$ independent and identically distributed~(i.i.d.) random variables. This is also distinguished from the OSD, for which the performance analysis becomes more involved due to the dependence of the MRB. See~\cite{Fossorier1995} for details.		 
 
\begin{figure}[!t]
\centering
\includegraphics[width=3.3in]{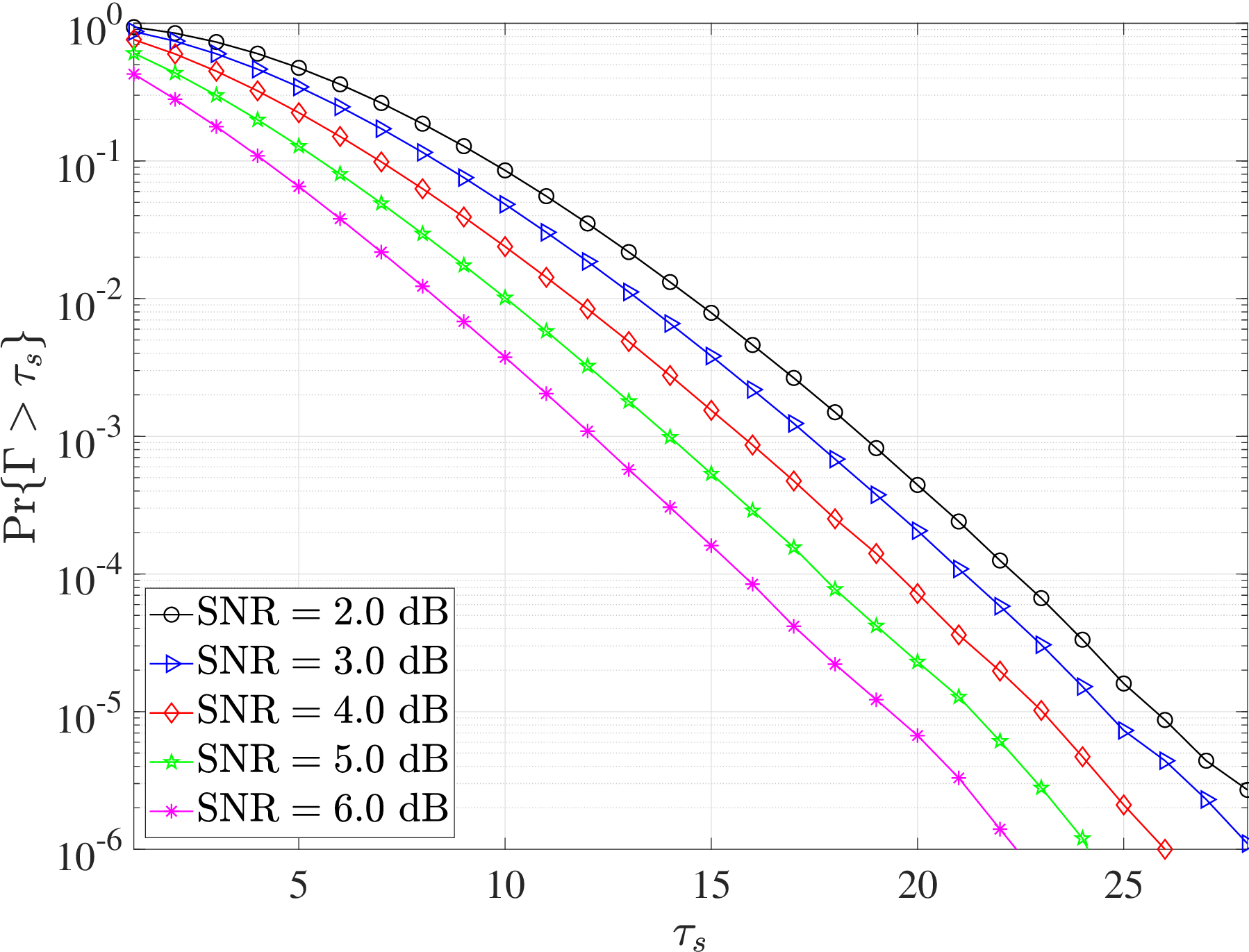}
\caption{The CCDF of $\Gamma$. Here, $K=42$.}
\label{Fig_Ts}
\end{figure}
\begin{figure}[!t]
  \centering
  \subfloat[Performance.\label{fer_soft_42}]{\includegraphics[width=3.1in]{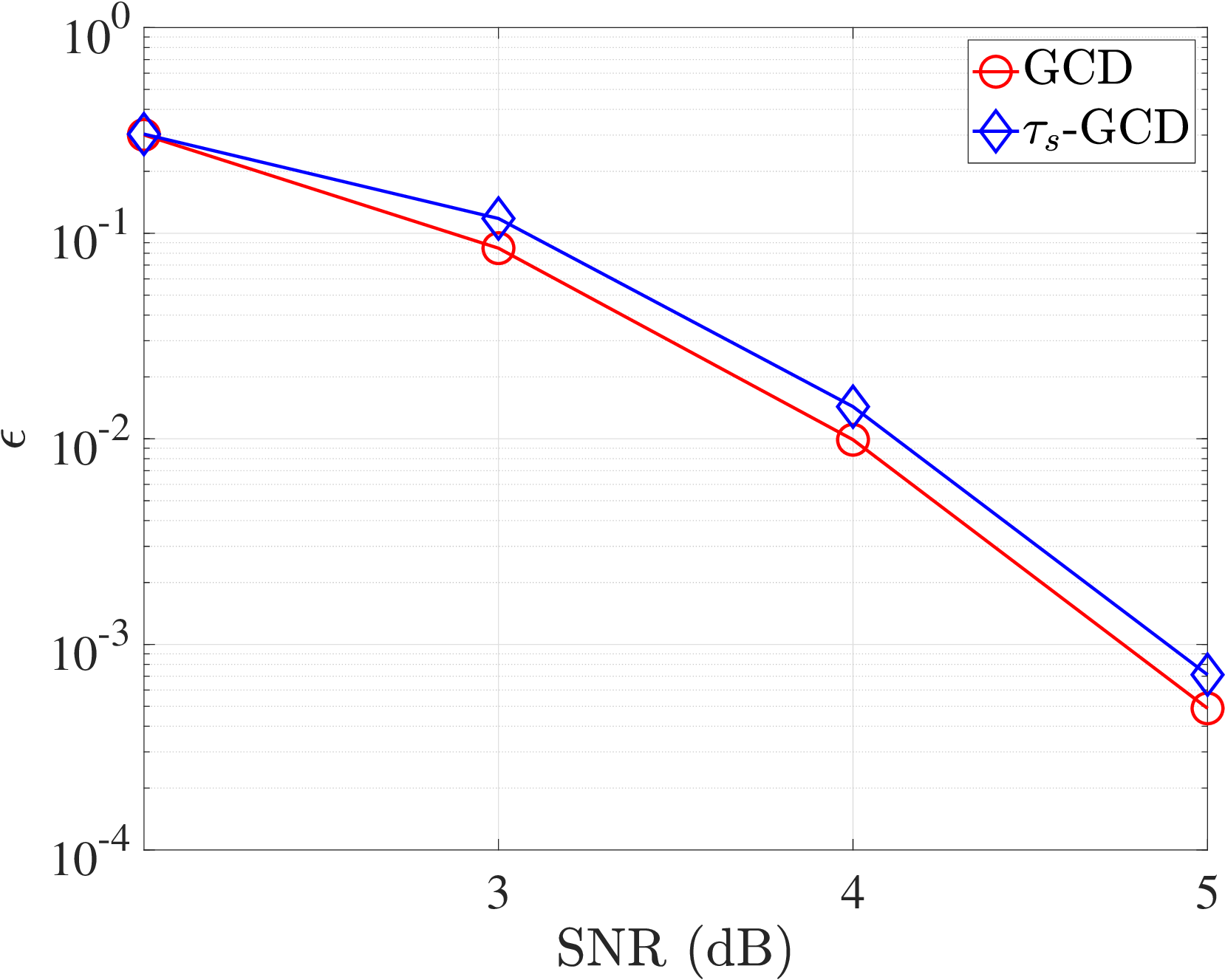}}
  \hfill
  \subfloat[Average number of queries.\label{query_soft_42}]{\includegraphics[width=3.1in]{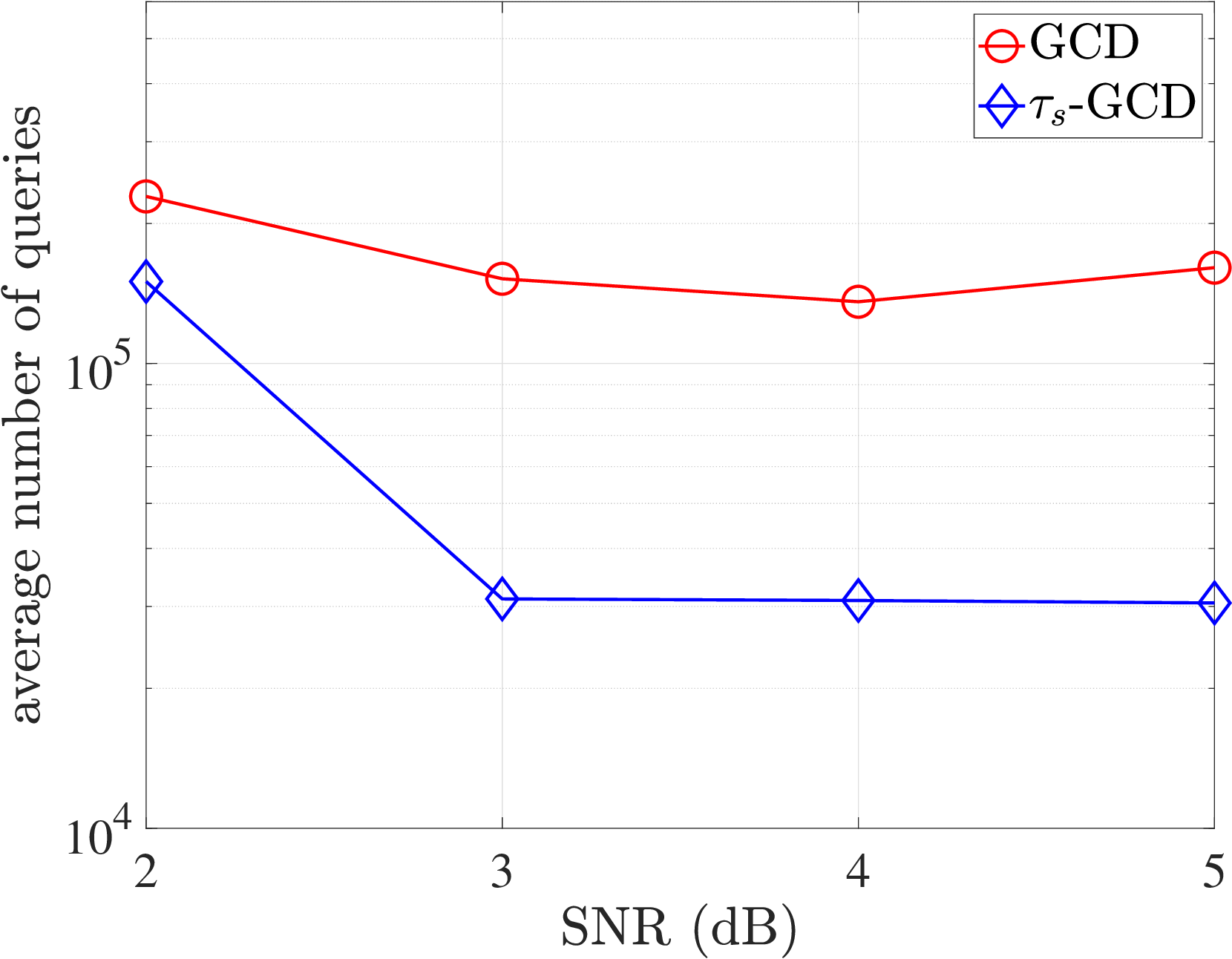}}
  \caption{
   Simulation results of the RM code $\mathscr{C}_{\rm{RM}}[64,42]$ for GCD and $\tau_s$-GCD. Here, $L=2$.
  }
   \label{fig_ufer_sotweight}  
\end{figure}

\begin{example}
Fig.~\ref{Fig_Ts} provides the CCDF of $\Gamma$ for $K=42$. Based on the simulation results from Fig.~\ref{Fig_Ts}, we compare the $\tau_s$-GCD with the GCD in terms of the performance and average number of queries for an RM code $\mathscr{C}_{\rm{RM}}[64,42]$ over BPSK-AWGN channels, as shown in Fig.~\ref{fig_ufer_sotweight}, from which we see that the $\tau_s$-GCD exhibits comparable performance to the GCD but requires a fewer average number of queries.
\end{example}

Notice that, to reduce the performance loss brought by the truncated GCD, we can carefully optimize the  parameters $\ell_{\rm{max}}$ and $\tau_s$ at a given SNR by simulating~(offline) the related CCDFs. This can be circumvented by employing the $\tau_p$-GCD, where the parameter $\tau_p$ is simply set to be lower than the target error probability by an order of magnitude. 




\subsubsection{$\tau_p$-GCD} We have the following proposition.
\begin{proposition}
The performance gap between $\tau_p$-GCD is upper bounded by 
 $\tau_p$. That is,
 \begin{equation}
\epsilon(L,\tau_p)\leq \epsilon(L,\rm{GCD}) +  \tau_p.
\end{equation}
\end{proposition}

\begin{IEEEproof}
From \eqref{TOESD-FER}, the error probability of the $\tau_p$-GCD is bounded  by
\begin{equation}
\begin{aligned}
\epsilon(L,\tau_p) &\leq \epsilon(L,\text{GCD}) +  \mathbb{P}\{\boldsymbol{e}_P \text{ is not queried}\} .
\end{aligned}
\end{equation}

Since 
\begin{equation}
\begin{aligned}
 \sum\limits_{i=1}^{2^K}P(\boldsymbol{e}_P^{(i)}|\boldsymbol{r}_P) = 1,
\end{aligned}
\end{equation}
the probability that $\boldsymbol{e}_P$ is not queried is bounded by $\tau_p$. That is,  
\begin{equation}
\begin{aligned}
\mathbb{P}\{\boldsymbol{e}_P \text{ is not queried}\}
 = \sum\limits_{i=j+1}^{2^K}P(\boldsymbol{e}_P^{(i)}|\boldsymbol{r}_P)
 \leq \tau_p,
\end{aligned}
\end{equation}
where the query process is terminated at the $j$-th query~($j< 2^K$). Hence, we have
\begin{equation}
\begin{aligned}
\epsilon(L,\tau_p) &\leq \epsilon(L,\rm{GCD}) +  \tau_p .
\end{aligned}
\end{equation}
\end{IEEEproof}





\begin{figure}[htbp]
  \centering
   \subfloat[The CCDF of $D$ approximated by saddlepoint approach.\label{uFER_51_L_saddle}]{\includegraphics[width=3.1in]{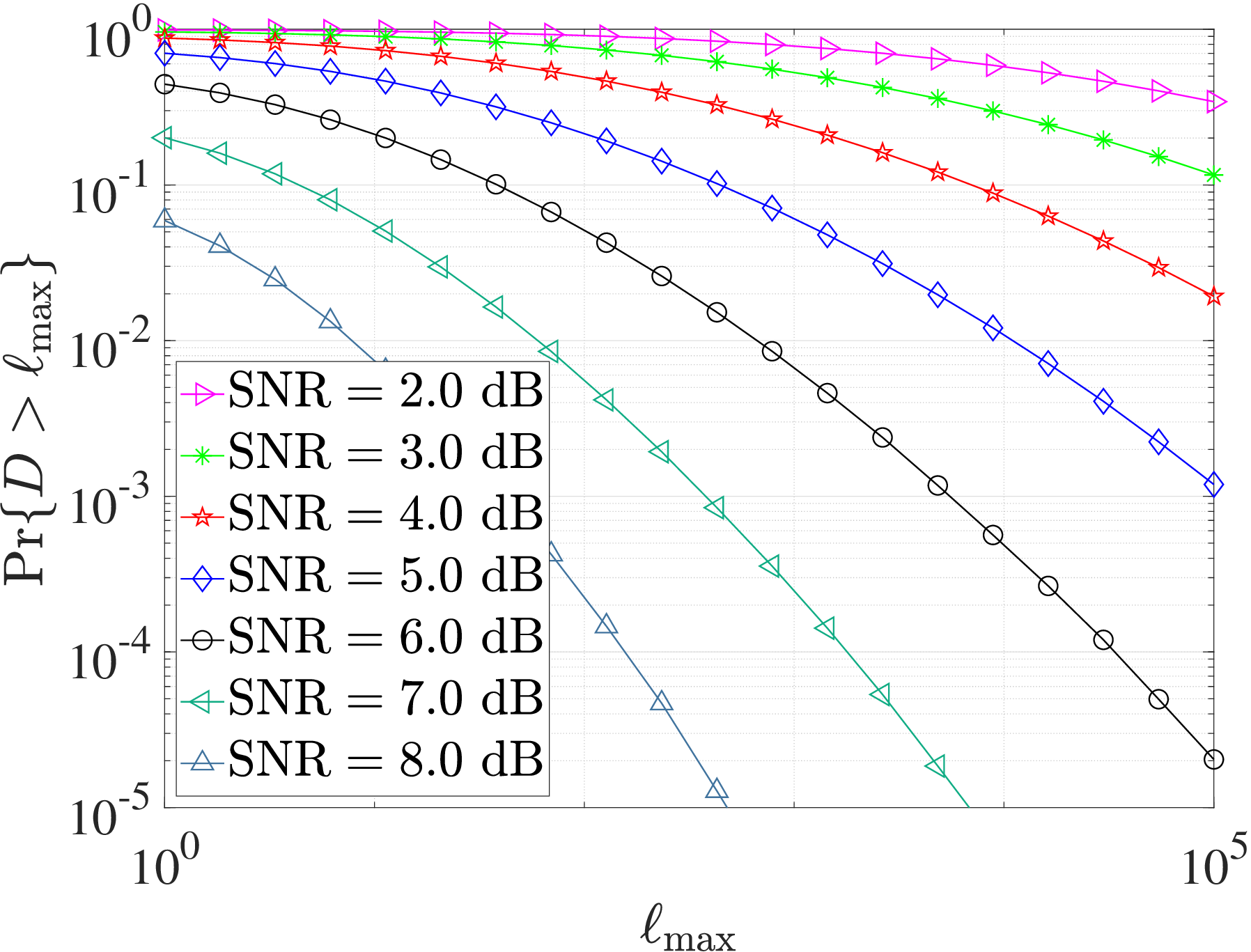}}
   \hfill
    \subfloat[The CCDF of $\Gamma$.\label{uFER_softweight}]{\includegraphics[width=3.1in]{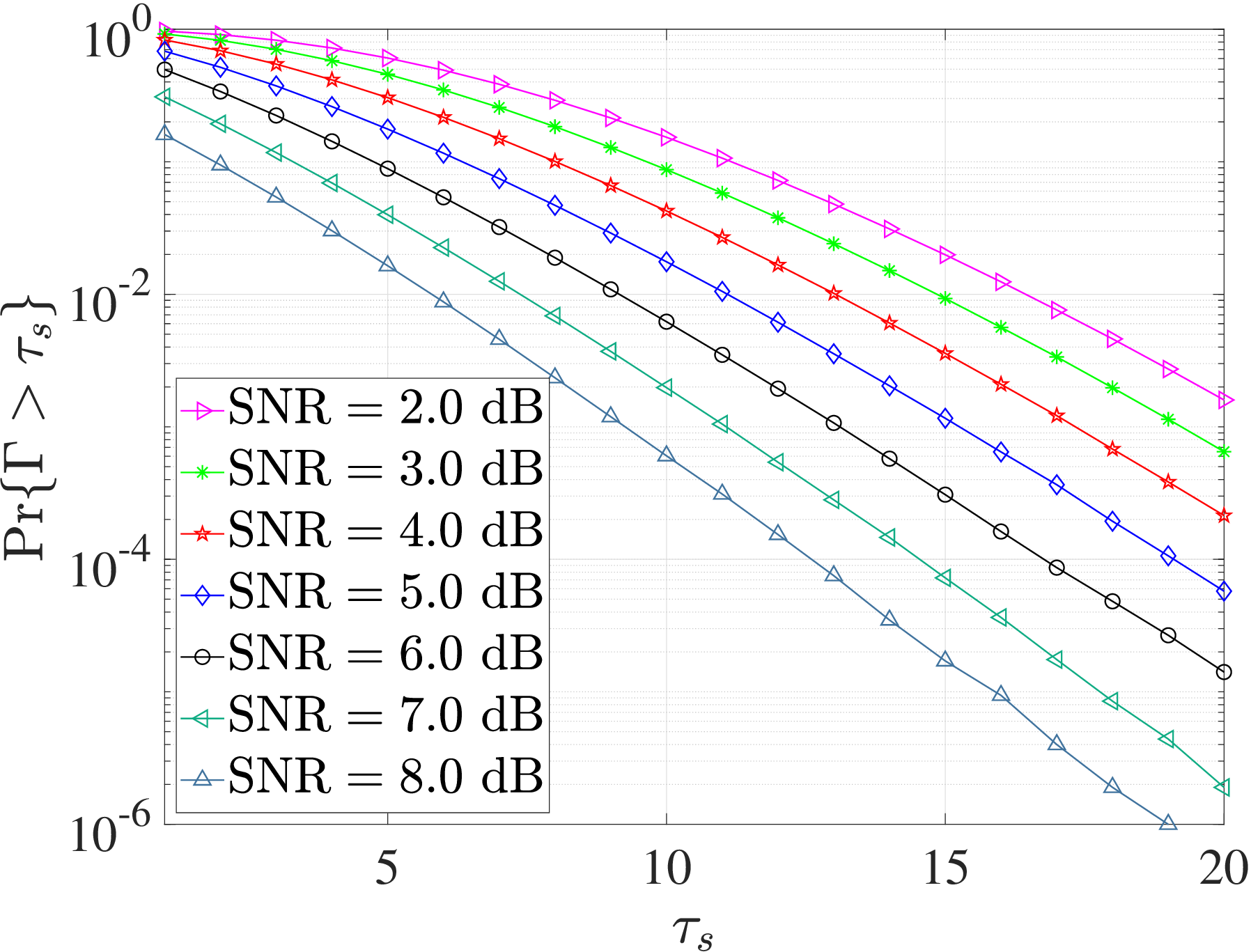}}
   \\
  \subfloat[Performance.\label{fer_eBCH51}]{\includegraphics[width=3.1in]{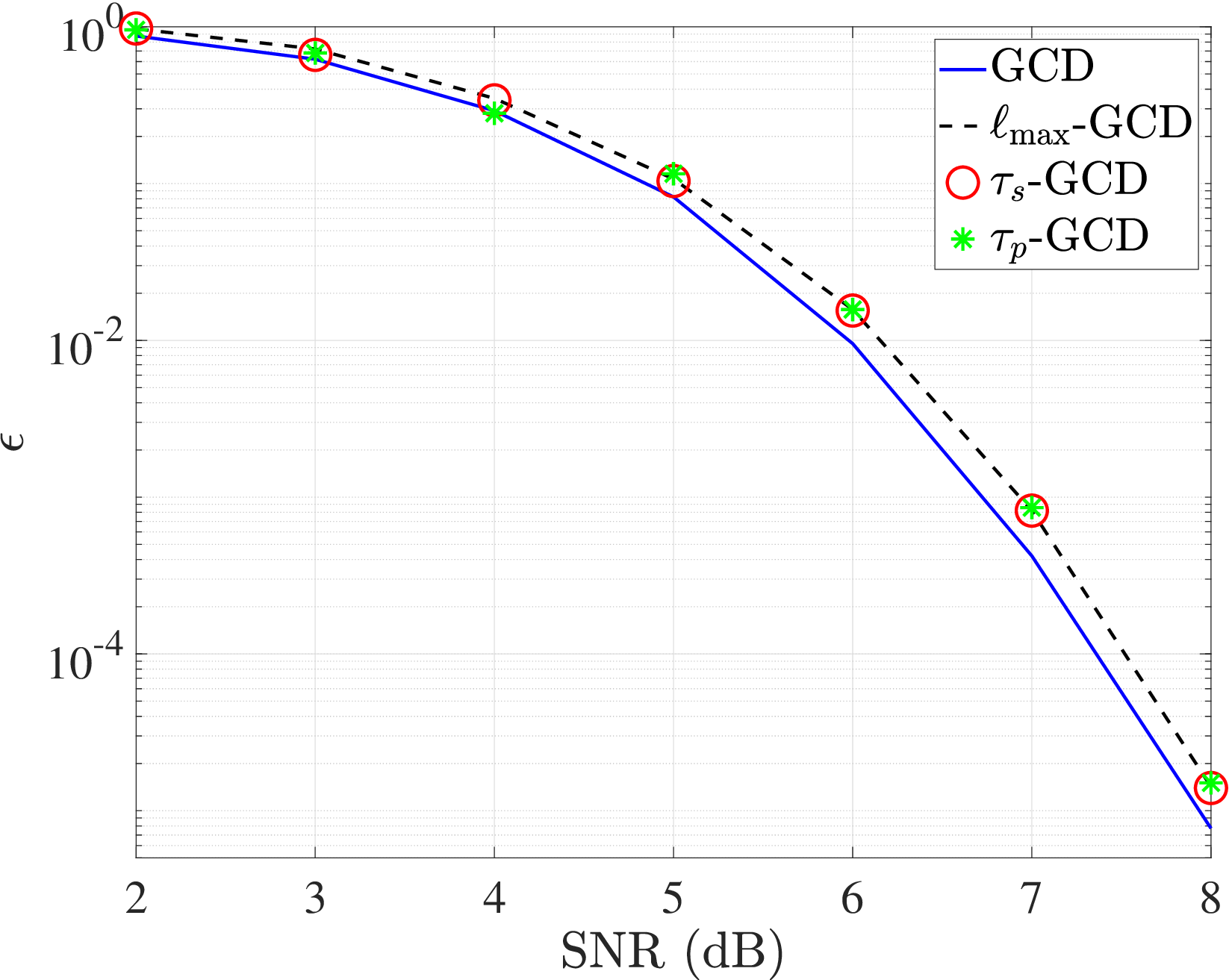}}
  \hfill
  \subfloat[Average number of queries. \label{Query_eBCH51}]{\includegraphics[width=3.1in]{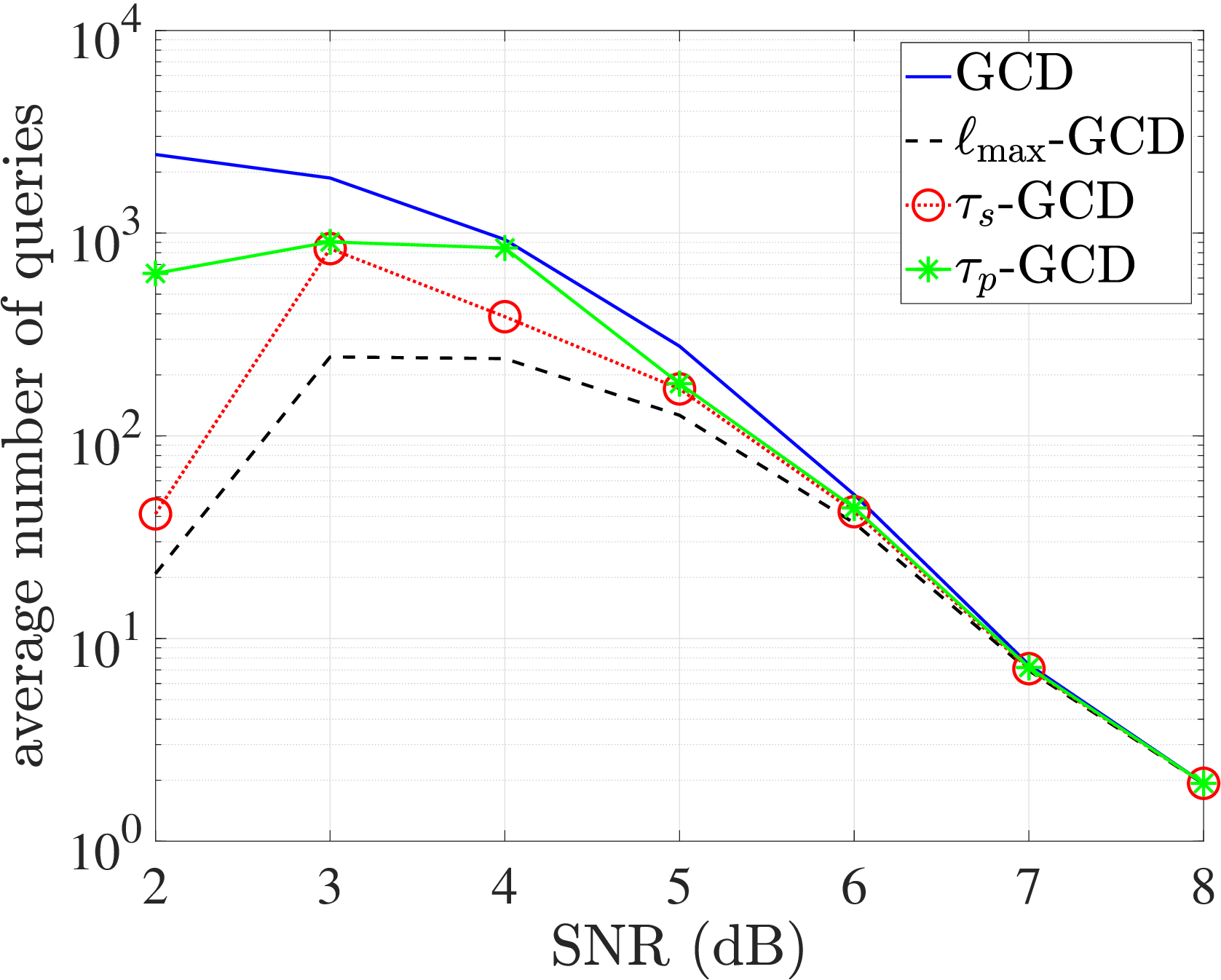}}
  \caption{
    Simulation results of the eBCH code $\mathscr{C}_{\rm{eBCH}}[64,51]$ for GCD and truncated GCD. Here, $L=1$.
  }
  \label{Fig_eBCH51_fer}
\end{figure}

\begin{example}
Consider an extended BCH~(eBCH) code $\mathscr{C}_{\rm{eBCH}}[64,51]$. We have presented the performance and the average number of queries for the GCD and three truncated GCD algorithms over BPSK-AWGN channels, along with the probabilities $ \mathbb{P}\{\boldsymbol{e}_P \text{ is not queried}\}$ for different $\ell_{\rm{max}}$ and $\tau_s$. The simulation results are shown in Figs.~\ref{Fig_eBCH51_fer}, from which we observe that the performance of the truncated GCD is similar to that of the  GCD but with a fewer average number of queries.


\end{example}


\textbf{Remarks.} It is worth pointing out that the upper bound of the performance gap does not depend on any specific code since the event that $\boldsymbol{e}_P$ is not queried is solely related to the condition to truncate the searches of partial TEPs. In other words, these three conditions are universally applicable to all linear block codes of dimension $K$~(over the same channel), as confirmed by the following example.

\begin{figure}[!t]
  \centering
  \subfloat[Performance.\label{FERk64}]{\includegraphics[width=3.1in]{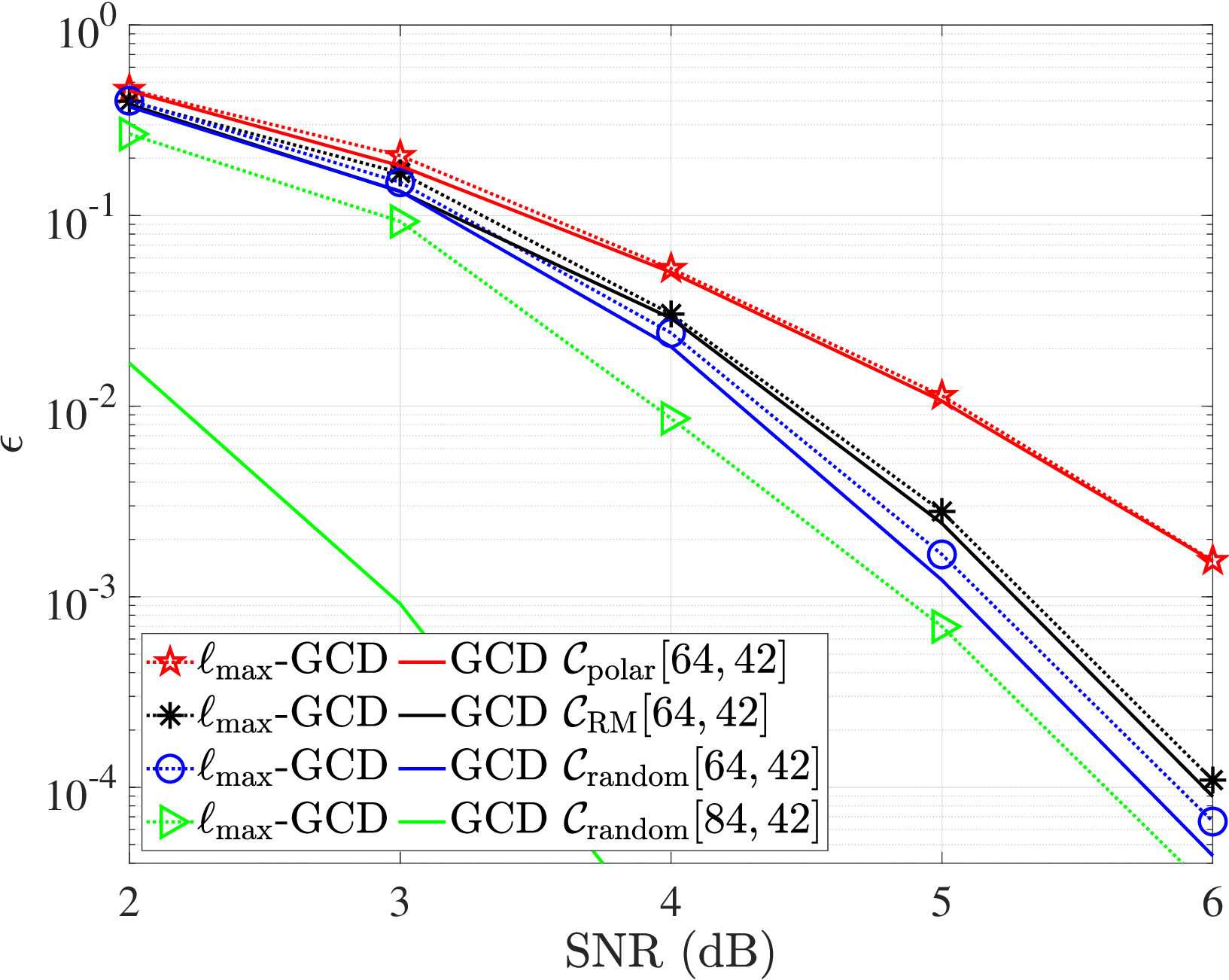}}
  \hfill
  \subfloat[Performance gaps.\label{gapk64}]{\includegraphics[width=3.1in]{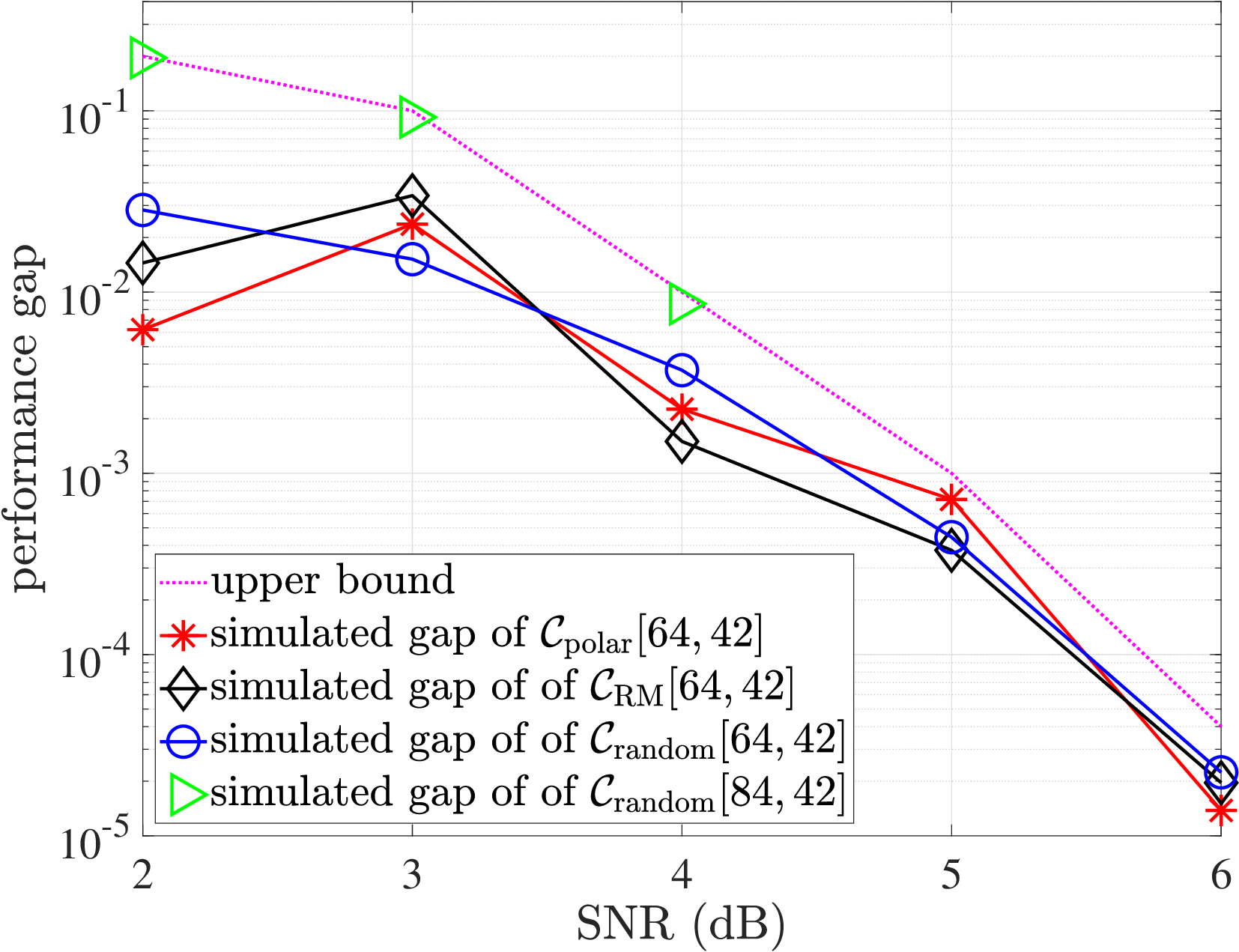}}
  \caption{
   Simulation results of the RM, polar, and random codes with the same code dimension $K=42$ for $\ell_{\rm{max}}$-GCD and GCD. Here, $L=1$.
  }
   \label{fer_K64}  
\end{figure}

\begin{example}
In this example, we present in Fig.~\ref{fer_K64} the performance of the $\ell_{\rm{max}}$-GCD and the GCD for three types of codes: RM, polar, and random codes, over BPSK-AWGN channels, along with the upper bound of the performance gaps. Specifically, given a specific SNR, $\ell_{\rm{max}}$ is set based on the optimal list decoding performance of the random code $\mathscr{C}_{\rm{random}}[64,42]$ and used to decode all three types of codes with different code lengths. The results show that the random code $\mathscr{C}_{\rm{random}}[64,42]$ performs better than the RM code $\mathscr{C}_{\rm{RM}}[64,42]$ and the polar code $\mathscr{C}_{\rm{polar}}[64,42]$. From Fig.~\ref{gapk64}, we observe that the codes with the same dimension $K=42$ have the same upper bound of the performance gap and the simulated performance gap matches well with the upper bound in the high SNR region.
\end{example}

\textbf{Remarks.} Notice that the condition $\gamma\left(\boldsymbol{e}_P\right)\geq \gamma_L$ in GCD is sufficient to identify the lightest TEPs but is typically too strong, which might be relaxed to $\gamma\left(\boldsymbol{e}_P\right) + \Delta \geq \gamma_L$  by
introducing a positive tolerance $\Delta$. If so, the GCD can be terminated earlier, resulting in a lower complexity. It has been illustrated in~\cite{LC_OSDljf2023}~\cite{2024jifangTIT} that setting 
$\Delta = \mathbb{E}[\gamma(\boldsymbol{e}_I)]$, the expection of the true partial TEP $\boldsymbol{e}_I$, incurs negligible performance loss. Also, notice that all truncations and early termination criteria can be integrated for further complexity reduction.

\subsection{Parallel GCD}



We consider implementing the~(truncated) GCD in parallel to reduce the decoding latency without compromising performance, referred to as the parallel GCD.

\begin{figure}[!t]
\centering
\includegraphics[width=5.6in]{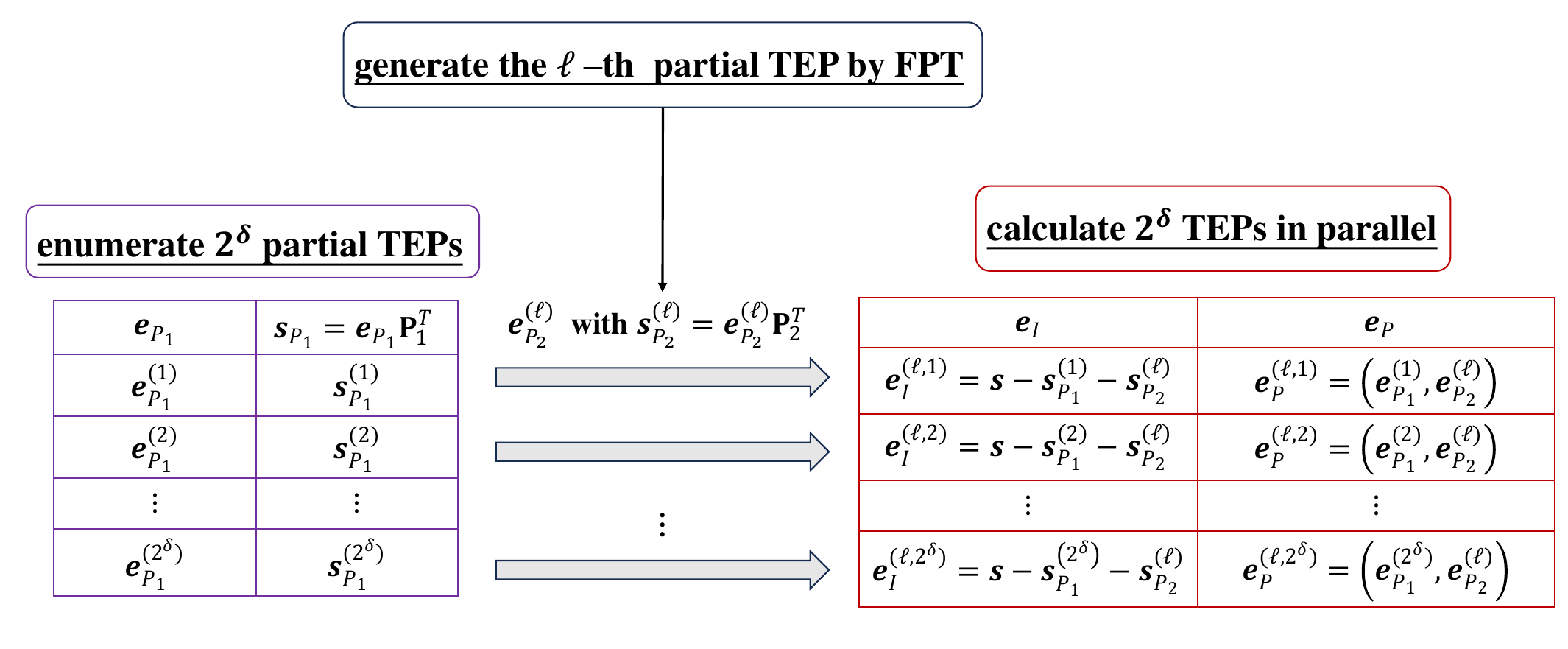}
\caption{The parallel implementation at the $\ell$-th query, where  the partial TEP $\boldsymbol{e}_{P_2}^{(\ell)}$ is delivered by the FPT algorithm and juxtaposed to $2^\delta$ TEPs in parallel.}
\label{parallelOESD_example}
\end{figure}

Without loss of generality, we assume that the reliability of the information part  is non-decreasing. That is, $$|\boldsymbol{r}_P[1]|\leq |\boldsymbol{r}_P[2]|\leq \cdots \leq |\boldsymbol{r}_P[K]|.$$ 
Given a non-negative integer $\delta \leq K$, the matrix $\mathbf{P}$ can be written as $\mathbf{P}=[\mathbf{P}_1,\mathbf{P}_2]$. Similarly, the partial TEP $\boldsymbol{e}_P$ can be written as $\boldsymbol{e}_P=\left(\boldsymbol{e}_{P_1},\boldsymbol{e}_{P_2}\right)$, in which $\boldsymbol{e}_{P_1} = \boldsymbol{e}_{P}[1:\delta]\in \mathbb{F}_2^{\delta}$ contains the first $\delta$ components in $\boldsymbol{e}_{P}$ and $\boldsymbol{e}_{P_2}=\boldsymbol{e}_{P}[\delta+1 : K]\in\mathbb{F}_2^{K-\delta}$ contains the remaining $K-\delta$ components in $\boldsymbol{e}_{P}$. 
Different from the query process over $\boldsymbol{e}_P$ in the GCD algorithm, $2^\delta$ partial TEPs $\boldsymbol{e}_{P_1}^{(i)}$~($1\leq i\leq 2^\delta$) are first enumerated, and then the query process are implemented over $\boldsymbol{e}_{P_2}$. That is, at the $\ell$-th query, $\boldsymbol{e}_{P_2}^{(\ell)}$ is produced by the FPT and juxtaposed to $\boldsymbol{e}_{P_1}$, resulting in $2^\delta$ parallel TEPs $\boldsymbol{e}^{(\ell,i)}$~($1\leq i \leq 2^\delta$), as illustrated in Fig.~\ref{parallelOESD_example}. Specifically, the juxtaposition to the $i$-th partial TEP  $\boldsymbol{e}_{P_1}^{(i)}$ is not necessary whenever 
\begin{equation}\label{parallelOESD}
\gamma\left(\boldsymbol{e}_{P_1}^{(i)}\right) + \gamma\left(\boldsymbol{e}_{P_2}^{(\ell)}\right) \geq \gamma_L,   
\end{equation}
 where $\gamma_L$ represents the current $L$-th most lightest TEP among all the queried TEPs. Thus the parallel GCD, as summarized in Algorithm~\ref{pOESD},  is terminated when the condition~\eqref{parallelOESD} is satisfied for all $2^\delta$ partial TEPs $\boldsymbol{e}_{P_1}^{(i)}$~($1\leq i\leq 2^\delta$). 

\begin{figure}[!t]
  \centering
  \includegraphics[width=3.1in]{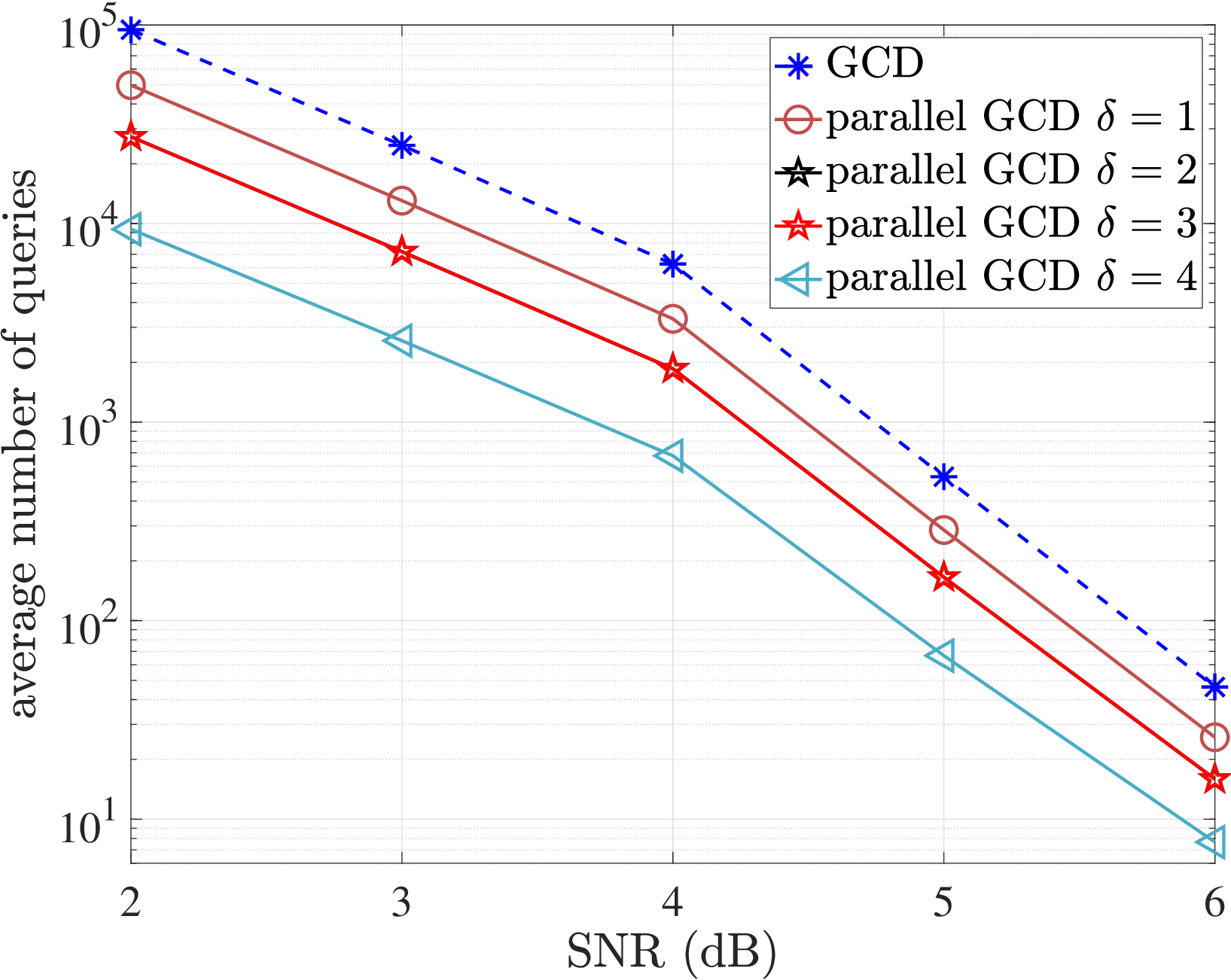}
  \caption{
   The average number of queries of the parallel GCD
and the original GCD for the RM code $\mathscr{C}_{\rm{RM}}[64,42]$. Here, $L=1$.
  }
  \label{fig_parallel}
\end{figure}

\begin{example}
Consider the RM code $\mathscr{C}_{\rm{RM}}[64,42]$. We compare the parallel GCD with the GCD over BPSK-AWGN channels in this example. The comparisons on the average number of queries are shown in Fig.~\ref{fig_parallel}, from which we see that the parallel GCD has a significant reduction in the number of queries, compared with the GCD, indicating a much lower decoding latency. In addition, as the $\delta$ increases, the reduction becomes more significant.
\end{example}

\newpage
\begin{breakablealgorithm}
\renewcommand{\algorithmicrequire}{\textbf{Input:}}
    \renewcommand{\algorithmicensure}{\textbf{Output:}}
\caption{Parallel GCD}\label{pOESD}
\begin{algorithmic}[1]
\Require The parity-check matrix $[\mathbf{I}\quad \mathbf{P}]$, the LLR vector $\boldsymbol{r}$, the list size $L$, an integer $\delta$,  and the maximum query number $\ell_{\rm{max}}=2^K$.
\State Initialization: $\mathcal{L}=\{\boldsymbol{f}^{(i)},1\leq i \leq L\}$ is a linked list of size $L$ which is maintained in order~(with non-decreasing soft weight) during the query process and the initial $L$ elements in $\mathcal{L}$ are set to NULL TEPs with soft weight $\infty$.
\State $\mathcal{N}=\{\boldsymbol{e}_{P_1}^{(i)}, 1\leq i\leq 2^\delta \}$. \Comment{Enumerate $2^\delta$ partial TEPs.}
\State $\boldsymbol{s}_{P_1}^{(i)}=\boldsymbol{e}_{P_1}^{(i)}\mathbf{P}_1^T$, $t_i \gets 0$, for $ 1\leq i\leq 2^\delta$.
\State $m_t\gets 0$.
\For{$\ell = 1,\dots,\ell_{\rm{max}}$}
\If{ certain conditions for truncation are satisfied}
\State \textbf{break.}
\EndIf
\If{$m_t == 2^\delta$}
\State \textbf{break}.
\Else
\State Generate the $\ell$-th lightest partial TEP $\boldsymbol{e}_{P_2}^{(\ell)}$.
\For{$i = 1,\dots, 2^\delta$ in parallel }
\If{$t_i==1$} 
\State \textbf{continue}.
\EndIf
\If{$ \gamma\left(\boldsymbol{e}_{P_1}^{(i)}\right)+ \gamma\left(\boldsymbol{e}_{P_2}^{(\ell)}\right) \geq \gamma\left(\boldsymbol{f}^{(L)}\right)$}
\State $t_i \gets 1$, $m_t \gets m_t + 1$.
\State \textbf{continue}.
\Else
\State $\boldsymbol{e}_P^{(\ell,i)}=\left(\boldsymbol{e}_{P_1}^{(i)},\boldsymbol{e}_{P_2}^{(\ell)}\right)$.
\State $\boldsymbol{e}_I^{(\ell,i)} = \boldsymbol{s}- \boldsymbol{s}_{P_1}^{(i)}-\boldsymbol{e}_{P_2}^{(\ell,i)}\mathbf{P}_2^T$.
\State $\boldsymbol{e}^{(\ell,i)} = \left(\boldsymbol{e}_I^{(\ell,i)} , \boldsymbol{e}_P^{(\ell,i)}\right)$.
\If{$\gamma\left(\boldsymbol{e}^{(\ell,i)}\right)<\gamma\left(\boldsymbol{f}^{(L)}\right)$}
\State Update the list $\mathcal{L}$ by removing $\boldsymbol{f}^{(L)}$ and inserting $\boldsymbol{e}^{(\ell,i)}$.
\EndIf
\EndIf
\EndFor
\EndIf
\EndFor
\Ensure The list $\mathcal{L}$ and the corresponding codewords $\{\boldsymbol{c}|\boldsymbol{c}=\boldsymbol{z}-\boldsymbol{e}, \boldsymbol{e}\in \mathcal{L} \}$.
\end{algorithmic}
\label{alg3}
\end{breakablealgorithm}

\vspace{0.5cm}

\section{Application to Polar Decoding} 
Although the GCD is developed for short codes and only efficient for high-rate or low-rate codes, it can be embedded in the decoders for long codes that are constructed from short codes, including polar codes~\cite{ref1Arikan}, block oriented unit memory convolutional codes~(BOUMCCs)~\cite{BOMCC2021} and twisted-pair superposition transmission~(TPST) codes~\cite{TPSTTCOM}, among others. In this section, we apply the GCD to polar codes with a wide range of code lengths.



We consider a polar code $\mathscr{C}[N,K]$ with length $N=2^m$. We select $K$ reliable subchannels indexed by the subset $\mathcal{A} \subseteq\{1,2,\dots, N-1 \}$ to transmit the information bits $\boldsymbol{u}_{\mathcal{A}}$ and the remaining $N-K$ subchannels indexed by the complementary set $\mathcal{A}_c$ to transmit frozen bits $\boldsymbol{u}_{\mathcal{A}_c}$
(set to zeros here). Then the codeword is obtained by $\boldsymbol{c} = \boldsymbol{u} \mathbf{G}_m$, where\footnote{Notice that the notation $\boldsymbol{u} = (\boldsymbol{u}_{\mathcal{A}}, \boldsymbol{u}_{\mathcal{A}_c})$ does not mean that the active bits $\boldsymbol{u}_{\mathcal{A}}$ are located at the left  to the frozen bits $\boldsymbol{u}_{\mathcal{A}_c}$. Instead, it means that $\boldsymbol{u}_{\mathcal{A}} = (u_i:i\in \mathcal{A})$ denotes $K$ active bits indexed by $\mathcal{A}$ and $\boldsymbol{u}_{\mathcal{A}_c} = (u_i:i\in \mathcal{A}_c)$ denotes the $N-K$ frozen bits indexed by $\mathcal{A}_c$.} $\boldsymbol{u} = (\boldsymbol{u}_{\mathcal{A}}, \boldsymbol{u}_{\mathcal{A}_c})$ and $\mathbf{G}_m \in \mathbf{F}_2^{N \times N}$ is the Arikan matrix. The parity-check matrix $\mathbf{H}$ is formed by the columns of $\mathbf{G}_m$ with indices in $\mathcal{A}_c$~\cite{2010GoelaLP}. Suppose that $\boldsymbol{c}$ is modulated by BPSK into $\boldsymbol{x}$ with $x_i = (-1)^{c_i}$, $1 \leq i \leq N$ and then transmitted through an AWGN channel with zero mean and variance $\sigma^2$, resulting in $\boldsymbol{y} = \boldsymbol{x} + \boldsymbol{n}$, where $\boldsymbol{n} \sim \mathcal{N}(\textbf{0}, \sigma^2\mathbf{I}_n)$.

\subsection{Polar Decoding Tree}
The conventional polar decoding can be implemented over a full binary tree with $m$ levels and $2^m$ leaf nodes at the $m$-th level. We index the leaf nodes from left to right by $\{1,2,\dots,N\}$. Given an inner node $v$,  we denote its left child node and right child node by $v_l$ and $v_r$, respectively. Let $\mathcal{I}_v$ be the set of the indices of all leaf nodes which are the descendents of the node $v$. Each node $v$ at the $i$-th level, $0\leq i \leq m$, is associated with a polar sub-code $\mathscr{C}[n,k]$ of length $n = 2^{m-i}$ and dimension $k = |\mathcal{I}_v \cap \mathcal{A}|$. Specifically, the root node is associated with the polar code $\mathscr{C}[N,K]$, and the leaf node is associated with either an active bit or a frozen bit. 
Upon receiving $\boldsymbol{y}$, each node is also associated with an LLR vector $\boldsymbol{r} \in \mathbb{R}^{n}$ and a vector $\boldsymbol{\beta} \in \mathbb{F}_2^n$. 

The SC decoding algorithm starts by initializing the root node with $ \boldsymbol{r} = (r_1,r_2,\dots,r_{N})$ corresponding to the received vector $\boldsymbol{y}$ and then performs recursively according to the following rules.
\begin{enumerate}
    \item For an inner node $v$ at the $i$-th level,
    \begin{enumerate}
        \item when its LLR vector is available, the LLR vector associated with its left child $\boldsymbol{r}_{v_l}$ is computed by
    \begin{equation}\label{ffun}
        \boldsymbol{r}_{v_l}[j] = f(\boldsymbol{r}_{v}[j],\boldsymbol{r}_{v}[j+2^{m-i-1}])
    \end{equation}
    for $1 \leq j \leq 2^{m-i-1}$, where 
    \begin{equation}
         f(a,b)=\rm{sgn}(a)\rm{sgn}(b)\min (|a|,|b|).
    \end{equation}
        \item when $\boldsymbol{\beta}_{v_l}$ of its left child node $v_l$ is available,
         \begin{equation}\label{gfun}
        \boldsymbol{r}_{v_r}[j]= g(\boldsymbol{r}_{v}[j],\boldsymbol{r}_{v}[j+2^{m-i-1}], \boldsymbol{\beta}_{v_l}[j])
    \end{equation}
    for $1 \leq j \leq 2^{m-i-1}$, where 
    \begin{equation}
        g(a,b,c)=(-1)^c a + b.
    \end{equation}
        \item when $\boldsymbol{\beta}_{v_l}$ and $\boldsymbol{\beta}_{v_r}$ are available,  its associated vector $\boldsymbol{\beta}_{v}$ is computed via
        \begin{equation}\label{xor}
    \boldsymbol{\beta}_{v} = \left( \boldsymbol{\beta}_{v_l}\oplus \boldsymbol{\beta}_{v_r},\boldsymbol{\beta}_{v_r} \right),
\end{equation}
    where $\oplus$ denotes the bit-wise XOR.
    \end{enumerate}
    \item For a leaf node $v$ at the $m$-th level, if $|\mathcal{I}_v\cap \mathcal{A}| = 0$, $z = 0$. If $|\mathcal{I}_v\cap \mathcal{A}| = 1$, then
    \begin{equation}\label{SCLOESDhard}
      \begin{cases}
        z = 0, & r \geq 0\\
        z = 1, & \rm{otherwise}
    \end{cases},  
    \end{equation}
    where $r$ is the~(available) associated LLR. For the SC decoding, set $\beta = z$, while, for the SCL decoding, set $\beta = 0$ and $1$ for path extension.
\end{enumerate}

Instead of making hard decisions at each active leaf node $v$ satisfying $|\mathcal{I}_v\cap \mathcal{A}| = 1$, the SCL decoding algorithm extends each survival path at an active leaf node into two paths by setting $\beta = 0$ and $1$, and retains $L$ best paths at each decoding step based on the path metric.



\begin{figure}[!t]
\centering
\subfloat[]{\includegraphics[width=1.5in]{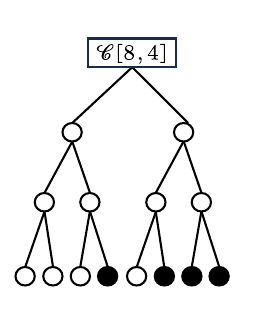}\label{SCLtree}
}
\hfil
\subfloat[]{\includegraphics[width=2.0in]{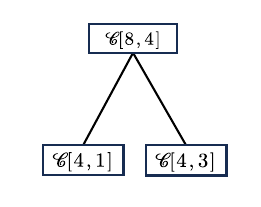}\label{twonodes}
}
\hfil
\subfloat[]{\includegraphics[width=2.0in]{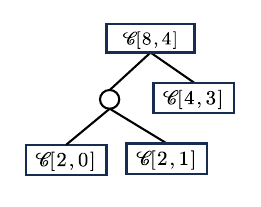}\label{fournodes}
}
\caption{Three polar trees of the code $\mathscr{C}[8,4]$. (a) The conventional tree with $3$ levels and $8$ leaf nodes,  where black circles denote the active bits and white circles at the leaf nodes denote the frozen bits. (b) The pruned tree with one level and $2$ leaf nodes. (c) The pruned tree with $2$ levels and $3$ leaf nodes.}
\label{TREE}
\end{figure}

The polar code $\mathscr{C}[N,K]$ is conventionally represented by a polar tree consisting of $m=\log N$ levels and $N$ leaf nodes. The polar tree can also be pruned to reduce the number of levels and the leaf nodes. As an example, Fig.~\ref{TREE} illustrates three different polar trees of the polar code $\mathscr{C}[8,4]$, where Fig.~\ref{SCLtree} is the conventional polar decoding tree and Figs.~\ref{twonodes} and~\ref{fournodes} are two pruned polar trees. For the traditional polar tree, we use the SCL decoding algorithm to make decisions bit by bit in serial. While, for the pruned polar tree, we naturally use the GCD algorithm to make multiple-bit-wise decisions for the leaves with low-rate and high-rate codes. In the following subsection, we will discuss how to prune the polar tree such that the multiple-bit-wise SCL decoding algorithm has relatively low complexity.

\subsection{Strategy for Pruning the Polar Decoding Tree}
Let $v$ be a node of the polar tree and  $\mathscr{C} [n,k]$ be the associated sub-code. For the list decoding algorithms with a list size $L$, the node $v$
is associated with  $L$ LLR vectors~(if any), which correspond to $L$ partial paths up to $v$. These partial paths are then extended by appending $n$ bits according to certain rules. This can be done at least by one of the two list decoding algorithms, namely, the SCL decoding algorithm,  and the GCD algorithm. 
To extend a partial path, the SCL decoding algorithm requires a pre-order traversal of the subtree to make choices bit by bit at $n$ leaf nodes and a post-order traversal of the subtree to obtain the final $L$ candidates associated with node $v$. This means that the SCL algorithm visits twice all the $2n-1$ nodes of the subtree stemming from $v$. The complexity of the SCL algorithm can be evaluated as
\begin{equation}\label{SCL}
C^{\rm{SCL}} = \underbrace{\mathcal{O}\left(2kL\log(2L)\right)}_\text{sorting} 
+ \underbrace{\mathcal{O}\left(Ln\log n\right)}_{\text{f-g calculation}}
+ \underbrace{\mathcal{O}\left(L\frac{n}{2}\log n\right)}_\text{transformation}.
\end{equation}
In contrast, the GCD algorithm visits the node $v$ only once without going into the subtree. Let $\ell_{\rm{avg}}$ be the average query number per LLR vector. From~\eqref{LIST-GRAND},  the complexity of the $L$ parallel GCD processors can be evaluated by
 \begin{equation}\label{LOESD}
\begin{aligned}
\centering
 C_{\rm{avg}}^{L \text{ GCD}} 
 &=\underbrace{\mathcal{O}\left(Lk\log k\right)}_{\text{sorting}}
+\underbrace{\mathcal{O}\left(\ell_{\text{avg}} \left(L\log \ell_{\rm{avg}}+\log L + Lk\right)\right)}_{\text{searching}} +
\underbrace{\mathcal{O}\left({L\ell}_\text{avg}(n-k)\right)}_\text{re-encoding}.   
\end{aligned}
\end{equation}

Based on the complexity shown in~\eqref{SCL}-\eqref{LOESD}, we consider the strategy for pruning the polar tree in the following. The main motivation and the basic idea of the strategy is to prune the conventional polar tree such that the GCD algorithm for each resulting leaf node of the pruned polar tree has lower complexity than the conventional~(bit-wise) SCL algorithm. This finally leads to a pruned tree with leaf nodes, referred to as the GCD nodes,  that perform the GCD algorithm.

\begin{figure}[!t]
\centering
\includegraphics[width=2.3in]{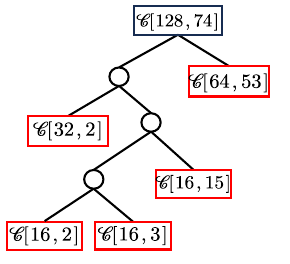}
\caption{A pruned polar tree of $\mathscr{C}[128,74]$, where the red box denotes the GCD node.}
\label{partitioning_128}
\end{figure}

We propose to visit the nodes over a polar tree in a pre-order traversal. To visit a node $v$~(initially being the root node), we perform the genie-aided GCD algorithm\footnote{The genie-aided GCD algorithm knows the correct partial path preceded the current node.} with a list size $L$ and a maximum query number $\ell_{\rm{max}}$ to estimate~(by simulations) the average query number $\ell_{\rm{avg}}$. Whenever the complexity of the GCD algorithm is less than that of the SCL decoding algorithm on the node, the node $v$ serves as a GCD node and the sub-tree stemming from the node $v$ is pruned. Notice that the offline strategy can be refined to achieve further complexity and decoding reduction by merging or splitting nodes into relatively high-rate and low-rate codes. As an example, the pruned polar tree of a polar code $\mathscr[128,74]$ has $4$ levels and $5$ leaves, as shown in Fig.~\ref{partitioning_128}.


We now compare our strategy with the existing approaches that are based on special bit patterns to prune the polar decoding tree. The commonly used nodes for fast SCL~\cite{SarkisFSSC,HanifFSSC,CondoFSSC,Zheng2020,ZhengSRnodes2021} are
tabulated in Table~\ref{tab_nodes}, from which we see that these node types depend on the bit patterns, i.e., the specific frozen bit positions or the specific information bit positions. In contrast, we derive the pruned tree based on the complexity analysis, where the leaf nodes are identified into GCD nodes with a relatively high~(or low) rate. By the pruning strategy, a GCD node decoded by the GCD has lower complexity compared with the SCL decoding.

\begin{table*}
\begin{center}
\caption{The commonly used nodes for fast SCL algorithms}
\label{tab_nodes}
\begin{tabular*}{0.95\linewidth}{@{}l|l@{}}
\hline
Node Type & Definition \\
\hline
Rate-0 node & all bits are frozen.  \\
Rate-1 node & all bits are information bits. \\
REP node & all bits are frozen except the last one. \\
SPC node & all bits are information bits except the first one.\\
Type-I node & all bits are frozen bits except the last two.\\
Type-II node & all bits are frozen bits except the last three. \\
Type-III node & all bits are information bits except the first two.\\
Type-IV node & all bits are information bits except the first three. \\
Type-V node &  all bits are frozen bits except the last three and the fifth-to-last. \\
G-REP node & all its descendants are rate-0 nodes, except the rightmost one at a certain level that is a generic node.\\
G-PC node & all its descendants are rate-1 nodes, except the leftmost one at a certain level that is rate-0.\\
SR node & (rate-0 node/REP node, $\cdots$, rate-0 node/REP node, a generic node). \\
\hline
\end{tabular*}
\end{center}
\end{table*}

\subsection{Successive Cancellation List Decoding Algorithm by GCD}
In this subsection, we present a multiple-bit-wise SCL decoding algorithm by GCD. Given a pruned polar tree of the polar code $\mathscr{C}[N,K]$, our proposed decoding algorithm is similar to the SCL decoding algorithm, with the main difference lying in the decoding process of the leaf nodes. For the SCL decoding algorithm, a leaf node is associated with a single bit and the decoding is performed bit by bit. In contrast, for the proposed decoding algorithm, a leaf node is associated with a short code and the decoding can be performed sub-block by sub-block by employing the GCD algorithm, which extend each partial path into $L$ best candidates, resulting in $L^2$ extended partial paths. From the perspective of performance, the proposed algorithm has the following property.

\begin{proposition}\label{lemma2}
 Given $L$ partial paths, for decoding the sub-code $\mathscr{C}[n,k]$ corresponding to some leaf node in the pruned polar tree, the $L$ extended paths delivered by the proposed algorithm are no worse~(in terms of paths metrics) than those delivered by the SCL decoding algorithm.
\end{proposition}

\begin{IEEEproof}
Given $L$ partial paths, the GCD algorithm  extends each partial path into  $L$ best candidates, resulting in $L^2$ extended partial paths, from which $L$ optimal candidates from the code $\mathscr{C}[n,k]$ can be selected by sorting.

The SCL algorithm is a greedy algorithm, which extends each partial path into two paths at each active leaf and keeps $L$ best paths from $2L$ extended paths according to the local path metric. Evidently, some partial paths are discarded halfway and do not survive to the final sorting processing. As a result, the final output of the conventional SCL algorithm may not be the $L$ most likely ones from the code $\mathscr{C}[n,k]$.
\end{IEEEproof}

\begin{example}
To illustrate the difference between the above two list decoding algorithms for decoding the sub-code $\mathscr{C}[n,k]$, we present the list decoding process of the $\mathscr{C}[3,3]$ with $L=2$, as shown in Fig.~\ref{Fig__SCLproof_example}. The GCD algorithm outputs the two optimal paths selected from the final $16$ extended paths at step $3$. In contrast, the SCL decoding algorithm only retains two extended paths at each step and outputs the two paths selected from the four extended paths at step $3$.
\end{example}

\begin{figure}[!t]
\centering
\includegraphics[width=3.0in]{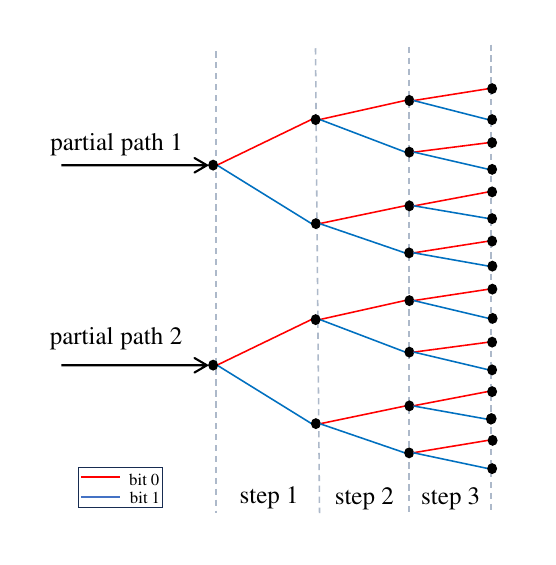}
\caption{The list decoding process of the $\mathscr{C}[3,3]$ with $L=2$ to illustrate the difference between the two list decoding algorithms, the SCL decoding algorithm and the GCD algorithm.}
\label{Fig__SCLproof_example}
\end{figure}

\textbf{Remarks.} Notice that Proposition~\ref{lemma2} says that, given the same  partial paths, the performance of any leaf node decoded by the proposed decoding algorithm is not worse than that with the conventional SCL decoding algorithm.  This is also intuitively true for the whole code, but we have no rigorous proof.

\subsection{Parallel GCD Algorithm with the Early Stopping Criteria}
Given a pruned polar decoding tree that has $q$ leaves, in this subsection, we show how to extend partial paths at a GCD node, denoted by $v_i$ and associated with a sub-code $\mathscr{C}_i[n_i,k_i]$~($1\leq i \leq q$). We assume that, for $i \geq 1$, the partial paths generated from the leaf node $v_{1}$ to $v_{i}$ are collectively denoted by $\mathcal{L}_{i}$ and that $\mathcal{L}_0 = \emptyset$ for convention. The objective of decoding at the GCD node $v_i$ is to extend each partial path in $\mathcal{L}_{i-1}$ into (at most) the $L_i$ most likely valid paths~(validated by the frozen bits and the possibly existing CRC bits of the sub-code) and then select the $L_i$ most likely valid paths from $L_{i-1}\cdot L_i$ valid paths. Here, $L_i = |\mathcal{L}_i| \geq 1$ and we allow $L_i \neq L_{i-1}$ for more general use cases. For each partial path in $\mathcal{L}_{i-1}$ indexed by $p\in\{1,2,\dots,L_{i-1}\}$, denote by $\boldsymbol{\beta}_{1 \rightarrow (i-1)}^{(p)} = \left(\boldsymbol{\beta}_1^{(p)},\boldsymbol{\beta}_2^{(p)},\dots,\boldsymbol{\beta}_{i-1}^{(p)}\right)$ the bit sequence associated with the $p$-th partial path through the first $(i-1)$ leaf nodes such that $\boldsymbol{\beta}_j^{(p)} \in \mathscr{C}_j$~($1\leq j<i$). Then the LLR vector at the node $v_i$ associated with the partial path $\boldsymbol{\beta}_{1 \rightarrow (i-1)}^{(p)}$, denoted as $\boldsymbol{r}_i^{(p)}$, can be derived from  \eqref{ffun}-\eqref{xor}. Given the LLR vector $\boldsymbol{r}_i^{(p)}$ and the corresponding hard-decision $\boldsymbol{z}_i^{(p)}$, the GCD generates a list of valid TEPs and remains the $L_i$ lightest ones $\left\{\boldsymbol{e}_i^{(p,\ell)},1\leq \ell \leq L_i\right\}$, corresponding to $L_i$ sub-codewords $\left\{\boldsymbol{\beta}_i^{(p,\ell)} = \boldsymbol{e}_i^{(p,\ell)}+\boldsymbol{z}_i^{(p)},\right.$ $\left. 1\leq \ell \leq L_i\right\} \subseteq \mathscr{C}_i$. Then the $p$-th partial path $\boldsymbol{\beta}_{1 \rightarrow (i-1)}^{(p)}$ can be extended to $L_i $ partial paths $\left\{\boldsymbol{\beta}_{1 \rightarrow i}^{(p,\ell)} = \left(\boldsymbol{\beta}_{1 \rightarrow (i-1)}^{(p)}, \boldsymbol{\beta}_i^{(p,\ell)} \right), 1\leq \ell \leq L_i\right\}$. For $i \geq 1$, to a path $\boldsymbol{\beta}_{1 \rightarrow i}^{(p,\ell)} = \left(\boldsymbol{\beta}_{1 \rightarrow (i-1)}^{(p)}, \boldsymbol{\beta}_i^{(p,\ell)} \right)$, we assign recursively a path metric $\lambda\left(\boldsymbol{\beta}_{1 \rightarrow i}^{(p,\ell)}\right)= \lambda\left(\boldsymbol{\beta}_{1 \rightarrow (i-1)}^{(p)}\right)+ \gamma\left(\boldsymbol{e}_{i}^{(p,\ell)}\right)$ with
$\lambda\left(\boldsymbol{\beta}_{1\rightarrow 0}\right)=0$. It can be verified that the SCL decoding attempts to find  $L$ paths with minimum path metrics.



To reduce the latency, we may perform $L_{i-1}$ GCD processors in parallel, each for one preceded partial path in $\mathcal{L}_{i-1}$. If so, the $L_{i-1}$  GCD may require different numbers of queries to identify the most likely $L_{i}$ valid extended paths, indicating that the “speed” of generating the most likely $L_{i}$ valid paths can be different for different preceded partial paths. To terminate some GCD earlier and avoid unnecessary queries, we propose the early stopping criteria at the GCD nodes without sacrificing the performance. The basic idea is to stop as early as possible the GCD for those preceded paths that cannot generate any candidates better
than the worst valid paths in $\mathcal{L}_i$ if $\mathcal{L}_i$ already has $L_i$ members, i.e., $|\mathcal{L}_i| = L_i$. The parallel GCD algorithm with the early stopping criteria for the GCD node $v_{i}$, whose systematic parity-check matrix is denoted as $\mathbf{H}=[\mathbf{I}\quad \mathbf{P}]$, is summarized in Algorithm~\ref{alg:alg3}.
Specifically, for the $q$-th GCD node, the goal is to find the optimal valid path. In this case, we keep only the up-to-date best valid path in $\mathcal{L}_{q}$ during the decoding process in Algorithm~\ref{alg:alg3}, i.e., $L_{q} = 1$.

\begin{algorithm}[!t]
\renewcommand{\algorithmicrequire}{\textbf{Input:}}
    \renewcommand{\algorithmicensure}{\textbf{Output:}}
\caption{Parallel GCD Algorithm with the Early Stopping Criteria}\label{alg:alg3}
\begin{algorithmic}[1]
\Require The partial paths $\mathcal{L}_{i-1} = \left\{\boldsymbol{\beta}_{1 \rightarrow (i-1)}^{(p)}, 1\leq p \leq L_{i-1} \right\}$, the corresponding path metrics $\mathcal{P}_{{i-1}} = \left\{\lambda\left(\boldsymbol{\beta}_{1 \rightarrow (i-1)}^{(p)}\right), 1\leq p \leq L_{i-1}\right\}$, the maximum query number $\ell_{\rm{max}}\leq 2^{k_i}$, and the parity-check matrix $\mathbf{H}=[\mathbf{I}\quad \mathbf{P}]$ associated with the node $v_i$.
\State Initializations:

1.a: $\mathcal{L}_i=\{\boldsymbol{f}^{(j)},1\leq j \leq L_i\}$ is a linked list of size $L_i$ which is maintained in order 

\quad\  (non-decreasing path metric) and initialized as $L_i$ NULL paths with metric $\infty$.

1.b: $t_p \gets 0$, for $1 \leq p \leq L_{i-1}$.

1.c: $m_t \gets 0$.
\State Calculations of LLRs: 
Given a partial path $\boldsymbol{\beta}_{1 \rightarrow (i-1)}^{(p)} \in \mathcal{L}_{i-1}$ for $1 \leq p \leq L_{i-1}$, calculate $\boldsymbol{r}_{i}^{(p)}$ following \eqref{ffun}-\eqref{xor}, $\boldsymbol{z}_i^{(p)}$ following \eqref{xor}-\eqref{SCLOESDhard}, and  $\boldsymbol{s}_i^{(p)} = \boldsymbol{z}_i^{(p)}\mathbf{H}^T$.
\For{$\ell = 1,2,\dots,\ell_{\text{max}}$} \Comment{The $\tau_p$-truncation and the $\tau_s$-truncation along with other early termination criteria can also be integrated here.}
\For{$p =1,2,\dots, L_{i-1}$ in parallel}
\If{$t_{p} == 0$}
\State $\boldsymbol{e}_{P}^{(p,\ell)} \in \mathbb{F}_2^{k_i} \gets$ the $\ell$-th partial TEP generated from the $p$-th GCD processor.
\If{$\gamma\left(\boldsymbol{e}_{P}^{(p,\ell)}\right) + \lambda\left(\boldsymbol{\beta}_{1 \rightarrow (i-1)}^{(p)}\right) \geq \lambda\left(\boldsymbol{f}^{(L_i)}\right)$}
\State $t_{p} \gets 1$, $m_t \gets m_t + 1$.
\Else
\State $\boldsymbol{e}_{I}^{(p,\ell)}=\boldsymbol{s}_i^{(p)}-\boldsymbol{e}_{P}^{(p,\ell)}\mathbf{P}^T$.
\State $\boldsymbol{\beta}_{i}^{(p,\ell)}=\boldsymbol{z}_i^{(p)}-\left(\boldsymbol{e}_{I}^{(p,\ell)},\boldsymbol{e}_{P}^{(p,\ell)}\right)$.
\State Extend the partial path $\boldsymbol{\beta}_{1 \rightarrow (i-1)}^{(p)}$ to $\boldsymbol{\beta}_{1 \rightarrow i}^{(p,\ell)}$, where $\boldsymbol{\beta}_{1 \rightarrow i}^{(p,\ell)} = \left(\boldsymbol{\beta}_{1 \rightarrow (i-1)}^{(p)}, \boldsymbol{\beta}_i^{(p,\ell)}\right)$.
\If{$\lambda\left(\boldsymbol{\beta}_{1 \rightarrow i}^{(p,\ell)}\right)<\lambda\left(\boldsymbol{f}^{(L_i)}\right)$}
\State Update the list $\mathcal{L}_i$ by removing $\boldsymbol{f}^{(L_i)}$ and inserting $\boldsymbol{\beta}_{1 \rightarrow i}^{(p,\ell)}$.
\EndIf
\EndIf
\EndIf
\EndFor
\EndFor
\Ensure The list $\mathcal{L}_i$ and the corresponding path metrics $\mathcal{P}_{i}$.
\end{algorithmic}
\label{alg1}
\end{algorithm}
\vspace{0.25cm}

\subsection{Numerical Results}
In this subsection, we simulate 5G polar codes~\cite{ref2ebmm} over BPSK-AWGN channels and validate our theoretical analysis in Section~IV-C. Furthermore, with the proposed early stopping criteria, we show the decoding latency reduction of the proposed decoding algorithm compared with
existing SCL decoding algorithms. Specifically, the GCD node $\mathscr{C}[n,k]$ with relatively low rate $k\leq \log L$ is simply decoded by ESD, which does not require the in-order generation of $\boldsymbol{e}_P$.

\subsubsection{Performance Comparisons}
\begin{figure}[!t]
\centering
\includegraphics[width=4.2in]{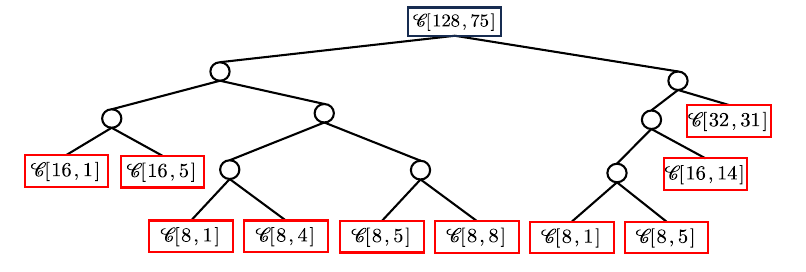}
\caption{A pruned polar tree of the 5G polar code $\mathscr{C}[128,64]$ with an 11-bit CRC, where the red box denotes the GCD node. The root is for $\mathscr{C}[128,75]$ due to the existence of the CRC.}
\label{Fig_partition_example128_64_10nodes}
\end{figure}

\begin{figure}[!t]
\centering
\includegraphics[width=3.4in]{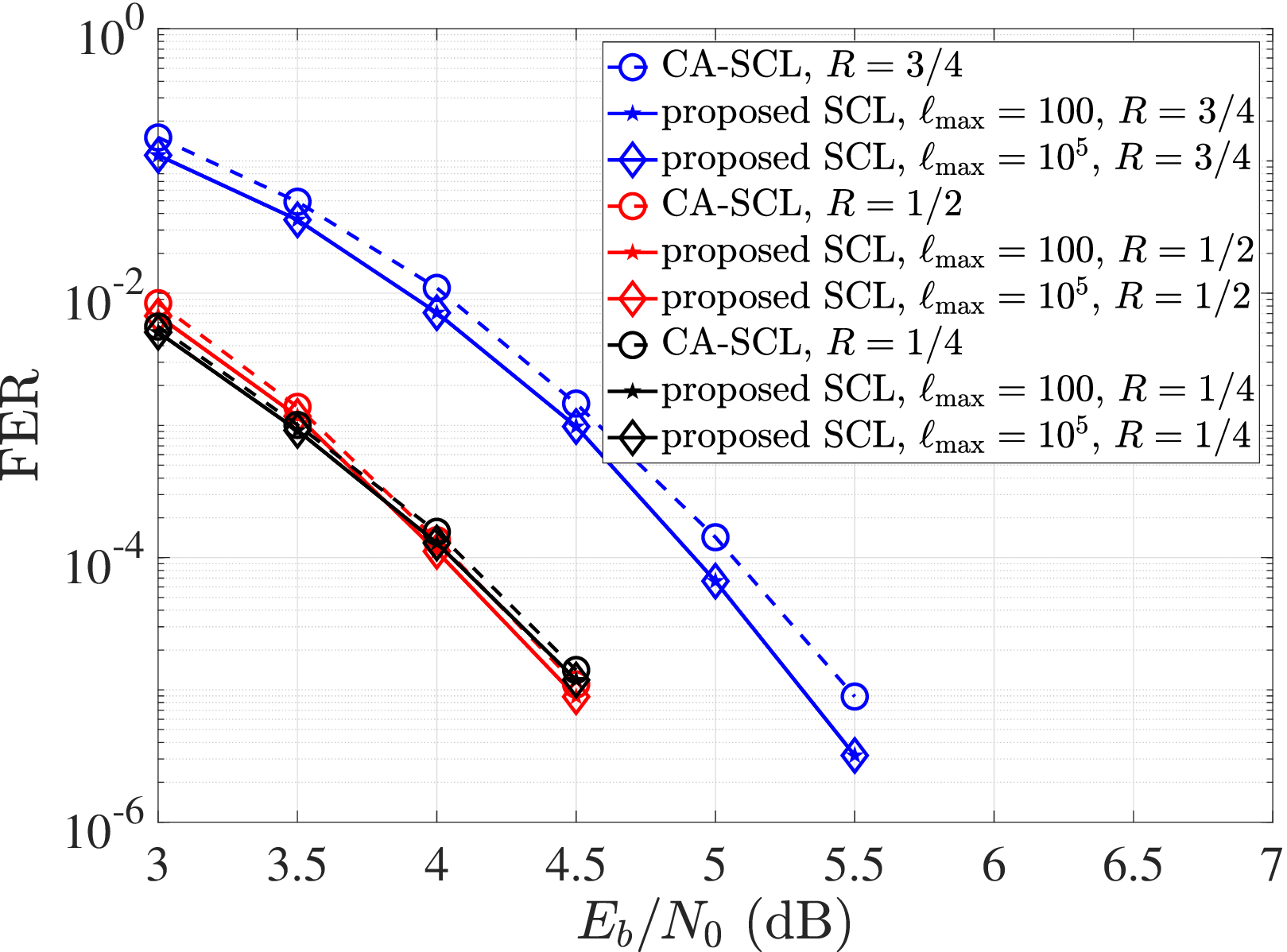}
\caption{Performance comparisons between the proposed SCL decoding and the CA-SCL decoding for the 5G polar codes of length $N=128$. Here, the list size $L=8$, $R \in \{ 1/4$, $1/2$, $3/4 \}$ and the maximum query number $\ell_{\text{max}}$ is $100$ or $10^5$.}
\label{fer_N128}
\end{figure}

\begin{figure}[!t]
\centering
\includegraphics[width=3.4in]{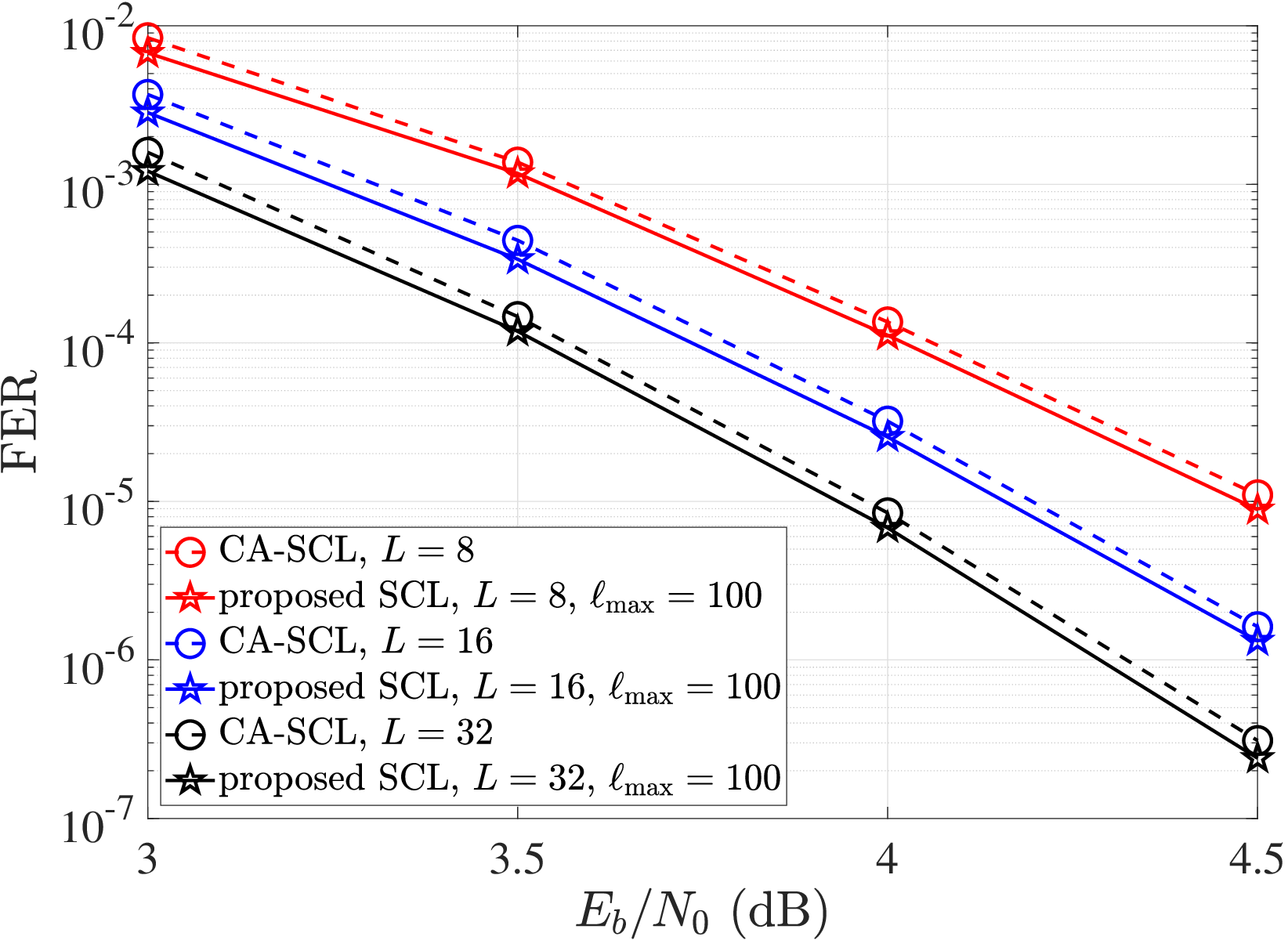}
\caption{Performance comparisons between the proposed SCL decoding and the CA-SCL decoding for the 5G polar code $\mathscr{C}[128,64]$ with an 11-bit CRC. Here, the list size $L \in \{8$, $16$, $32 \}$ and the maximum query number $\ell_{\text{max}}$ is $100$.}
\label{fer_N128K64}
\end{figure}

\begin{example}
Consider the 5G polar codes with an 11-bit CRC
$\mathscr{C}[128,32]$, $\mathscr{C}[128,64]$, and $\mathscr{C}[128,96]$, where the generator polynomial for the CRC is $x^{11} + x^{10} + x^9 + x^5 + 1$.  Fig.~\ref{Fig_partition_example128_64_10nodes} shows the pruned tree of $\mathscr{C}[128,64]$ with an 11-bit CRC, which has $10$ GCD nodes. For $\mathscr{C}[128,96]$ with an 11-bit CRC, the pruned tree has $7$ GCD nodes. For $\mathscr{C}[128,32]$ with an 11-bit CRC, the pruned tree has $11$ GCD nodes. We compare the proposed multiple-bit-wise SCL decoding algorithm and the CA-SCL decoding algorithm under a list size $L\in \{8,16,32 \}$. For $N=128$ and $R \in \{ 1/4$, $1/2$, $3/4 \}$, the performance comparisons are shown in Fig.~\ref{fer_N128}. For $\mathscr{C}[128,64]$ under different list sizes,  the performance comparisons are shown in Fig.~\ref{fer_N128K64}. As shown in Figs.~\ref{fer_N128}-\ref{fer_N128K64}, we can see that the performance of the proposed decoding algorithm is no worse than~(even slightly better than) that of the SCL decoding algorithm for different code rates under different list sizes. We also notice that the performance cannot be improved further by increasing the maximum number of queries $\ell_{\text{max}}$ from $100$ to $10^5$. Thus $\ell_{\text{max}} = 100$ is sufficient and we fix $\ell_{\text{max}} = 100$ in the following simulations.
\end{example}

\begin{figure}[!t]
\centering
\includegraphics[width=3.4in]{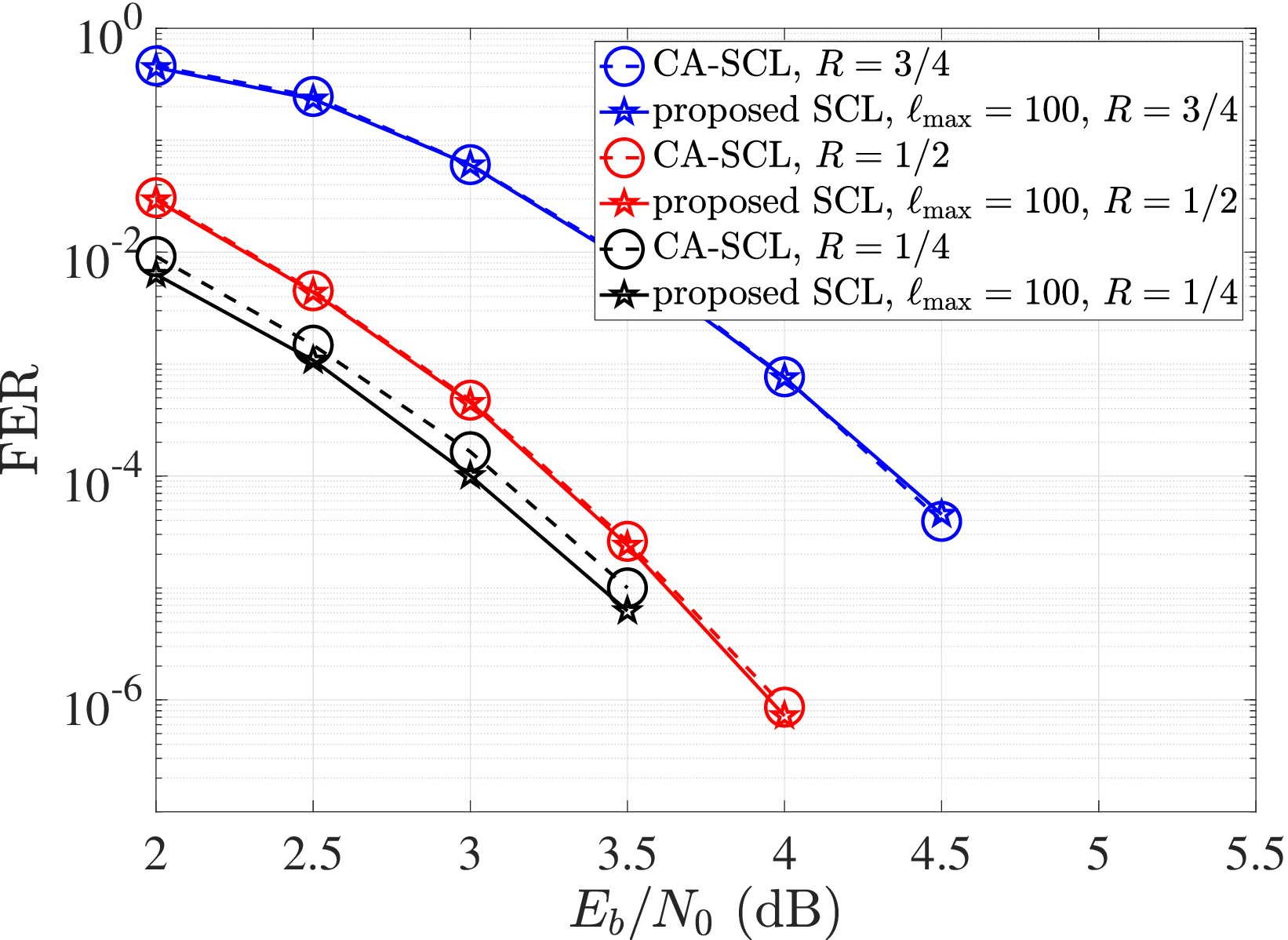}
\caption{Performance comparisons between the proposed SCL decoding and the conventional SCL decoding for the 5G polar codes of length $N=256$. Here, the list size $L=8$, $R \in \{ 1/4$, $1/2$, $3/4 \}$ and the maximum query number $\ell_{\text{max}}$ is $100$.}
\label{fer_N256}
\end{figure}

\begin{figure}[!t]
\centering
\includegraphics[width=3.4in]{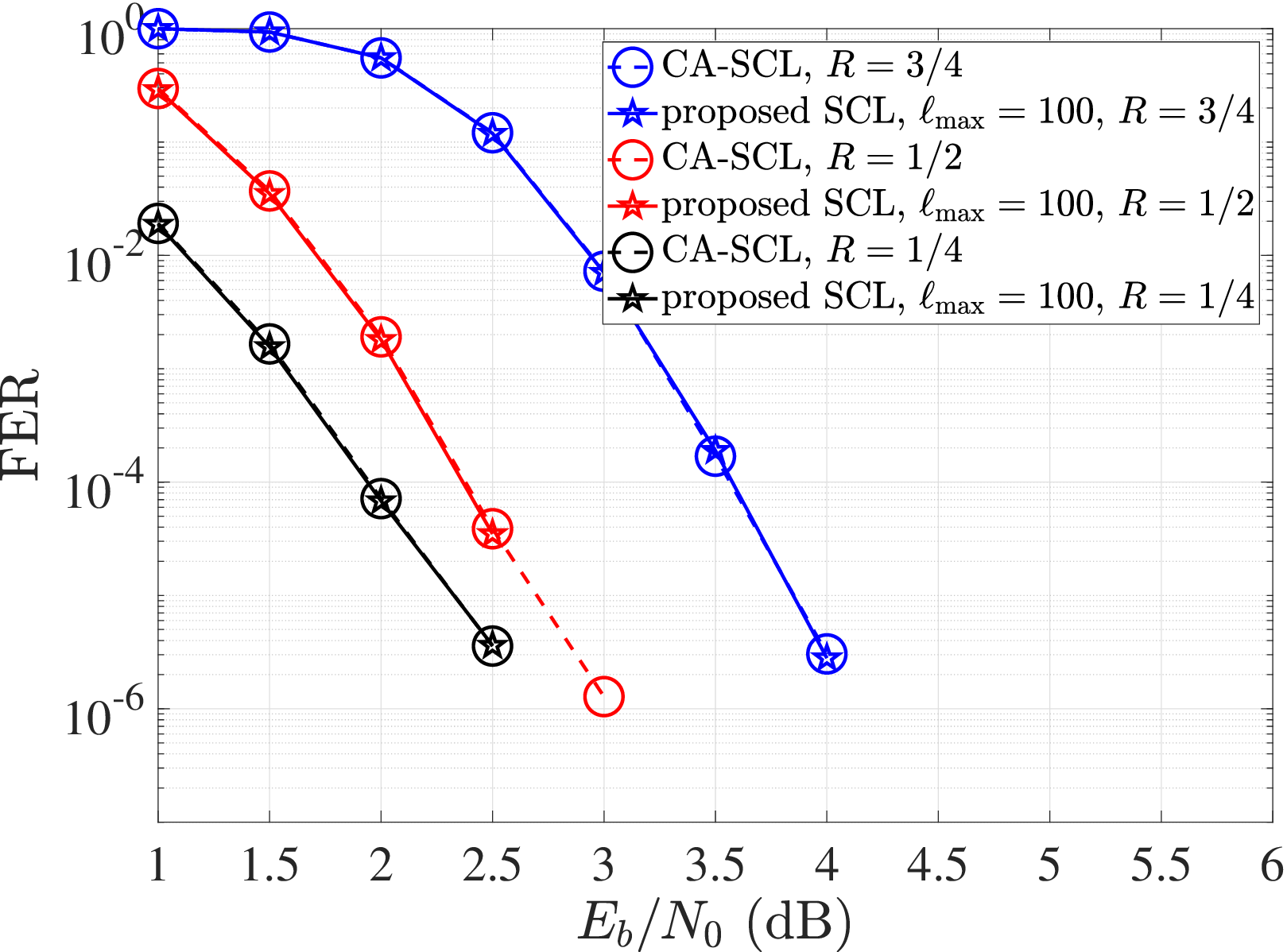}
\caption{Performance comparisons between the proposed SCL decoding and the conventional SCL decoding for the 5G polar codes of length $N=1024$. Here, the list size $L=8$, $R \in \{ 1/4$, $1/2$, $3/4 \}$ and  the maximum query number $\ell_{\text{max}}$ is $100$.}
\label{fer_N512}
\end{figure}

\begin{example}
Consider the 5G polar codes with an 11-bit CRC for $N \in \{256, 1024\}$ and $R \in \{ 1/4$, $1/2$, $3/4 \}$. The pruned tree of the 5G polar code $\mathscr{C}[256,193]$ consists of $15$ GCD nodes, where the right sub-tree contains $3$ GCD nodes $\mathscr{C}[32,26]$, $\mathscr{C}[32,32]$, and $\mathscr{C}[64,64]$. The simulation results regarding the performance are shown in Figs.~\ref{fer_N256}-\ref{fer_N512}.
It can be seen that, for these code rates and code lengths, the performance of the proposed SCL decoding algorithm by GCD is no worse than~(even slightly better than) that of the CA-SCL decoding algorithm. This suggests that the proposed algorithm is universal, as expected. 
\end{example}

\subsubsection{Decoding Latency Comparisons}
We now compare the decoding latency of the proposed decoding algorithm with that of the CA-SCL decoding~\cite{ref3SCL} and the fast SCL decoding~\cite{2019ArdakaniFastSCL}. Specifically, we compute the required total number of time steps to decode different nodes under the following assumptions similar to~\cite{HashemiFSSCL,2019ArdakaniFastSCL}. First, we assume there is no resource limitation so that all the parallelizable instructions are performed in one clock cycle. Second, the LLR calculations in \eqref{ffun} and \eqref{gfun} consume one time step, while the hard decision on LLRs and bit operations in \eqref{xor} and \eqref{SCLOESDhard} are carried out instantaneously. Third, the decoder extends all $L$ paths to $2L$ paths, sorts
the corresponding $2L$ paths based on the PMs, and selects the smallest $L$ ones during one time step.
For our proposed decoding algorithm, the GCD node $\mathscr{C} [n,k]$ with $k \leq \log L$ requires $k+1$ time steps similar to the commonly used nodes like REP, Type-I and Type-II nodes, and the GCD node $\mathscr{C} [n,k]$ with relatively high rate requires $ \ell_{\rm{avg}}  + \max\left(1,\frac{n}{2L}\right)$ time steps, where sorting $n$ bits in an order of increasing reliabilities requires $\max(1,\frac{n}{2L})$ time steps and each query requires one time step. Table \ref{tab1} shows the numbers of decoding time steps for $N = 128$ with a list size $L = 32$.  Table \ref{tab2} shows the numbers of decoding time steps for $N = 256$ with a list size $L = 32$.  Table \ref{tab3} shows the numbers of decoding time steps for $N = 1024$ with a list size $L = 32$.

As we can see, the proposed decoding algorithm requires fewer time steps than the SCL decoding algorithm~\cite{ref3SCL} and has a significant reduction in decoding latency compared with the fast SCL decoding algorithm~\cite{2019ArdakaniFastSCL} for the 5G polar codes with $N\in \{128,256,1024\}$ and $R \in \{1/4,1/2,3/4 \}$ under the list size $L = 32$. The number of decoding time steps for high-rate nodes in~\cite{2019ArdakaniFastSCL} is related to the list size and the code length. In contrast, the number of decoding time steps for GCD nodes in the proposed decoding algorithm mainly relies on the query number which is small in the high SNR region. For example, under a list size $L = 32$, the rate-1 node $\mathscr{C}[128,128]$ requires $32$ time steps in~\cite{2019ArdakaniFastSCL} while in the proposed decoding algorithm, as a GCD node, it requires $\ell_{\rm{avg}}  + 2$ time steps which are generally less than $10$ in the high SNR region.


\textbf{Remark.} It is worth pointing out that the decoding latency of the conventional SCL decoding algorithm is irrelevant to the SNR, while that of the proposed decoding algorithm varies with the SNR due to the trial-and-error nature of the GCD. For applications in 6G with reliability requirement at the level of $99.99999\%$ where the SNR is relatively high and the errors are typically sparse in the noisy received vector, one or two queries for the GCD is sufficient, suggesting that the proposed decoding algorithm can be very efficient.

\begin{table}[!t]
\begin{center}
\caption{The numbers of decoding time steps for $N = 128$ with a list size $L = 32$.}
\label{tab1}
\begin{tabular}{ c | c  c  c }
\hline
R & $1/4$ & $1/2$ &  $3/4$ \\
\hline
SCL~\cite{ref3SCL} & 297  & 329 & 361 \\
Fast SCL~\cite{2019ArdakaniFastSCL} & 52 & 75 & 114\\
Our proposed decoder~(3.0~dB) & 52 & 58 & 87 \\
Our proposed decoder~(4.5~dB) & 50 & 51 & 84 \\
\hline 
\end{tabular}
\end{center}
\end{table}

\begin{table}[!t]
\begin{center}
\caption{The numbers of decoding time steps for $N = 256$ with a list size $L = 32$.}
\label{tab2}
\begin{tabular}{ c | c  c  c }
\hline
R & $1/4$ & $1/2$ &  $3/4$ \\
\hline
SCL~\cite{ref3SCL} & 585  &  649 & 713\\
Fast SCL~\cite{2019ArdakaniFastSCL} & 94 & 159 & 190 \\
Our proposed decoder~(3.0~dB) & 64 & 108 & 170 \\
Our proposed decoder~(4.5~dB) & 62 & 107 & 170 \\
\hline 
\end{tabular}
\end{center}
\end{table}

\begin{table}[!t]
\begin{center}
\caption{The numbers of decoding time steps for $N = 1024$ with a list size $L = 32$.}
\label{tab3}
\begin{tabular}{ c | c  c  c }
\hline
R & $1/4$ & $1/2$ &  $3/4$ \\
\hline
SCL~\cite{ref3SCL} & 2313 & 2569 & 2825 \\
Fast SCL~\cite{2019ArdakaniFastSCL} & 283 & 460 & 784 \\
Our proposed decoder~(3.0~dB) & 226 & 380 &  416 \\
Our proposed decoder~(4.5~dB) & 221 & 334 &  330 \\
\hline 
\end{tabular}
\end{center}
\end{table}

\subsubsection{A Rethinking of Construction for Polar Codes}
As we know, the construction of polar codes is almost equivalent to ranking the bit-channels according to their reliabilities, which depend heavily on the bit-by-bit SC decoding algorithm. An immediate question arises:  Given the proposed multiple-bit-wise decoding algorithm, is it feasible to improve the SC-based construction? The answer is intuitively positive, but we have not found any systematic approaches. Heuristically, we may move some active bits from the left sub-tree to the right sub-tree, increasing the rate of the GCD node and fully exploiting the potential of the GCD algorithm, since GCD is more effective for the low/high-rate codes. To illustrate this idea, we present the following example.

\begin{figure}[!t]
\centering
\includegraphics[width=4.2in]{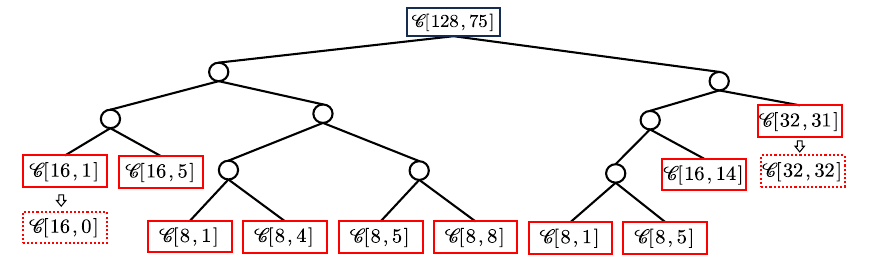}
\caption{A pruned polar tree and its improved counterpart of the 5G polar code $\mathscr{C}[128,64]$ with an 11-bit CRC, where the red solid box denotes the GCD node of the 5G polar code $\mathscr{C}[128,64]$ and the red dash box denotes the GCD node in the improved counterpart, respectively.}
\label{Fig_partition_example128_64}
\end{figure}

\begin{figure}[!t]
\centering
\includegraphics[width=3.4in]{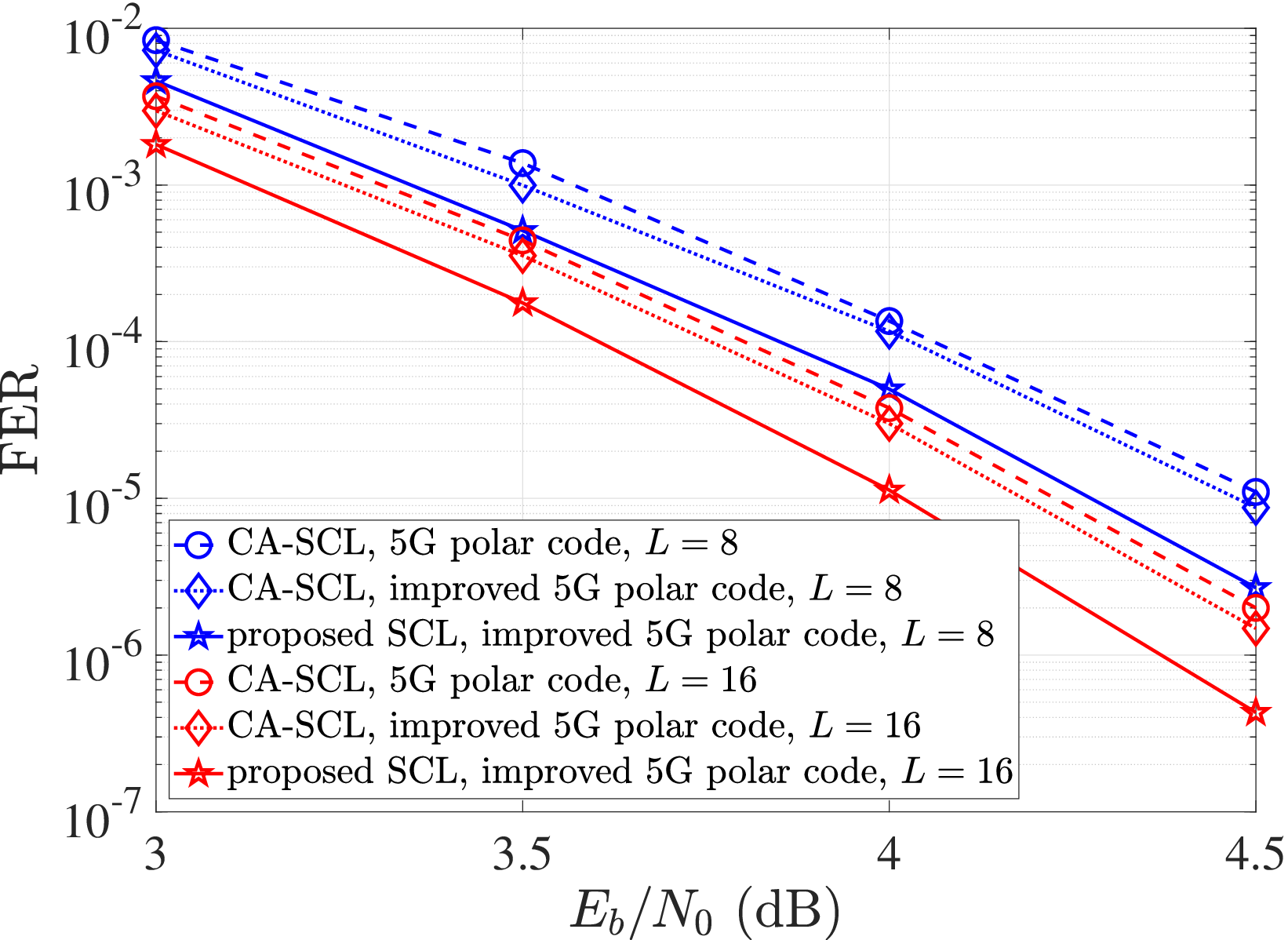}
\caption{Performance comparisons between the proposed SCL decoding and the CA-SCL decoding for the 5G polar code $\mathscr{C}[128,64]$ with an 11-bit CRC and the corresponding improved 5G polar code $\mathscr{C}'[128,64]$. Here, the list size $L \in \{ 8$, $16\}$.}
\label{fer_N128_redistribution}
\end{figure}

\begin{example}
Consider the 5G polar code with an 11-bit CRC
$\mathscr{C}[128,64]$, where the generator polynomial for the CRC is $x^{11} + x^{10} + x^9 + x^5 + 1$.  A pruned tree is illustrated in Fig.~\ref{Fig_partition_example128_64} 
which consists of $10$ GCD nodes. We then move the information bit allocated at the least reliable active channel~(belong to the leftmost leaf node) to the most reliable frozen channel~(belong to the rightmost leaf node). The resulting code is denoted as $\mathscr{C}'[128,64]$, whose pruned tree is slightly different from that of the 5G $\mathscr{C}[128,64]$. The simulation results are shown in Fig.~\ref{fer_N128_redistribution},
from which we can see that, the performance of $\mathscr{C}'[128,64]$ has nearly $0.2$ dB gain over that of the original 5G polar code $\mathscr{C}[128,64]$. 
\end{example}

\section{Conclusions}
In this paper, we have analyzed the GCD and presented three conditions for truncation, resulting in the truncated GCD. To analyze the upper bound on the performance gap between the truncated GCD and the GCD, we utilize the saddlepoint approach which does not depend on the specific code structure. Simulation results show that the approximation is accurate, which provides the guideline on the choice of the decoding parameters. It is also shown that the truncated GCD can reduce the complexity in terms of the number of TEPs, without noticeable performance loss in comparison with the GCD.  To further reduce the decoding latency, we have proposed a parallel framework for the (truncated) GCD. These decoding techniques can find applications not only to decoding of short block codes in ultra-reliable and low-latency communications but also to decoding of long codes constructed from short block codes. When applied to decoding of polar codes, the proposed multiple-bit-wise
SCL decoding by GCD incurs no performance loss but exhibits significant latency advantages compared with the existing SCL decoding.

\bibliographystyle{IEEEtran}
\bibliography{ref}

\end{document}